\documentclass[twocolumn,traditabstract,longauth]{aa}
\usepackage{graphicx,amsmath}
\usepackage{txfonts}
\usepackage[breaklinks, colorlinks, citecolor=blue]{hyperref}
\usepackage{url}
\usepackage{natbib}
\bibpunct{(}{)}{;}{a}{}{,} % to follow the A&A style

\def\setsymbol#1#2{\expandafter\def\csname #1\endcsname{#2}}
\def\getsymbol#1{\csname #1\endcsname}

\def\Planck{{\it Planck\/}}

\def\HeJT{$^4$He-JT}

\def\allearlypapers{\nocite{planck2011-1.1, planck2011-1.3, planck2011-1.4, planck2011-1.5, planck2011-1.6, planck2011-1.7, planck2011-1.10, planck2011-1.10sup, planck2011-5.1a, planck2011-5.1b, planck2011-5.2a, planck2011-5.2b, planck2011-5.2c, planck2011-6.1, planck2011-6.2, planck2011-6.3a, planck2011-6.4a, planck2011-6.4b, planck2011-6.6, planck2011-7.0, planck2011-7.2, planck2011-7.3, planck2011-7.7a, planck2011-7.7b, planck2011-7.12, planck2011-7.13}}

\newbox\tablebox    \newdimen\tablewidth
\def\leaderfil{\leaders\hbox to 5pt{\hss.\hss}\hfil}
\def\endPlancktable{\tablewidth=\columnwidth 
    $$\hss\copy\tablebox\hss$$
    \vskip-\lastskip\vskip -2pt}
\def\endPlancktablewide{\tablewidth=\textwidth 
    $$\hss\copy\tablebox\hss$$
    \vskip-\lastskip\vskip -2pt}
\def\tablenote#1 #2\par{\begingroup \parindent=0.8em
    \abovedisplayshortskip=0pt\belowdisplayshortskip=0pt
    \noindent
    $$\hss\vbox{\hsize\tablewidth \hangindent=\parindent \hangafter=1 \noindent
    \hbox to \parindent{\sup{\rm #1}\hss}\strut#2\strut\par}\hss$$
    \endgroup}
\def\doubleline{\vskip 3pt\hrule \vskip 1.5pt \hrule \vskip 5pt}

\def\L2{\ifmmode L_2\else $L_2$\fi}

\def\DeltaT{\ifmmode \Delta T\else $\Delta T$\fi}
\def\deltat{\ifmmode \Delta t\else $\Delta t$\fi}
\def\fknee{\ifmmode f_{\rm knee}\else $f_{\rm knee}$\fi}
\def\Fmax{\ifmmode F_{\rm max}\else $F_{\rm max}$\fi}
\def\solar{\ifmmode{\rm M}_{\mathord\odot}\else${\rm M}_{\mathord\odot}$\fi}

\def\inv{\ifmmode^{-1}\else$^{-1}$\fi}
\def\mo{\ifmmode^{-1}\else$^{-1}$\fi}
\def\sup#1{\ifmmode ^{\rm #1}\else $^{\rm #1}$\fi}
\def\expo#1{\ifmmode \times 10^{#1}\else $\times 10^{#1}$\fi}
\def\,{\thinspace}
\def\lsim{\mathrel{\raise .4ex\hbox{\rlap{$<$}\lower 1.2ex\hbox{$\sim$}}}}
\def\gsim{\mathrel{\raise .4ex\hbox{\rlap{$>$}\lower 1.2ex\hbox{$\sim$}}}}

\def\simprop{\mathrel{\raise .4ex\hbox{\rlap{$\propto$}\lower 1.2ex\hbox{$\sim$}}}}
\def\deg{\ifmmode^\circ\else$^\circ$\fi}
\def\pdeg{\ifmmode $\setbox0=\hbox{$^{\circ}$}\rlap{\hskip.11\wd0 .}$^{\circ}
          \else \setbox0=\hbox{$^{\circ}$}\rlap{\hskip.11\wd0 .}$^{\circ}$\fi}
\def\arcs{\ifmmode {^{\scriptstyle\prime\prime}}
          \else $^{\scriptstyle\prime\prime}$\fi}
\def\arcm{\ifmmode {^{\scriptstyle\prime}}
          \else $^{\scriptstyle\prime}$\fi}
\newdimen\sa  \newdimen\sb
\def\parcs{\sa=.07em \sb=.03em
     \ifmmode \hbox{\rlap{.}}^{\scriptstyle\prime\kern -\sb\prime}\hbox{\kern -\sa}
     \else \rlap{.}$^{\scriptstyle\prime\kern -\sb\prime}$\kern -\sa\fi}
\def\parcm{\sa=.08em \sb=.03em
     \ifmmode \hbox{\rlap{.}\kern\sa}^{\scriptstyle\prime}\hbox{\kern-\sb}
     \else \rlap{.}\kern\sa$^{\scriptstyle\prime}$\kern-\sb\fi}
\def\ra[#1 #2 #3.#4]{#1\sup{h}#2\sup{m}#3\sup{s}\llap.#4}
\def\dec[#1 #2 #3.#4]{#1\deg#2\arcm#3\arcs\llap.#4}
\def\deco[#1 #2 #3]{#1\deg#2\arcm#3\arcs}
\def\rra[#1 #2]{#1\sup{h}#2\sup{m}}

\def\dots{\relax\ifmmode \ldots\else $\ldots$\fi}
\def\WHzsr{\ifmmode $W\,Hz\mo\,sr\mo$\else W\,Hz\mo\,sr\mo\fi}
\def\mHz{\ifmmode $\,mHz$\else \,mHz\fi}
\def\GHz{\ifmmode $\,GHz$\else \,GHz\fi}
\def\mKs{\ifmmode $\,mK\,s$^{1/2}\else \,mK\,s$^{1/2}$\fi}
\def\muKs{\ifmmode \,\mu$K\,s$^{1/2}\else \,$\mu$K\,s$^{1/2}$\fi}
\def\muKRJs{\ifmmode \,\mu$K$_{\rm RJ}$\,s$^{1/2}\else \,$\mu$K$_{\rm RJ}$\,s$^{1/2}$\fi}
\def\muKHz{\ifmmode \,\mu$K\,Hz$^{-1/2}\else \,$\mu$K\,Hz$^{-1/2}$\fi}
\def\MJysr{\ifmmode \,$MJy\,sr\mo$\else \,MJy\,sr\mo\fi}
\def\MJysrmK{\ifmmode \,$MJy\,sr\mo$\,mK$_{\rm CMB}\mo\else \,MJy\,sr\mo\,mK$_{\rm CMB}\mo$\fi}
\def\microns{\ifmmode \,\mu$m$\else \,$\mu$m\fi}

\def\muK{\ifmmode \,\mu$K$\else \,$\mu$\hbox{K}\fi}
\def\microK{\ifmmode \,\mu$K$\else \,$\mu$\hbox{K}\fi}
\def\muW{\ifmmode \,\mu$W$\else \,$\mu$\hbox{W}\fi}
\def\kms{\ifmmode $\,km\,s$^{-1}\else \,km\,s$^{-1}$\fi}
\def\kmsMpc{\ifmmode $\,\kms\,Mpc\mo$\else \,\kms\,Mpc\mo\fi}
\setsymbol{LFI:center:frequency:70GHz:units}{70.3\,GHz}
\setsymbol{LFI:center:frequency:44GHz:units}{44.1\,GHz}
\setsymbol{LFI:center:frequency:30GHz:units}{28.5\,GHz}

\setsymbol{LFI:center:frequency:70GHz}{70.3}
\setsymbol{LFI:center:frequency:44GHz}{44.1}
\setsymbol{LFI:center:frequency:30GHz}{28.5}

\setsymbol{LFI:center:frequency:LFI18:Rad:M:units}{71.7\GHz}
\setsymbol{LFI:center:frequency:LFI19:Rad:M:units}{67.5\GHz}
\setsymbol{LFI:center:frequency:LFI20:Rad:M:units}{69.2\GHz}
\setsymbol{LFI:center:frequency:LFI21:Rad:M:units}{70.4\GHz}
\setsymbol{LFI:center:frequency:LFI22:Rad:M:units}{71.5\GHz}
\setsymbol{LFI:center:frequency:LFI23:Rad:M:units}{70.8\GHz}
\setsymbol{LFI:center:frequency:LFI24:Rad:M:units}{44.4\GHz}
\setsymbol{LFI:center:frequency:LFI25:Rad:M:units}{44.0\GHz}
\setsymbol{LFI:center:frequency:LFI26:Rad:M:units}{43.9\GHz}
\setsymbol{LFI:center:frequency:LFI27:Rad:M:units}{28.3\GHz}
\setsymbol{LFI:center:frequency:LFI28:Rad:M:units}{28.8\GHz}
\setsymbol{LFI:center:frequency:LFI18:Rad:S:units}{70.1\GHz}
\setsymbol{LFI:center:frequency:LFI19:Rad:S:units}{69.6\GHz}
\setsymbol{LFI:center:frequency:LFI20:Rad:S:units}{69.5\GHz}
\setsymbol{LFI:center:frequency:LFI21:Rad:S:units}{69.5\GHz}
\setsymbol{LFI:center:frequency:LFI22:Rad:S:units}{72.8\GHz}
\setsymbol{LFI:center:frequency:LFI23:Rad:S:units}{71.3\GHz}
\setsymbol{LFI:center:frequency:LFI24:Rad:S:units}{44.1\GHz}
\setsymbol{LFI:center:frequency:LFI25:Rad:S:units}{44.1\GHz}
\setsymbol{LFI:center:frequency:LFI26:Rad:S:units}{44.1\GHz}
\setsymbol{LFI:center:frequency:LFI27:Rad:S:units}{28.5\GHz}
\setsymbol{LFI:center:frequency:LFI28:Rad:S:units}{28.2\GHz}

\setsymbol{LFI:center:frequency:LFI18:Rad:M}{71.7}
\setsymbol{LFI:center:frequency:LFI19:Rad:M}{67.5}
\setsymbol{LFI:center:frequency:LFI20:Rad:M}{69.2}
\setsymbol{LFI:center:frequency:LFI21:Rad:M}{70.4}
\setsymbol{LFI:center:frequency:LFI22:Rad:M}{71.5}
\setsymbol{LFI:center:frequency:LFI23:Rad:M}{70.8}
\setsymbol{LFI:center:frequency:LFI24:Rad:M}{44.4}
\setsymbol{LFI:center:frequency:LFI25:Rad:M}{44.0}
\setsymbol{LFI:center:frequency:LFI26:Rad:M}{43.9}
\setsymbol{LFI:center:frequency:LFI27:Rad:M}{28.3}
\setsymbol{LFI:center:frequency:LFI28:Rad:M}{28.8}
\setsymbol{LFI:center:frequency:LFI18:Rad:S}{70.1}
\setsymbol{LFI:center:frequency:LFI19:Rad:S}{69.6}
\setsymbol{LFI:center:frequency:LFI20:Rad:S}{69.5}
\setsymbol{LFI:center:frequency:LFI21:Rad:S}{69.5}
\setsymbol{LFI:center:frequency:LFI22:Rad:S}{72.8}
\setsymbol{LFI:center:frequency:LFI23:Rad:S}{71.3}
\setsymbol{LFI:center:frequency:LFI24:Rad:S}{44.1}
\setsymbol{LFI:center:frequency:LFI25:Rad:S}{44.1}
\setsymbol{LFI:center:frequency:LFI26:Rad:S}{44.1}
\setsymbol{LFI:center:frequency:LFI27:Rad:S}{28.5}
\setsymbol{LFI:center:frequency:LFI28:Rad:S}{28.2}

\setsymbol{LFI:white:noise:sensitivity:70GHz:units}{134.7\muKs}
\setsymbol{LFI:white:noise:sensitivity:44GHz:units}{164.7\muKs}
\setsymbol{LFI:white:noise:sensitivity:30GHz:units}{143.4\muKs}

\setsymbol{LFI:white:noise:sensitivity:70GHz}{134.7}
\setsymbol{LFI:white:noise:sensitivity:44GHz}{164.7}
\setsymbol{LFI:white:noise:sensitivity:30GHz}{143.4}

\setsymbol{LFI:white:noise:sensitivity:LFI18:Rad:M:units}{512.0\muKs}
\setsymbol{LFI:white:noise:sensitivity:LFI19:Rad:M:units}{581.4\muKs}
\setsymbol{LFI:white:noise:sensitivity:LFI20:Rad:M:units}{590.8\muKs}
\setsymbol{LFI:white:noise:sensitivity:LFI21:Rad:M:units}{455.2\muKs}
\setsymbol{LFI:white:noise:sensitivity:LFI22:Rad:M:units}{492.0\muKs}
\setsymbol{LFI:white:noise:sensitivity:LFI23:Rad:M:units}{507.7\muKs}
\setsymbol{LFI:white:noise:sensitivity:LFI24:Rad:M:units}{462.2\muKs}
\setsymbol{LFI:white:noise:sensitivity:LFI25:Rad:M:units}{413.6\muKs}
\setsymbol{LFI:white:noise:sensitivity:LFI26:Rad:M:units}{478.6\muKs}
\setsymbol{LFI:white:noise:sensitivity:LFI27:Rad:M:units}{277.7\muKs}
\setsymbol{LFI:white:noise:sensitivity:LFI28:Rad:M:units}{312.3\muKs}
\setsymbol{LFI:white:noise:sensitivity:LFI18:Rad:S:units}{465.7\muKs}
\setsymbol{LFI:white:noise:sensitivity:LFI19:Rad:S:units}{555.6\muKs}
\setsymbol{LFI:white:noise:sensitivity:LFI20:Rad:S:units}{623.2\muKs}
\setsymbol{LFI:white:noise:sensitivity:LFI21:Rad:S:units}{564.1\muKs}
\setsymbol{LFI:white:noise:sensitivity:LFI22:Rad:S:units}{534.4\muKs}
\setsymbol{LFI:white:noise:sensitivity:LFI23:Rad:S:units}{542.4\muKs}
\setsymbol{LFI:white:noise:sensitivity:LFI24:Rad:S:units}{399.2\muKs}
\setsymbol{LFI:white:noise:sensitivity:LFI25:Rad:S:units}{392.6\muKs}
\setsymbol{LFI:white:noise:sensitivity:LFI26:Rad:S:units}{418.6\muKs}
\setsymbol{LFI:white:noise:sensitivity:LFI27:Rad:S:units}{302.9\muKs}
\setsymbol{LFI:white:noise:sensitivity:LFI28:Rad:S:units}{285.3\muKs}

\setsymbol{LFI:white:noise:sensitivity:LFI18:Rad:M}{512.0}
\setsymbol{LFI:white:noise:sensitivity:LFI19:Rad:M}{581.4}
\setsymbol{LFI:white:noise:sensitivity:LFI20:Rad:M}{590.8}
\setsymbol{LFI:white:noise:sensitivity:LFI21:Rad:M}{455.2}
\setsymbol{LFI:white:noise:sensitivity:LFI22:Rad:M}{492.0}
\setsymbol{LFI:white:noise:sensitivity:LFI23:Rad:M}{507.7}
\setsymbol{LFI:white:noise:sensitivity:LFI24:Rad:M}{462.2}
\setsymbol{LFI:white:noise:sensitivity:LFI25:Rad:M}{413.6}
\setsymbol{LFI:white:noise:sensitivity:LFI26:Rad:M}{478.6}
\setsymbol{LFI:white:noise:sensitivity:LFI27:Rad:M}{277.7}
\setsymbol{LFI:white:noise:sensitivity:LFI28:Rad:M}{312.3}
\setsymbol{LFI:white:noise:sensitivity:LFI18:Rad:S}{465.7}
\setsymbol{LFI:white:noise:sensitivity:LFI19:Rad:S}{555.6}
\setsymbol{LFI:white:noise:sensitivity:LFI20:Rad:S}{623.2}
\setsymbol{LFI:white:noise:sensitivity:LFI21:Rad:S}{564.1}
\setsymbol{LFI:white:noise:sensitivity:LFI22:Rad:S}{534.4}
\setsymbol{LFI:white:noise:sensitivity:LFI23:Rad:S}{542.4}
\setsymbol{LFI:white:noise:sensitivity:LFI24:Rad:S}{399.2}
\setsymbol{LFI:white:noise:sensitivity:LFI25:Rad:S}{392.6}
\setsymbol{LFI:white:noise:sensitivity:LFI26:Rad:S}{418.6}
\setsymbol{LFI:white:noise:sensitivity:LFI27:Rad:S}{302.9}
\setsymbol{LFI:white:noise:sensitivity:LFI28:Rad:S}{285.3}

\setsymbol{LFI:knee:frequency:70GHz:units}{29.5\mHz}
\setsymbol{LFI:knee:frequency:44GHz:units}{56.2\mHz}
\setsymbol{LFI:knee:frequency:30GHz:units}{113.7\mHz}

\setsymbol{LFI:knee:frequency:70GHz}{29.5}
\setsymbol{LFI:knee:frequency:44GHz}{56.2}
\setsymbol{LFI:knee:frequency:30GHz}{113.7}

\setsymbol{LFI:knee:frequency:LFI18:Rad:M:units}{16.3\mHz}
\setsymbol{LFI:knee:frequency:LFI19:Rad:M:units}{15.1\mHz}
\setsymbol{LFI:knee:frequency:LFI20:Rad:M:units}{18.7\mHz}
\setsymbol{LFI:knee:frequency:LFI21:Rad:M:units}{37.2\mHz}
\setsymbol{LFI:knee:frequency:LFI22:Rad:M:units}{12.7\mHz}
\setsymbol{LFI:knee:frequency:LFI23:Rad:M:units}{34.6\mHz}
\setsymbol{LFI:knee:frequency:LFI24:Rad:M:units}{46.2\mHz}
\setsymbol{LFI:knee:frequency:LFI25:Rad:M:units}{24.9\mHz}
\setsymbol{LFI:knee:frequency:LFI26:Rad:M:units}{67.6\mHz}
\setsymbol{LFI:knee:frequency:LFI27:Rad:M:units}{187.4\mHz}
\setsymbol{LFI:knee:frequency:LFI28:Rad:M:units}{122.2\mHz}
\setsymbol{LFI:knee:frequency:LFI18:Rad:S:units}{17.7\mHz}
\setsymbol{LFI:knee:frequency:LFI19:Rad:S:units}{22.0\mHz}
\setsymbol{LFI:knee:frequency:LFI20:Rad:S:units}{8.7\mHz}
\setsymbol{LFI:knee:frequency:LFI21:Rad:S:units}{25.9\mHz}
\setsymbol{LFI:knee:frequency:LFI22:Rad:S:units}{15.8\mHz}
\setsymbol{LFI:knee:frequency:LFI23:Rad:S:units}{129.8\mHz}
\setsymbol{LFI:knee:frequency:LFI24:Rad:S:units}{100.9\mHz}
\setsymbol{LFI:knee:frequency:LFI25:Rad:S:units}{38.9\mHz}
\setsymbol{LFI:knee:frequency:LFI26:Rad:S:units}{58.9\mHz}
\setsymbol{LFI:knee:frequency:LFI27:Rad:S:units}{104.4\mHz}
\setsymbol{LFI:knee:frequency:LFI28:Rad:S:units}{40.7\mHz}

\setsymbol{LFI:knee:frequency:LFI18:Rad:M}{16.3}
\setsymbol{LFI:knee:frequency:LFI19:Rad:M}{15.1}
\setsymbol{LFI:knee:frequency:LFI20:Rad:M}{18.7}
\setsymbol{LFI:knee:frequency:LFI21:Rad:M}{37.2}
\setsymbol{LFI:knee:frequency:LFI22:Rad:M}{12.7}
\setsymbol{LFI:knee:frequency:LFI23:Rad:M}{34.6}
\setsymbol{LFI:knee:frequency:LFI24:Rad:M}{46.2}
\setsymbol{LFI:knee:frequency:LFI25:Rad:M}{24.9}
\setsymbol{LFI:knee:frequency:LFI26:Rad:M}{67.6}
\setsymbol{LFI:knee:frequency:LFI27:Rad:M}{187.4}
\setsymbol{LFI:knee:frequency:LFI28:Rad:M}{122.2}
\setsymbol{LFI:knee:frequency:LFI18:Rad:S}{17.7}
\setsymbol{LFI:knee:frequency:LFI19:Rad:S}{22.0}
\setsymbol{LFI:knee:frequency:LFI20:Rad:S}{8.7}
\setsymbol{LFI:knee:frequency:LFI21:Rad:S}{25.9}
\setsymbol{LFI:knee:frequency:LFI22:Rad:S}{15.8}
\setsymbol{LFI:knee:frequency:LFI23:Rad:S}{129.8}
\setsymbol{LFI:knee:frequency:LFI24:Rad:S}{100.9}
\setsymbol{LFI:knee:frequency:LFI25:Rad:S}{38.9}
\setsymbol{LFI:knee:frequency:LFI26:Rad:S}{58.9}
\setsymbol{LFI:knee:frequency:LFI27:Rad:S}{104.4}
\setsymbol{LFI:knee:frequency:LFI28:Rad:S}{40.7}

\setsymbol{LFI:slope:70GHz:units}{$-1.03$\mHz}
\setsymbol{LFI:slope:44GHz:units}{$-0.89$\mHz}
\setsymbol{LFI:slope:30GHz:units}{$-0.87$\mHz}

\setsymbol{LFI:slope:70GHz}{$-1.03$}
\setsymbol{LFI:slope:44GHz}{$-0.89$}
\setsymbol{LFI:slope:30GHz}{$-0.87$}

\setsymbol{LFI:slope:LFI18:Rad:M:units}{$-1.04$\mHz}
\setsymbol{LFI:slope:LFI19:Rad:M:units}{$-1.09$\mHz}
\setsymbol{LFI:slope:LFI20:Rad:M:units}{$-0.69$\mHz}
\setsymbol{LFI:slope:LFI21:Rad:M:units}{$-1.56$\mHz}
\setsymbol{LFI:slope:LFI22:Rad:M:units}{$-1.01$\mHz}
\setsymbol{LFI:slope:LFI23:Rad:M:units}{$-0.96$\mHz}
\setsymbol{LFI:slope:LFI24:Rad:M:units}{$-0.83$\mHz}
\setsymbol{LFI:slope:LFI25:Rad:M:units}{$-0.91$\mHz}
\setsymbol{LFI:slope:LFI26:Rad:M:units}{$-0.95$\mHz}
\setsymbol{LFI:slope:LFI27:Rad:M:units}{$-0.87$\mHz}
\setsymbol{LFI:slope:LFI28:Rad:M:units}{$-0.88$\mHz}
\setsymbol{LFI:slope:LFI18:Rad:S:units}{$-1.15$\mHz}
\setsymbol{LFI:slope:LFI19:Rad:S:units}{$-1.00$\mHz}
\setsymbol{LFI:slope:LFI20:Rad:S:units}{$-0.95$\mHz}
\setsymbol{LFI:slope:LFI21:Rad:S:units}{$-0.92$\mHz}
\setsymbol{LFI:slope:LFI22:Rad:S:units}{$-1.01$\mHz}
\setsymbol{LFI:slope:LFI23:Rad:S:units}{$-0.95$\mHz}
\setsymbol{LFI:slope:LFI24:Rad:S:units}{$-0.73$\mHz}
\setsymbol{LFI:slope:LFI25:Rad:S:units}{$-1.16$\mHz}
\setsymbol{LFI:slope:LFI26:Rad:S:units}{$-0.79$\mHz}
\setsymbol{LFI:slope:LFI27:Rad:S:units}{$-0.82$\mHz}
\setsymbol{LFI:slope:LFI28:Rad:S:units}{$-0.91$\mHz}

\setsymbol{LFI:slope:LFI18:Rad:M}{$-1.04$}
\setsymbol{LFI:slope:LFI19:Rad:M}{$-1.09$}
\setsymbol{LFI:slope:LFI20:Rad:M}{$-0.69$}
\setsymbol{LFI:slope:LFI21:Rad:M}{$-1.56$}
\setsymbol{LFI:slope:LFI22:Rad:M}{$-1.01$}
\setsymbol{LFI:slope:LFI23:Rad:M}{$-0.96$}
\setsymbol{LFI:slope:LFI24:Rad:M}{$-0.83$}
\setsymbol{LFI:slope:LFI25:Rad:M}{$-0.91$}
\setsymbol{LFI:slope:LFI26:Rad:M}{$-0.95$}
\setsymbol{LFI:slope:LFI27:Rad:M}{$-0.87$}
\setsymbol{LFI:slope:LFI28:Rad:M}{$-0.88$}
\setsymbol{LFI:slope:LFI18:Rad:S}{$-1.15$}
\setsymbol{LFI:slope:LFI19:Rad:S}{$-1.00$}
\setsymbol{LFI:slope:LFI20:Rad:S}{$-0.95$}
\setsymbol{LFI:slope:LFI21:Rad:S}{$-0.92$}
\setsymbol{LFI:slope:LFI22:Rad:S}{$-1.01$}
\setsymbol{LFI:slope:LFI23:Rad:S}{$-0.95$}
\setsymbol{LFI:slope:LFI24:Rad:S}{$-0.73$}
\setsymbol{LFI:slope:LFI25:Rad:S}{$-1.16$}
\setsymbol{LFI:slope:LFI26:Rad:S}{$-0.79$}
\setsymbol{LFI:slope:LFI27:Rad:S}{$-0.82$}
\setsymbol{LFI:slope:LFI28:Rad:S}{$-0.91$}

\setsymbol{LFI:FWHM:70GHz:units}{13\parcm01}
\setsymbol{LFI:FWHM:44GHz:units}{27\parcm92}
\setsymbol{LFI:FWHM:30GHz:units}{32\parcm65}

\setsymbol{LFI:FWHM:70GHz}{13.01}
\setsymbol{LFI:FWHM:44GHz}{27.92}
\setsymbol{LFI:FWHM:30GHz}{32.65}

\setsymbol{LFI:FWHM:LFI18:units}{13\parcm39}
\setsymbol{LFI:FWHM:LFI19:units}{13\parcm01}
\setsymbol{LFI:FWHM:LFI20:units}{12\parcm75}
\setsymbol{LFI:FWHM:LFI21:units}{12\parcm74}
\setsymbol{LFI:FWHM:LFI22:units}{12\parcm87}
\setsymbol{LFI:FWHM:LFI23:units}{13\parcm27}
\setsymbol{LFI:FWHM:LFI24:units}{22\parcm98}
\setsymbol{LFI:FWHM:LFI25:units}{30\parcm46}
\setsymbol{LFI:FWHM:LFI26:units}{30\parcm31}
\setsymbol{LFI:FWHM:LFI27:units}{32\parcm65}
\setsymbol{LFI:FWHM:LFI28:units}{32\parcm66}

\setsymbol{LFI:FWHM:LFI18}{13.39}
\setsymbol{LFI:FWHM:LFI19}{13.01}
\setsymbol{LFI:FWHM:LFI20}{12.75}
\setsymbol{LFI:FWHM:LFI21}{12.74}
\setsymbol{LFI:FWHM:LFI22}{12.87}
\setsymbol{LFI:FWHM:LFI23}{13.27}
\setsymbol{LFI:FWHM:LFI24}{22.98}
\setsymbol{LFI:FWHM:LFI25}{30.46}
\setsymbol{LFI:FWHM:LFI26}{30.31}
\setsymbol{LFI:FWHM:LFI27}{32.65}
\setsymbol{LFI:FWHM:LFI28}{32.66}

\setsymbol{LFI:FWHM:uncertainty:LFI18:units}{0.170\arcm}
\setsymbol{LFI:FWHM:uncertainty:LFI19:units}{0.174\arcm}
\setsymbol{LFI:FWHM:uncertainty:LFI20:units}{0.170\arcm}
\setsymbol{LFI:FWHM:uncertainty:LFI21:units}{0.156\arcm}
\setsymbol{LFI:FWHM:uncertainty:LFI22:units}{0.164\arcm}
\setsymbol{LFI:FWHM:uncertainty:LFI23:units}{0.171\arcm}
\setsymbol{LFI:FWHM:uncertainty:LFI24:units}{0.652\arcm}
\setsymbol{LFI:FWHM:uncertainty:LFI25:units}{1.075\arcm}
\setsymbol{LFI:FWHM:uncertainty:LFI26:units}{1.131\arcm}
\setsymbol{LFI:FWHM:uncertainty:LFI27:units}{1.266\arcm}
\setsymbol{LFI:FWHM:uncertainty:LFI28:units}{1.287\arcm}

\setsymbol{LFI:FWHM:uncertainty:LFI18}{0.170}
\setsymbol{LFI:FWHM:uncertainty:LFI19}{0.174}
\setsymbol{LFI:FWHM:uncertainty:LFI20}{0.170}
\setsymbol{LFI:FWHM:uncertainty:LFI21}{0.156}
\setsymbol{LFI:FWHM:uncertainty:LFI22}{0.164}
\setsymbol{LFI:FWHM:uncertainty:LFI23}{0.171}
\setsymbol{LFI:FWHM:uncertainty:LFI24}{0.652}
\setsymbol{LFI:FWHM:uncertainty:LFI25}{1.075}
\setsymbol{LFI:FWHM:uncertainty:LFI26}{1.131}
\setsymbol{LFI:FWHM:uncertainty:LFI27}{1.266}
\setsymbol{LFI:FWHM:uncertainty:LFI28}{1.287}

\setsymbol{HFI:center:frequency:100GHz:units}{100\,GHz}
\setsymbol{HFI:center:frequency:143GHz:units}{143\,GHz}
\setsymbol{HFI:center:frequency:217GHz:units}{217\,GHz}
\setsymbol{HFI:center:frequency:353GHz:units}{353\,GHz}
\setsymbol{HFI:center:frequency:545GHz:units}{545\,GHz}
\setsymbol{HFI:center:frequency:857GHz:units}{857\,GHz}

\setsymbol{HFI:center:frequency:100GHz}{100}
\setsymbol{HFI:center:frequency:143GHz}{143}
\setsymbol{HFI:center:frequency:217GHz}{217}
\setsymbol{HFI:center:frequency:353GHz}{353}
\setsymbol{HFI:center:frequency:545GHz}{545}
\setsymbol{HFI:center:frequency:857GHz}{857}

\setsymbol{HFI:Ndetectors:100GHz}{8}
\setsymbol{HFI:Ndetectors:143GHz}{11}
\setsymbol{HFI:Ndetectors:217GHz}{12}
\setsymbol{HFI:Ndetectors:353GHz}{12}
\setsymbol{HFI:Ndetectors:545GHz}{3}
\setsymbol{HFI:Ndetectors:857GHz}{4}

\setsymbol{HFI:FWHM:Maps:100GHz:units}{9\parcm88}
\setsymbol{HFI:FWHM:Maps:143GHz:units}{7\parcm18}
\setsymbol{HFI:FWHM:Maps:217GHz:units}{4\parcm87}
\setsymbol{HFI:FWHM:Maps:353GHz:units}{4\parcm65}
\setsymbol{HFI:FWHM:Maps:545GHz:units}{4\parcm72}
\setsymbol{HFI:FWHM:Maps:857GHz:units}{4\parcm39}
\setsymbol{HFI:FWHM:Maps:100GHz}{9.88}
\setsymbol{HFI:FWHM:Maps:143GHz}{7.18}
\setsymbol{HFI:FWHM:Maps:217GHz}{4.87}
\setsymbol{HFI:FWHM:Maps:353GHz}{4.65}
\setsymbol{HFI:FWHM:Maps:545GHz}{4.72}
\setsymbol{HFI:FWHM:Maps:857GHz}{4.39}

\setsymbol{HFI:beam:ellipticity:Maps:100GHz}{1.15}
\setsymbol{HFI:beam:ellipticity:Maps:143GHz}{1.01}
\setsymbol{HFI:beam:ellipticity:Maps:217GHz}{1.06}
\setsymbol{HFI:beam:ellipticity:Maps:353GHz}{1.05}
\setsymbol{HFI:beam:ellipticity:Maps:545GHz}{1.14}
\setsymbol{HFI:beam:ellipticity:Maps:857GHz}{1.19}

\setsymbol{HFI:FWHM:Mars:100GHz:units}{9\parcm37}
\setsymbol{HFI:FWHM:Mars:143GHz:units}{7\parcm04}
\setsymbol{HFI:FWHM:Mars:217GHz:units}{4\parcm68}
\setsymbol{HFI:FWHM:Mars:353GHz:units}{4\parcm43}
\setsymbol{HFI:FWHM:Mars:545GHz:units}{3\parcm80}
\setsymbol{HFI:FWHM:Mars:857GHz:units}{3\parcm67}

\setsymbol{HFI:FWHM:Mars:100GHz}{9.37}
\setsymbol{HFI:FWHM:Mars:143GHz}{7.04}
\setsymbol{HFI:FWHM:Mars:217GHz}{4.68}
\setsymbol{HFI:FWHM:Mars:353GHz}{4.43}
\setsymbol{HFI:FWHM:Mars:545GHz}{3.80}
\setsymbol{HFI:FWHM:Mars:857GHz}{3.67}

\setsymbol{HFI:beam:ellipticity:Mars:100GHz}{1.18}
\setsymbol{HFI:beam:ellipticity:Mars:143GHz}{1.03}
\setsymbol{HFI:beam:ellipticity:Mars:217GHz}{1.14}
\setsymbol{HFI:beam:ellipticity:Mars:353GHz}{1.09}
\setsymbol{HFI:beam:ellipticity:Mars:545GHz}{1.25}
\setsymbol{HFI:beam:ellipticity:Mars:857GHz}{1.03}

\setsymbol{HFI:CMB:relative:calibration:100GHz}{$\lsim 1\%$}
\setsymbol{HFI:CMB:relative:calibration:143GHz}{$\lsim 1\%$}
\setsymbol{HFI:CMB:relative:calibration:217GHz}{$\lsim 1\%$}
\setsymbol{HFI:CMB:relative:calibration:353GHz}{$\lsim 1\%$}
\setsymbol{HFI:CMB:relative:calibration:545GHz}{}
\setsymbol{HFI:CMB:relative:calibration:857GHz}{}

\setsymbol{HFI:CMB:absolute:calibration:100GHz}{$\lsim 2\%$}
\setsymbol{HFI:CMB:absolute:calibration:143GHz}{$\lsim 2\%$}
\setsymbol{HFI:CMB:absolute:calibration:217GHz}{$\lsim 2\%$}
\setsymbol{HFI:CMB:absolute:calibration:353GHz}{$\lsim 2\%$}
\setsymbol{HFI:CMB:absolute:calibration:545GHz}{}
\setsymbol{HFI:CMB:absolute:calibration:857GHz}{}

\setsymbol{HFI:FIRAS:gain:calibration:accuracy:statistical:100GHz}{}
\setsymbol{HFI:FIRAS:gain:calibration:accuracy:statistical:143GHz}{}
\setsymbol{HFI:FIRAS:gain:calibration:accuracy:statistical:217GHz}{}
\setsymbol{HFI:FIRAS:gain:calibration:accuracy:statistical:353GHz}{2.5\%}
\setsymbol{HFI:FIRAS:gain:calibration:accuracy:statistical:545GHz}{1\%}
\setsymbol{HFI:FIRAS:gain:calibration:accuracy:statistical:857GHz}{0.5\%}

\setsymbol{HFI:FIRAS:gain:calibration:accuracy:systematic:100GHz}{}
\setsymbol{HFI:FIRAS:gain:calibration:accuracy:systematic:143GHz}{}
\setsymbol{HFI:FIRAS:gain:calibration:accuracy:systematic:217GHz}{}
\setsymbol{HFI:FIRAS:gain:calibration:accuracy:systematic:353GHz}{}
\setsymbol{HFI:FIRAS:gain:calibration:accuracy:systematic:545GHz}{7\%}
\setsymbol{HFI:FIRAS:gain:calibration:accuracy:systematic:857GHz}{7\%}

\setsymbol{HFI:FIRAS:zero:point:accuracy:100GHz:units}{0.8\MJysr}
\setsymbol{HFI:FIRAS:zero:point:accuracy:143GHz:units}{}
\setsymbol{HFI:FIRAS:zero:point:accuracy:217GHz:units}{}
\setsymbol{HFI:FIRAS:zero:point:accuracy:353GHz:units}{1.4\MJysr}
\setsymbol{HFI:FIRAS:zero:point:accuracy:545GHz:units}{2.2\MJysr}
\setsymbol{HFI:FIRAS:zero:point:accuracy:857GHz:units}{1.7\MJysr}

\setsymbol{HFI:FIRAS:zero:point:accuracy:100GHz}{0.8}
\setsymbol{HFI:FIRAS:zero:point:accuracy:143GHz}{}
\setsymbol{HFI:FIRAS:zero:point:accuracy:217GHz}{}
\setsymbol{HFI:FIRAS:zero:point:accuracy:353GHz}{1.4}
\setsymbol{HFI:FIRAS:zero:point:accuracy:545GHz}{2.2}
\setsymbol{HFI:FIRAS:zero:point:accuracy:857GHz}{1.7}

\setsymbol{HFI:unit:conversion:100GHz:units}{0.2415\MJysrmK}
\setsymbol{HFI:unit:conversion:143GHz:units}{0.3694\MJysrmK}
\setsymbol{HFI:unit:conversion:217GHz:units}{0.4811\MJysrmK}
\setsymbol{HFI:unit:conversion:353GHz:units}{0.2883\MJysrmK}
\setsymbol{HFI:unit:conversion:545GHz:units}{0.05826\MJysrmK}
\setsymbol{HFI:unit:conversion:857GHz:units}{0.002238\MJysrmK}

\setsymbol{HFI:unit:conversion:100GHz}{0.2415}
\setsymbol{HFI:unit:conversion:143GHz}{0.3694}
\setsymbol{HFI:unit:conversion:217GHz}{0.4811}
\setsymbol{HFI:unit:conversion:353GHz}{0.2883}
\setsymbol{HFI:unit:conversion:545GHz}{0.05826}
\setsymbol{HFI:unit:conversion:857GHz}{0.002238}

\setsymbol{HFI:colour:correction:alpha=-2:V1.01:100GHz}{0.9893}
\setsymbol{HFI:colour:correction:alpha=-2:V1.01:143GHz}{0.9759}
\setsymbol{HFI:colour:correction:alpha=-2:V1.01:217GHz}{1.0007}
\setsymbol{HFI:colour:correction:alpha=-2:V1.01:353GHz}{1.0028}
\setsymbol{HFI:colour:correction:alpha=-2:V1.01:545GHz}{1.0019}
\setsymbol{HFI:colour:correction:alpha=-2:V1.01:857GHz}{0.9889}

\setsymbol{HFI:colour:correction:alpha=0:V1.01:100GHz}{1.0008}
\setsymbol{HFI:colour:correction:alpha=0:V1.01:143GHz}{1.0148}
\setsymbol{HFI:colour:correction:alpha=0:V1.01:217GHz}{0.9909}
\setsymbol{HFI:colour:correction:alpha=0:V1.01:353GHz}{0.9888}
\setsymbol{HFI:colour:correction:alpha=0:V1.01:545GHz}{0.9878}
\setsymbol{HFI:colour:correction:alpha=0:V1.01:857GHz}{1.0014}

\def\HeJT{$^4$He-JT}

\def\TJT4K{\ifmmode T_{\rm JT\,4K}\else $T_{\rm JT\,4K}$\fi}
\def\HLMax{\ifmmode {\rm HL}_{\rm max}\else HL$_{\rm max}$\fi}
\def\strokeamp{\ifmmode \Delta S\else $\Delta S$\fi}
\def\fcomp{\ifmmode f_{\rm comp}\else $f_{\rm comp}$\fi}
\def\Tcomp{\ifmmode T_{\rm comp}\else $T_{\rm comp}$\fi}
\def\fdiv{\ifmmode f_{\rm div}\else $f_{\rm div}$\fi}
\def\NS{\ifmmode N_{\rm S}\else $N_{\rm S}$\fi}
\def\Pfill{\ifmmode P_{\rm fill}\else $P_{\rm fill}$\fi}
\def\Tpc{\ifmmode T_{\rm pc}\else $T_{\rm pc}$\fi}
\def\Tvg3{\ifmmode T_{\rm vg3}\else $T_{\rm vg3}$\fi}
\def\Pregulation{\ifmmode P_{\rm regulation}\else $P_{\rm regulation}$\fi}

\def\HeFlow{\ifmmode \hbox{He}_{\rm flow}\else He$_{\rm flow}$\fi}
\def\mumols{\ifmmode \mu\hbox{mol}\,\hbox{s}\mo\else $\mu$mol\,s\mo\fi}
\def\TDCCU{\ifmmode T_{\rm DCCU}\else $T_{\rm DCCU}$\fi}
\def\Tbolo{\ifmmode T_{\rm bolo}\else $T_{\rm bolo}$\fi}
\def\T16{\ifmmode T_{\rm 1.4K}\else $T_{\rm 1.4K}$\fi}
\def\Tdilu{\ifmmode T_{\rm dilu}\else $T_{\rm dilu}$\fi}

\def\Planck{\textit{Planck}}

\begin{document}

\title{\Planck\ Collaboration: \Planck\ Early Results. II. The thermal performance of \Planck}

%This author list corresponds to \title{Author list for Proj. Ref. 1.3: The thermal performance of Planck}
%Prepared by R. Leonardi (rleonardi@sciops.esa.int), ESAC/ESA, on 10MAY2011
%This version is from 16 May 2011 at 12:00 CET
%\subtitle{There are 231 co-authors in this list}
\author{\small
Planck Collaboration:
P.~A.~R.~Ade\inst{76}
\and
N.~Aghanim\inst{49}
\and
M.~Arnaud\inst{62}
\and
M.~Ashdown\inst{60, 4}
\and
J.~Aumont\inst{49}
\and
C.~Baccigalupi\inst{74}
\and
M.~Baker\inst{34}
\and
A.~Balbi\inst{29}
\and
A.~J.~Banday\inst{81, 8, 67}
\and
R.~B.~Barreiro\inst{56}
% ...
}
\institute{\small
Aalto University Mets\"{a}hovi Radio Observatory, Mets\"{a}hovintie 114, FIN-02540 Kylm\"{a}l\"{a}, Finland\\
\and
Agenzia Spaziale Italiana Science Data Center, c/o ESRIN, via Galileo Galilei, Frascati, Italy\\
\and
Astroparticule et Cosmologie, CNRS (UMR7164), Universit\'{e} Denis Diderot Paris 7, B\^{a}timent Condorcet, 10 rue A. Domon et L\'{e}onie Duquet, Paris, France\\
\and
Astrophysics Group, Cavendish Laboratory, University of Cambridge, J J Thomson Avenue, Cambridge CB3 0HE, U.K.\\
\and
Atacama Large Millimeter/submillimeter Array, ALMA Santiago Central Offices, Alonso de Cordova 3107, Vitacura, Casilla 763 0355, Santiago, Chile\\
\and
CITA, University of Toronto, 60 St. George St., Toronto, ON M5S 3H8, Canada\\
\and
CNES, 18 avenue Edouard Belin, 31401 Toulouse Cedex 9, France\\
\and
CNRS, IRAP, 9 Av. colonel Roche, BP 44346, F-31028 Toulouse cedex 4, France\\
\and
California Institute of Technology, Pasadena, California, U.S.A.\\
\and
Centre of Mathematics for Applications, University of Oslo, Blindern, Oslo, Norway\\
\and
DAMTP, University of Cambridge, Centre for Mathematical Sciences, Wilberforce Road, Cambridge CB3 0WA, U.K.\\
\and
DSM/Irfu/SPP, CEA-Saclay, F-91191 Gif-sur-Yvette Cedex, France\\
\and
DTU Space, National Space Institute, Juliane Mariesvej 30, Copenhagen, Denmark\\
\and
Departamento de F\'{\i}sica, Universidad de Oviedo, Avda. Calvo Sotelo s/n, Oviedo, Spain\\
\and
Department of Astronomy and Astrophysics, University of Toronto, 50 Saint George Street, Toronto, Ontario, Canada\\
\and
Department of Physics \& Astronomy, University of British Columbia, 6224 Agricultural Road, Vancouver, British Columbia, Canada\\
\and
Department of Physics, Gustaf H\"{a}llstr\"{o}min katu 2a, University of Helsinki, Helsinki, Finland\\
\and
Department of Physics, Princeton University, Princeton, New Jersey, U.S.A.\\
\and
Department of Physics, Purdue University, 525 Northwestern Avenue, West Lafayette, Indiana, U.S.A.\\
\and
Department of Physics, University of California, Berkeley, California, U.S.A.\\
\and
Department of Physics, University of California, One Shields Avenue, Davis, California, U.S.A.\\
\and
Department of Physics, University of California, Santa Barbara, California, U.S.A.\\
\and
Department of Physics, University of Illinois at Urbana-Champaign, 1110 West Green Street, Urbana, Illinois, U.S.A.\\
\and
Dipartimento di Fisica G. Galilei, Universit\`{a} degli Studi di Padova, via Marzolo 8, 35131 Padova, Italy\\
\and
Dipartimento di Fisica, Universit\`{a} La Sapienza, P. le A. Moro 2, Roma, Italy\\
\and
Dipartimento di Fisica, Universit\`{a} degli Studi di Milano, Via Celoria, 16, Milano, Italy\\
\and
Dipartimento di Fisica, Universit\`{a} degli Studi di Trieste, via A. Valerio 2, Trieste, Italy\\
\and
Dipartimento di Fisica, Universit\`{a} di Ferrara, Via Saragat 1, 44122 Ferrara, Italy\\
\and
Dipartimento di Fisica, Universit\`{a} di Roma Tor Vergata, Via della Ricerca Scientifica, 1, Roma, Italy\\
\and
Discovery Center, Niels Bohr Institute, Blegdamsvej 17, Copenhagen, Denmark\\
\and
Dpto. Astrof\'{i}sica, Universidad de La Laguna (ULL), E-38206 La Laguna, Tenerife, Spain\\
\and
European Southern Observatory, ESO Vitacura, Alonso de Cordova 3107, Vitacura, Casilla 19001, Santiago, Chile\\
\and
European Space Agency, ESAC, Planck Science Office, Camino bajo del Castillo, s/n, Urbanizaci\'{o}n Villafranca del Castillo, Villanueva de la Ca\~{n}ada, Madrid, Spain\\
\and
European Space Agency, ESOC, Robert-Bosch-Str. 5, Darmstadt, Germany\\
\and
European Space Agency, ESTEC, Keplerlaan 1, 2201 AZ Noordwijk, The Netherlands\\
\and
Helsinki Institute of Physics, Gustaf H\"{a}llstr\"{o}min katu 2, University of Helsinki, Helsinki, Finland\\
\and
INAF - Osservatorio Astronomico di Padova, Vicolo dell'Osservatorio 5, Padova, Italy\\
\and
INAF - Osservatorio Astronomico di Roma, via di Frascati 33, Monte Porzio Catone, Italy\\
\and
INAF - Osservatorio Astronomico di Trieste, Via G.B. Tiepolo 11, Trieste, Italy\\
\and
INAF/IASF Bologna, Via Gobetti 101, Bologna, Italy\\
\and
INAF/IASF Milano, Via E. Bassini 15, Milano, Italy\\
\and
INRIA, Laboratoire de Recherche en Informatique, Universit\'{e} Paris-Sud 11, B\^{a}timent 490, 91405 Orsay Cedex, France\\
\and
INSU, Institut des sciences de l'univers, CNRS, 3, rue Michel-Ange, 75794 Paris Cedex 16, France\\
\and
IPAG: Institut de Plan\'{e}tologie et d'Astrophysique de Grenoble, Universit\'{e} Joseph Fourier, Grenoble 1 / CNRS-INSU, UMR 5274, Grenoble, F-38041, France\\
\and
ISDC Data Centre for Astrophysics, University of Geneva, ch. d'Ecogia 16, Versoix, Switzerland\\
\and
Imperial College London, Astrophysics group, Blackett Laboratory, Prince Consort Road, London, SW7 2AZ, U.K.\\
\and
Infrared Processing and Analysis Center, California Institute of Technology, Pasadena, CA 91125, U.S.A.\\
\and
Institut N\'{e}el, CNRS, Universit\'{e} Joseph Fourier Grenoble I, 25 rue des Martyrs, Grenoble, France\\
\and
Institut d'Astrophysique Spatiale, CNRS (UMR8617) Universit\'{e} Paris-Sud 11, B\^{a}timent 121, Orsay, France\\
\and
Institut d'Astrophysique de Paris, CNRS UMR7095, Universit\'{e} Pierre \& Marie Curie, 98 bis boulevard Arago, Paris, France\\
\and
Institut de Ci\`{e}ncies de l'Espai, CSIC/IEEC, Facultat de Ci\`{e}ncies, Campus UAB, Torre C5 par-2, Bellaterra 08193, Spain\\
\and
Institute of Astronomy and Astrophysics, Academia Sinica, Taipei, Taiwan\\
\and
Institute of Astronomy, University of Cambridge, Madingley Road, Cambridge CB3 0HA, U.K.\\
\and
Institute of Theoretical Astrophysics, University of Oslo, Blindern, Oslo, Norway\\
\and
Instituto de Astrof\'{\i}sica de Canarias, C/V\'{\i}a L\'{a}ctea s/n, La Laguna, Tenerife, Spain\\
\and
Instituto de F\'{\i}sica de Cantabria (CSIC-Universidad de Cantabria), Avda. de los Castros s/n, Santander, Spain\\
\and
Istituto di Fisica del Plasma, CNR-ENEA-EURATOM Association, Via R. Cozzi 53, Milano, Italy\\
\and
Jet Propulsion Laboratory, California Institute of Technology, 4800 Oak Grove Drive, Pasadena, California, U.S.A.\\
\and
Jodrell Bank Centre for Astrophysics, Alan Turing Building, School of Physics and Astronomy, The University of Manchester, Oxford Road, Manchester, M13 9PL, U.K.\\
\and
Kavli Institute for Cosmology Cambridge, Madingley Road, Cambridge, CB3 0HA, U.K.\\
\and
LERMA, CNRS, Observatoire de Paris, 61 Avenue de l'Observatoire, Paris, France\\
\and
Laboratoire AIM, IRFU/Service d'Astrophysique - CEA/DSM - CNRS - Universit\'{e} Paris Diderot, B\^{a}t. 709, CEA-Saclay, F-91191 Gif-sur-Yvette Cedex, France\\
\and
Laboratoire Traitement et Communication de l'Information, CNRS (UMR 5141) and T\'{e}l\'{e}com ParisTech, 46 rue Barrault F-75634 Paris Cedex 13, France\\
\and
Laboratoire de Physique Subatomique et de Cosmologie, CNRS/IN2P3, Universit\'{e} Joseph Fourier Grenoble I, Institut National Polytechnique de Grenoble, 53 rue des Martyrs, 38026 Grenoble cedex, France\\
\and
Laboratoire de l'Acc\'{e}l\'{e}rateur Lin\'{e}aire, Universit\'{e} Paris-Sud 11, CNRS/IN2P3, Orsay, France\\
\and
Lawrence Berkeley National Laboratory, Berkeley, California, U.S.A.\\
\and
Max-Planck-Institut f\"{u}r Astrophysik, Karl-Schwarzschild-Str. 1, 85741 Garching, Germany\\
\and
MilliLab, VTT Technical Research Centre of Finland, Tietotie 3, Espoo, Finland\\
\and
National University of Ireland, Department of Experimental Physics, Maynooth, Co. Kildare, Ireland\\
\and
Niels Bohr Institute, Blegdamsvej 17, Copenhagen, Denmark\\
\and
Observational Cosmology, Mail Stop 367-17, California Institute of Technology, Pasadena, CA, 91125, U.S.A.\\
\and
Optical Science Laboratory, University College London, Gower Street, London, U.K.\\
\and
Rutherford Appleton Laboratory, Chilton, Didcot, U.K.\\
\and
SISSA, Astrophysics Sector, via Bonomea 265, 34136, Trieste, Italy\\
\and
SUPA, Institute for Astronomy, University of Edinburgh, Royal Observatory, Blackford Hill, Edinburgh EH9 3HJ, U.K.\\
\and
School of Physics and Astronomy, Cardiff University, Queens Buildings, The Parade, Cardiff, CF24 3AA, U.K.\\
\and
Space Sciences Laboratory, University of California, Berkeley, California, U.S.A.\\
\and
Stanford University, Dept of Physics, Varian Physics Bldg, 382 Via Pueblo Mall, Stanford, California, U.S.A.\\
\and
Thales Alenia Space France, 100 Boulevard du Midi, Cannes la Bocca, France\\
\and
Universit\"{a}t Heidelberg, Institut f\"{u}r Theoretische Astrophysik, Albert-\"{U}berle-Str. 2, 69120, Heidelberg, Germany\\
\and
Universit\'{e} de Toulouse, UPS-OMP, IRAP, F-31028 Toulouse cedex 4, France\\
% ...
}

 \abstract{The performance of the \Planck's instruments in space is enabled by their low operating temperatures, 20\,K for LFI and 0.1\,K for HFI, achieved through a combination of passive radiative cooling and three active mechanical coolers.   The scientific requirement for very broad frequency coverage led to two detector technologies with widely different temperature and cooling needs.  Active coolers could satisfy these needs; a helium cryostat, as used by previous  cryogenic space missions (\textit{IRAS, COBE, ISO, Spitzer, Akari}), could not.  Radiative cooling is provided by three V-groove radiators and a large telescope baffle.  The active coolers are a hydrogen sorption cooler ($<20$\,K), a $^4$He Joule-Thomson cooler (4.7\,K), and a $^3$He-$^4$He dilution cooler (1.4\,K and 0.1\,K).  The flight system was at ambient temperature at launch and cooled in space to operating conditions.  The HFI bolometer plate reached 93\,mK on 3 July 2009, 50 days after launch.  The solar panel always faces the Sun, shadowing the rest of \Planck, and operates at a mean temperature of 384\,\hbox{K}.  At the other end of the spacecraft, the telescope baffle operates at 42.3\,K and the telescope primary mirror operates at 35.9\,\hbox{K}.  The temperatures of key parts of the instruments are stabilized by both active and passive methods.  Temperature fluctuations are driven by changes in the distance from the Sun, sorption cooler cycling and fluctuations in gas-liquid flow, and fluctuations in cosmic ray flux on the dilution and bolometer plates.  These fluctuations do not compromise the science data.}

\titlerunning{The thermal performance of \textit{Planck}}
\authorrunning{Planck Collaboration}

\maketitle

\allearlypapers

\section{Introduction \Planck}
\label{sec:intro}

\Planck\footnote{\Planck\ (\url{http://www.esa.int/Planck}) is a project of the European Space Agency (ESA) with instruments provided by two scientific consortia funded by ESA member states (in particular the lead countries France and Italy), with contributions from NASA (USA) and telescope reflectors provided by a collaboration between ESA and a scientific consortium led and funded by Denmark.} \citep{tauber2010a, planck2011-1.1} is the third generation space mission to measure the anisotropy of the cosmic microwave background (CMB).  It observes the sky in nine frequency bands covering 30--857\,GHz with high sensitivity and angular resolution from 31\arcmin\ to 5\arcmin.  The Low Frequency Instrument (LFI; \citealt{Mandolesi2010, Bersanelli2010, planck2011-1.4}) covers the 30, 44, and 70\,GHz bands with amplifiers cooled to 20\,\hbox{K}.  The High Frequency Instrument (HFI; \citealt{Lamarre2010, planck2011-1.5}) covers the 100, 143, 217, 353, 545, and 857\,GHz bands with bolometers cooled to 0.1\,\hbox{K}.  Polarisation is measured in all but the two highest bands \citep{Leahy2010, Rosset2010}.  A combination of radiative cooling and three mechanical coolers produces the temperatures needed for the detectors and optics \citep{planck2011-1.3}.  Two data processing centres (DPCs) check and calibrate the data and make maps of the sky \citep{planck2011-1.7, planck2011-1.6}.  \Planck's sensitivity, angular resolution, and frequency coverage make it a powerful instrument for Galactic and extragalactic astrophysics as well as cosmology.  Early astrophysics results are given in Planck Collaboration (2011d--u).

The unprecedented performance of the \Planck\ instruments in space is enabled by their low operating temperatures, 20\,K for LFI and 0.1\,K for HFI, achieved through a combination of passive radiative cooling and three active coolers.  This architecture is unlike that of the previous CMB space missions, the \textit{Cosmic Background Explorer} (\textit{COBE\/}; \citealt{Boggess1992}) and the \textit{Wilkinson Microwave Anisotropy Probe} (\textit{WMAP\/}; \citealt{Bennett2003}). 
\textit{COBE} used a liquid helium cryostat to enable cooling of the bolometers on its Far-Infrared Absolute spectrophotometer (FIRAS) instrument \citep{Mather1990} to 1.5\,\hbox{K}.  This approach was not adopted for \Planck\ as it would restrict the on-orbit lifetime for the HFI, require additional coolers to reach sub-Kelvin temperatures, and be entirely infeasible for cooling the active heat load from the LFI.  \textit{WMAP} relied on passive radiative cooling alone which, while simpler, resulted in a higher operating temperature for its amplifiers and a higher noise temperature.  Additionally, purely passive cooling is unable to reach the sub-kelvin operating temperatures required by HFI's high-sensitivity bolometers. 

In this paper we describe the design and in-flight performance of the mission-enabling \Planck\ thermal system.

\section{Thermal Design}
\label{sec:design}

\subsection{Overview, philosophy, requirements, and redundancy}

\begin{figure*}
\begin{center}
\leavevmode
\includegraphics[width=9.9cm]{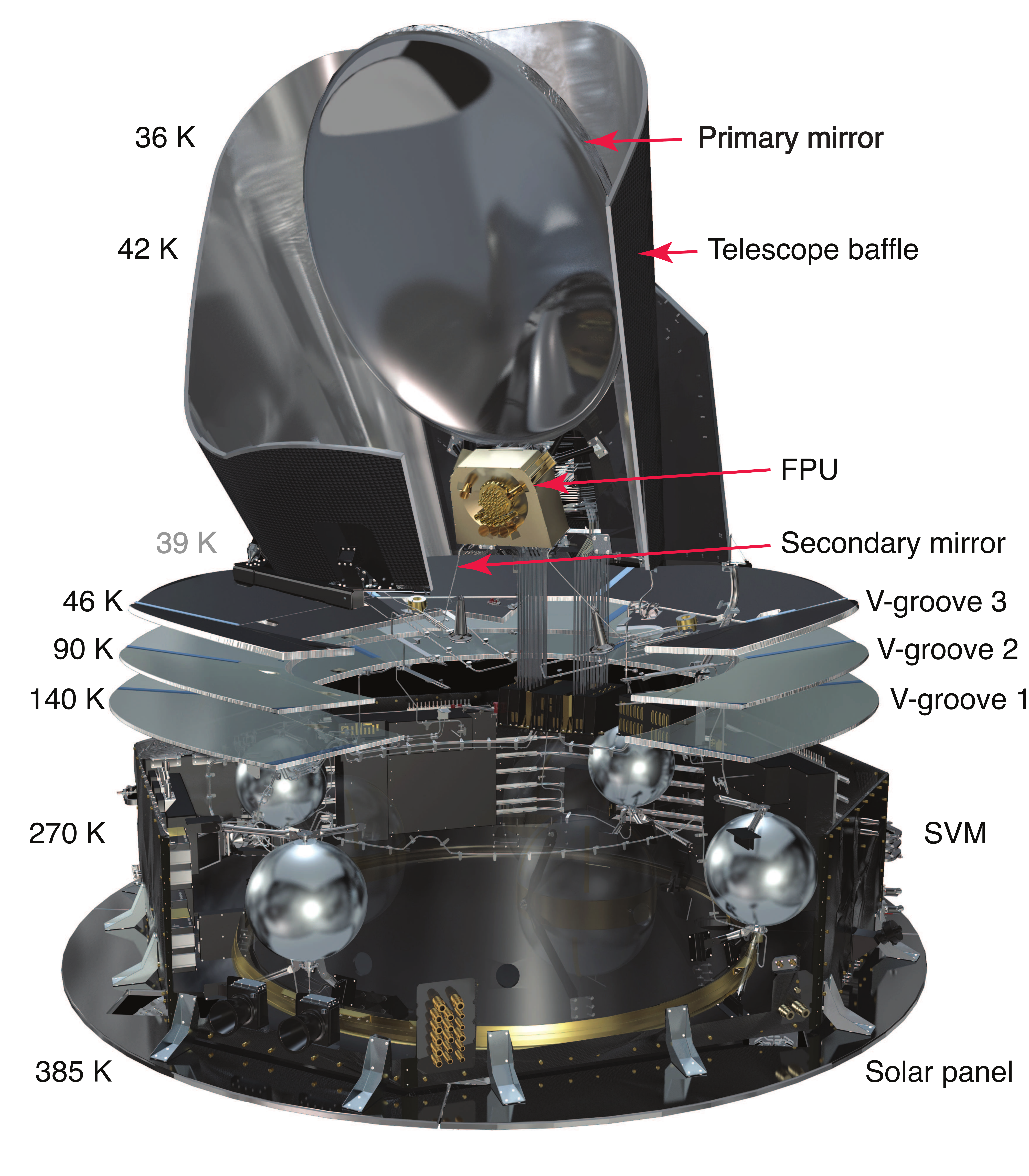}
\caption{Cutaway view of \Planck, with the temperatures of key components in flight.  The solar panel at the bottom always faces the Sun and the Earth, and is the only part of the flight system illuminated by the Sun, the Earth, and the Moon.  Temperature decreases steadily towards the telescope end, due to low-conductivity mechanical connections and aggressive use of radiative cooling.  The focal plane detectors are actively cooled to 20\,K and 0.1\,\hbox{K}.}
\label{fig:PlanckCAD}
\end{center}
\end{figure*}

The thermal design of \Planck\ is driven by the scientific requirement of  very broad frequency coverage, implying two instruments with detector technologies requiring different cooling temperatures (20\,K and 0.1\,K) and heatlifts (0.5\,W and 1\muW). This, combined with a scanning strategy based on a spinner satellite, led to the choice of an active cooling system. The overall architecture can be understood from Fig.~\ref{fig:PlanckCAD}.  The solar panel always faces the Sun and the Earth, the only two significant sources of heat in the sky, and operates at 385\,\hbox{K}.  The service vehicle module (SVM) operates at room temperature.  The telescope, at the opposite end of the flight system, operates below 40\,\hbox{K}.  The detectors at the focus of the telescope are actively cooled to 20\,K (amplifiers) or 0.1\,K (bolometers).  Between the SVM and the ``cold end,'' aggressive measures are taken to minimize heat conduction and to maximize the radiation of heat to cold space.  These measures include low-conductivity support elements, three V-groove radiators, and a telescope baffle with low emissivity inside and high emissivity outside.

\subsubsection{Mission design, scientific requirements, and thermal architecture}

\Planck\ is designed to extract all information in the temperature anisotropies of the CMB down to angular scales of 5\arcm, and to provide a major advance in the measurement of polarisation anisotropies.  This requires both extremely low noise and broad frequency coverage  from tens to hundreds of gigahertz to separate foreground sources of radiation from the \hbox{CMB}.  The necessary noise level can be reached only with cryogenically-cooled detectors.  The lowest noise is achieved with amplifiers cooled to $\leq20$\,K and bolometers cooled to $\sim0.1$\,\hbox{K}.  Temperature fluctuations must not compromise the sensitivity.  
Additional constraints on \Planck\ that affect the thermal design include: 1)~no deployables (e.g., a shield that could block the Sun over a large solid angle);  2)~no optical elements such as windows with warm edges between the feed horns and telescope; 3)~1.5\,yr minimum total lifetime;  4)~a spinning spacecraft;  5)~an off-axis telescope below 60\,K; 6)~feed horns for the bolometers below 5\,K and a bolometer environment below 2\,K; 7)~reference targets (loads) for the pseudo-correlation amplifier radiometers below 5\,K to minimize $1/f$ noise; and 8)~0.5\,W heatlift for the 20\,K amplifiers.

\begin{figure*}
\begin{center}
\includegraphics[width=11.2cm]{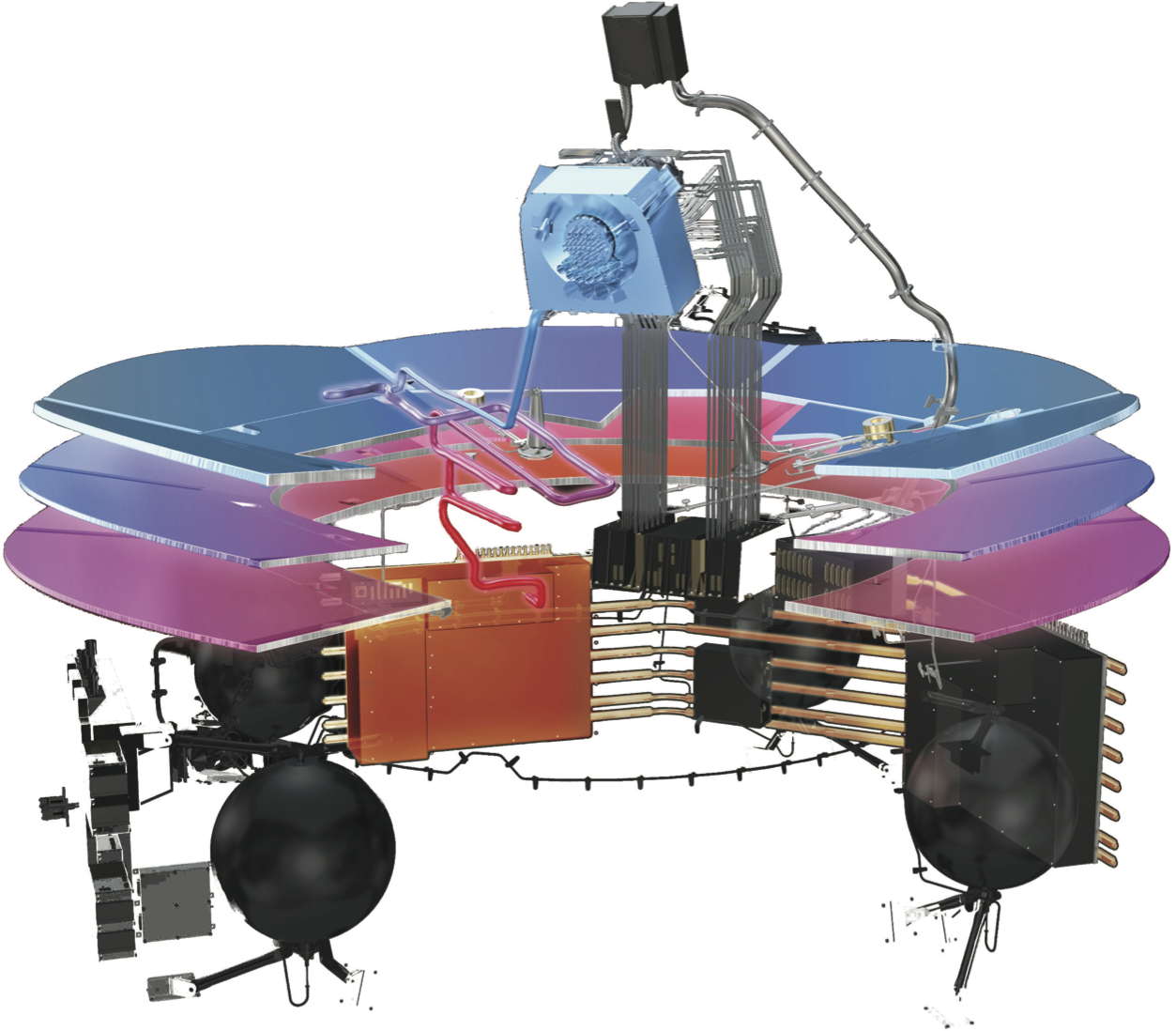}
\caption{Sorption cooler system.  The system is fully redundant.  One of the compressor assemblies, mounted on one of the warm radiator panels, which faces cold space, is highlighted in orange.  Heat pipes run horizontally connecting the radiators on three sides of the service vehicle  octagon.  The second compressor assembly is the black box on the right.  A tube-in-tube heat exchanger carries high pressure hydrogen gas from the compressor assembly to the focal plane assembly and low pressure hydrogen back to the compressor, with heat-exchanging attachments to each of the three V-grooves.  Colours indicate temperature, from warm (red, orange, purple) to cold (blue).}
\label{figure:20KcoolerCAD}
\end{center}
\end{figure*}

\begin{figure*}[htbp]
 \centering
\includegraphics[width=11.2cm]{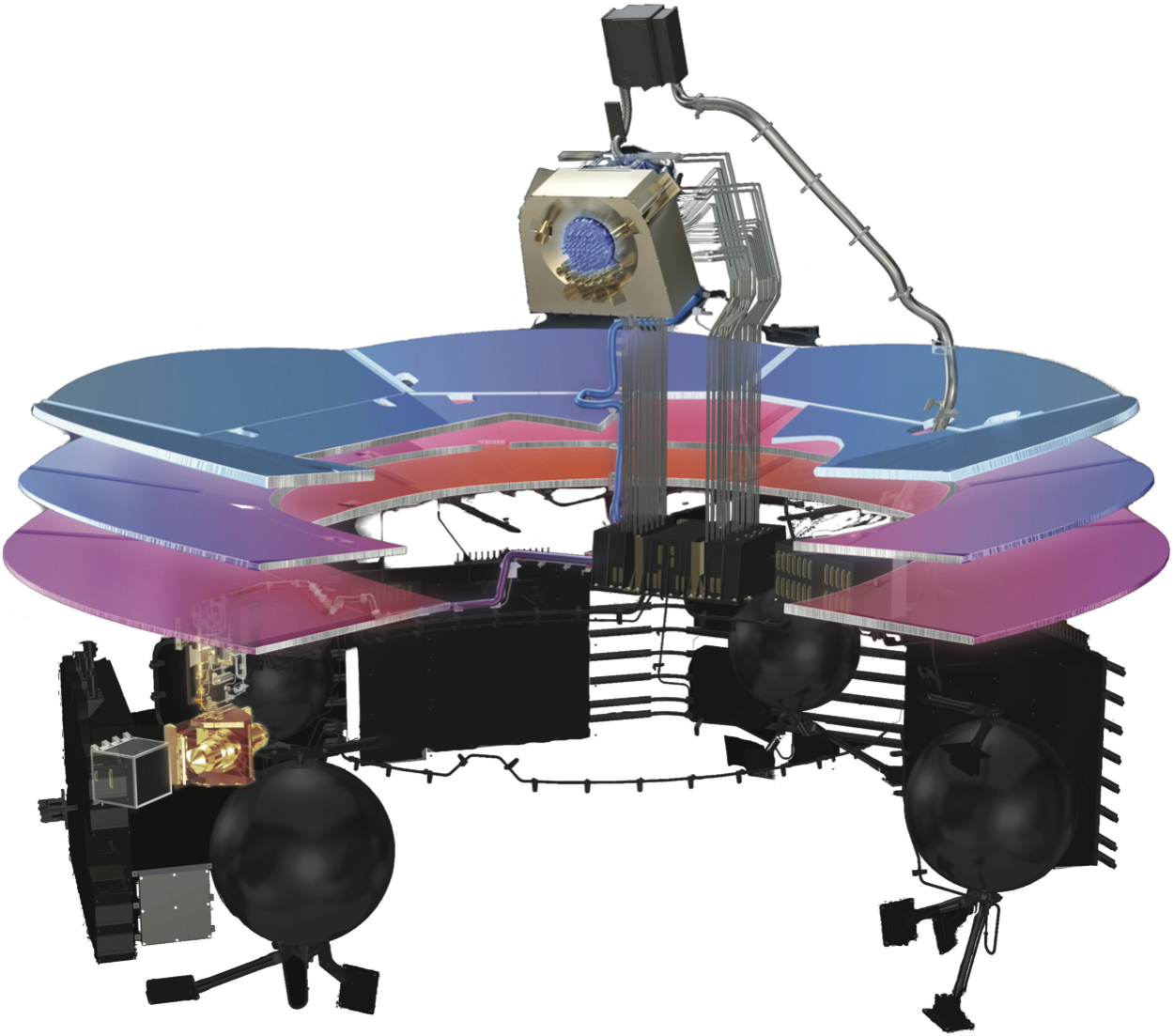}
  \caption{\HeJT\ cooler system.  The back-to-back compressors are highlighted in gold on the left-hand side of the service vehicle, adjacent to control electronics boxes.  The high and low pressure $^4$He gas tubes connecting the compressors with the JT valve in the focal plane are coloured from purple to blue, indicating temperature as in Fig.~\ref{figure:20KcoolerCAD}.} 
  \label{fig:4Kcooler}
\end{figure*}

\begin{figure*}[htbp]
  \centering
\includegraphics[width=11.3cm]{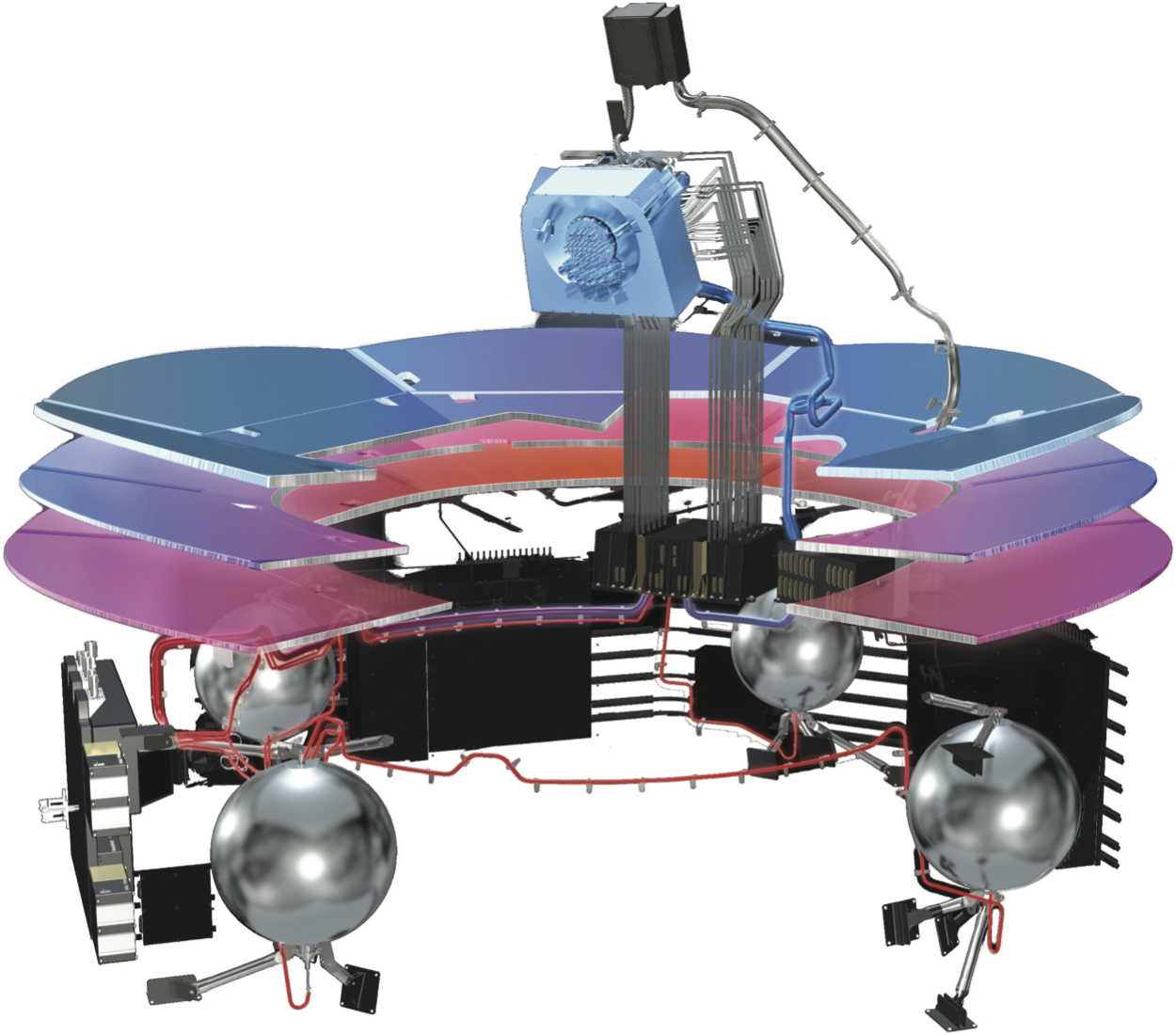}
  \caption{Dilution cooler system.  Four high pressure tanks of $^3$He and $^4$He are highlighted in silver.  The dilution cooler control unit (DCCU), to which piping from the tanks and to the focal plane unit is attached, is highlighted on the left.} 
  \label{fig:100mK}
\end{figure*}

These requirements (stated precisely in Tables~\ref{table:SCS_requirements}, \ref{table:4Kcooler}, and \ref{table:HFIstability}, and in Sect.~2.1.2) led to a design that includes the following:

\begin{itemize}

\item A ``warm launch'' scenario, in which the entire flight system is at ambient temperature for launch.  This allows a very clean environment from the straylight point of view.

\item The overall thermal architecture shown in Fig.~\ref{fig:PlanckCAD}, with the solar panel acting as a Sun shield, and temperature decreasing along the spin axis toward the cold end and the passively cooled telescope.

\item Extensive use of passive (radiative) cooling, especially the V-groove radiators and the telescope and telescope baffle.

\item Detectors based on amplifiers at 30, 44, and 70\,GHz, and on bolometers at 100, 143, 217, 353, 545, and 857\,GHz.  

\item An active cooling chain with three mechanical coolers.  No large helium cryostat had ever been flown on a spinning spacecraft.  A focal plane with detectors at 0.1\,K and 20\,K and reference loads at less than 5\,K would have been  difficult to accomodate in a cryostat.  A heatlift of 0.5\,W at 20\,K would have required a huge cryostat of many thousands of liters.

\begin{enumerate}

\item The ``sorption cooler'' (Fig.~\ref{figure:20KcoolerCAD}), a closed-cycle sorption cooler using hydrogen as the working fluid with a Joule-Thomson (JT) expansion, which produces temperatures below 20\,\hbox{K} and a heat lift close to 1\,\hbox{W}.  The sorption cooler cools the LFI focal plane to $<20$\,K and provides precooling to lower temperature stages.  Although this cooler was a new development, it was the only one that could provide such a large heatlift at the required temperature.

\item The ``\HeJT\ cooler'' (Fig.~\ref{fig:4Kcooler}), a closed-cycle cooler using a Stirling cycle compressor and $^4$He as the working fluid with a JT expansion, which produces temperatures below 5\,\hbox{K}.  The \HeJT\ cooler cools the structure hosting the HFI focal plane and the LFI reference loads to $<5$\,K and provides precooling to the dilution cooler.  The chosen technology had a long flight heritage for the compressors and a specific development to minimize microvibrations.

\item The ``dilution cooler'' (Fig.~\ref{fig:100mK}), a $^3$He-$^4$He dilution cooler that vents combined $^3$He and $^4$He to space, and which produces temperatures of  $1.4$\,K through JT expansion of the $^3$He and $^4$He, and $\sim0.1$\,\hbox{K} for the bolometers through the dilution of $^3$He into $^4$He.  Although this was a new technology, it was chosen for its simple architecture, continuous operation, and temperature stability.  An ADR suitable for HFI would have to operate from a 4\,K heat sink rather than the 1.5\,K heat sink available on Astro-E, and would have to be scaled up in mass by a factor of 10 to lift {$8\,\mu$W} at 0.1\,K \citep{planckOCDR06}.   This is because the dilution lifts heat not only at 100\,mK where the $^3$He and $^4$He lines combine, but also all along the return line (which is attached to the struts and wiring) as the mixture warms, reducing the required lift at 100\,mK from about 8\muW\ to only 1\muW. If an ADR similar to that of astro-e were used for Planck, it would have to lift the entire 8\muW\ at 100\,mK.  Furthermore, ADRs require high magnetic fields and cycling, not easily compatible with continuous measurements and the very high stability requirements of \Planck. 

\end{enumerate}

\end{itemize}

The above design results in a complicated architecture and test philosophy.  The two instruments, passive radiators, and active coolers cannot be separated easily from the spacecraft, either mechanically or thermally.  The components of the flight system are highly interdependent, and difficult to integrate.  All of this made the cryogenic chain the most challenging element in the \Planck\ mission.

% 2.1.2
\subsubsection{Requirements on the coolers}

The three coolers were required to deliver temperatures of $<20$\,K, $<5$\,K, and 0.1\,K, continuously.  Although use of proven technology is preferred in space missions, the special requirements of  \Planck\ led to the choice of two new-technology coolers.  

The first is a hydrogen sorption cooler developed by the Jet Propulsion Laboratory in California, which provides a large heat lift with no mechanical compressors (avoiding vibration). The  sorption cooler requires precooling of the hydrogen to $\leq60$\,\hbox{K}.  The precooling temperature required by the \HeJT\ cooler, which is supplied by the sorption cooler, is $\leq20$\,\hbox{K}.  This interface temperature is critical in the cooling chain: the \HeJT\ cooler heat load increases with the interface temperature when, at the same time, its heat lift decreases.  Thus the goal for the precool temperature supplied by the sorption cooler was 18\,K, with a strict requirement of $\leq19.5$\,K to leave adequate margin in the cooling chain. 

The second new cooler is a $^3$He-$^4$He dilution cooler.  The microgravity dilution cooler principle was invented and tested by A. Beno{\^\i}t \citep{Benoit1997} and his team at Institut N\'eel, Grenoble, and developed into a space qualified system by DTA Air Liquide \citep{Triqueneaux2006} under a contract led by Guy Guyot and the system group at \hbox{IAS} .  It provides two stages of cooling at 1.4\,K and 100\,mK. The total heat lift requirement at 100\,mK was 0.6\,$\mu$W, and could be achieved with an open circuit system carrying enough $^3$He and $^4$He for the mission. The precooling temperature required was less than 4.5\,\hbox{K}.

This precool is provided by the third cooler, a closed-circuit $^4$He JT expansion cooler driven by two mechanical compressors in series \citep[p.465]{Bradshaw1997} developed by RAL and EADS Astrium in Stevenage, \hbox{UK} (formerly British Aerospace).  The drive electronics were designed and built by a consortium of RAL and Systems Engineering and Assessment (SEA) in Bristol.  The pre-charge regulator was built by CRISA in Madrid with oversight from ESA and the University of Granada.  The net heat lift requirement for this cooler was 15\,mW at a precooling temperature of 20\,\hbox{K}.  Since bolometers are sensitive to microvibrations, this cooler includes a new vibration control system.  The major components of the system are shown in Fig.~\ref{fig:4Kcooler}, and the basic characteristics are summarized in Table~\ref{table:4Kcooler}.

%2.1.3
\subsubsection{Redundancy philosophy}

A redundant sorption cooler system was required because both instruments depend on it.  Furthermore the hydride used in the sorption beds ages.  The other two coolers are needed only by the HFI, and were not required to be redundant.   The critical elements of the dilution cooler are passive.  Although the flow of $^3$He and $^4$He is adjustable, a minimum flow is always available. The only single point failure is the opening of the valves on the high pressure tanks of $^3$He and $^4$He at the start of the mission. Making those redundant would increase the risk of leaks without improving reliability significantly. The compressors used in the \HeJT\ cooler have good flight heritage.  The vibration control system was not considered a single point failure for the high frequency instrument, although its failure would degrade performance significantly.  The \HeJT\ cooler provides the reference loads for the LFI, and is thus a quasi-single point failure; however, it was not made redundant because of its good flight heritage and in view of the extra resources it would have required from the spacecraft.

%2.1.4
\subsubsection{The critical importance of passive cooling}

Nearly 14\,kW of solar power illuminates the solar panel.  Of this, less than 1\,W reaches the focal plane, a result of careful thermal isolation of components, extremely effective passive cooling from the V-grooves and the telescope baffle, and the overall geometry.  The low temperatures achieved passively have a dramatic effect on the design of the active coolers.  In particular, the efficiency and heat lift of the sorption cooler increase rapidly as the precooling temperature provided by the V-grooves, especially V-groove 3, decreases.  The \HeJT\ cooler heat lift also increases as its precool temperature --- the $\sim$18\,K provided by the sorption cooler --- decreases.   The low temperatures of the primary and secondary mirrors mean that their thermal emission contributes negligibly to the overall noise of the instruments (see Sect.~4.1.), and the radiative heat load from the baffle and mirrors on the focal plane unit (FPU) is extremely low.  None of this could work as it does without the passive cooling.

\subsection{Passive components}

\subsubsection{V-grooves}

The V-groove radiators are cones built from flat wedges of carbon fibre honeycomb panel covered with aluminium face sheets.  All surfaces are low emissivity except the exposed top of V-groove 3, which is painted black for good radiative coupling to cold space.  The vertex angle of successive cones decreases byabout 7\deg, thus facing cones are not parallel, and photons between them are redirected to cold space in a few reflections.  V-grooves are extremely effective at both thermal isolation and radiative cooling, with many advantages over multilayer insulation (MLI), including negligible outgassing after launch.  Three V-grooves are required to achieve the $\leq60$\,K requirement on precool temperature for the sorption cooler, with margin.   More were unnecessary, but would have added cost and mass.

\subsubsection{Telescope baffle}

The telescope baffle provides both radiative shielding and passive cooling.  Its interior is covered with polished aluminium for low emissivity, while its outside is covered with open hex-cells painted black, for high emissivity.

%2.3
\subsection{Active components}

\subsubsection{Sorption cooler and warm radiator}

Each of the two hydrogen sorption coolers comprises a compressor assembly, warm radiator, piping assembly (including heat exchangers on three V-groove radiators), a JT expander, and control electronics.  The major components of the system are highlighted in Fig.~\ref{figure:20KcoolerCAD}.  The only moving parts are passive check valves.  Six ``compressor elements'' containing a La$_{1.0}$Ni$_{4.78}$Sn$_{0.22}$ alloy absorb hydrogen at 270\,K and 1/3\,atmosphere, and desorb it at 460\,K and 30\,atmospheres.  By varying the temperature of the six beds sequentially with resistance heaters and thermal connections to the warm radiator, a continuous flow of high-pressure hydrogen is produced.  

Sorption coolers provide vibration-free cooling with no active moving parts, along with great flexibility in integration of the cooler to the cold payload (instrument, detectors, and telescope mirrors) and the warm spacecraft.   No heat is rejected in or near the focal plane.  The refrigerant fluid in the \Planck\ sorption coolers is hydrogen, selected for operation at a temperature of $\sim$17\,\hbox{K}.  The \Planck\ sorption coolers are the first continuous cycle sorption coolers to be used in space. 

Table~\ref{table:SCS_requirements} gives the requirements on the sorption cooler system.  The temperature stability requirement listed is an inadequate simplification of a complicated reality.  Fluctuations in the temperatures of the sorption cooler interfaces to the LFI and HFI have no intrinsic significance.  What matters is the effect of temperature fluctuations on the science results.  Fluctuations at the cooler interfaces with HFI and LFI (LVHX1 and LVHX2, respectively) propagate to the detectors themselves through complicated conductive and radiative paths (quite different for the two instruments).  Temperature controls, passive components of varying emissivities in the optical paths, the structure of the detectors and the effect of thermal fluctuations on their output, and the effects of the spinning scan strategy and data processing all must be taken into account.  Fluctuations at frequencies well below the spin frequency (16.67\,mHz) cannot be due to the sky, and are easily removed by the spin and data processing.  Fluctuations at frequencies well above the spin frequency are heavily damped by the front-end structure of the instruments.  The impact of these factors could not be calculated accurately at the time a cooler fluctuation requirement had to be devised, and therefore it was not possible to derive a power spectral density limit curve --- the only kind of specification that could capture the true requirements --- with high fidelity.  We will return to this point in Sect.7.

\begin{table}                    
\begingroup
\newdimen\tblskip \tblskip=5pt
\caption{Requirements on the sorption cooler system.}  
\label{table:SCS_requirements}
\nointerlineskip
\vskip -3mm
\footnotesize
\setbox\tablebox=\vbox{
   \newdimen\digitwidth 
   \setbox0=\hbox{\rm 0} 
   \digitwidth=\wd0 
   \catcode`*=\active 
   \def*{\kern\digitwidth}
   \newdimen\signwidth 
   \setbox0=\hbox{+} 
   \signwidth=\wd0 
   \catcode`!=\active 
   \def!{\kern\signwidth}
\halign{\hbox to 1.4in{#\leaderfil}\tabskip=0.5em&
        \hfil#\hfil\tabskip=0pt\cr                             % Template goes here.
\noalign{\doubleline}
\omit\hfil Item\hfil&Requirement\cr
\noalign{\vskip 3pt\hrule\vskip 5pt}
Cold end temperature&17.5\,K $<$ LVHX1 $<$ 19.02\,K\cr
\omit&               17.5\,K $<$ LVHX2 $<$ 22.50\,K\cr
\noalign{\vskip 6pt}
Cooling power&       at LVHX1 $>$ 190\,mW\cr
\omit&               at LVHX2 $>$ 646\,mW\cr
\noalign{\vskip 6pt}
Input power&         $<$ 426\,W, beginning of life\cr
\noalign{\vskip 6pt}
Cold end temperature&$\Delta T$ at LVHX1 $<$ 450\,mK\cr
\omit \hglue 2em fluctuations&$\Delta T$ at TSA $<$ 100\,mK\cr                          
\noalign{\vskip 5pt\hrule\vskip 3pt}}}
\endPlancktable
\endgroup
\end{table}

Each compressor element is connected to both the high pressure and low pressure sides of the piping system through check valves that  allow gas flow in a single direction only.  The high pressure is stabilized by a 4\,litre ballast tank, the high-pressure stabilization tank (HPST).  On the low pressure side, the low pressure storage bed (LPSB), filled with hydride and maintained at a temperature near that of the warm radiator, stores a large fraction of the H$_2$ required to operate the cooler during flight and ground testing, while minimizing the pressure in the non-operational cooler during launch and transportation. 

A single compressor element comprises a cylinder supported at its ends by low thermal conductivity tubes connected to a larger semi-cylinder with a flat side. The inner  cylinder contains the La$_{1.0}$Ni$_{4.78}$Sn$_{0.22}$; the outer semi-cylinder creates a volume around the inner cylinder, and its flat side is bolted to the ``warm radiator.'' The volume between the two is evacuated or filled with low pressure hydrogen by a gas-gap heat switch using a second metal hydride, ZrNi.  When filled with low pressure hydrogen, there is a good thermal connection from the inner hydride bed to the warm radiator.  When hydrogen is evacuated, the inner hydride bed is thermally isolated, and can be heated up efficiently.

The compressor elements are taken sequentially through four steps: {\it heat up\/} to pressurize; {\it desorb\/}; {\it cool down\/} to depressurize; {\it absorb\/}.  At a given instant, one of the six compressor elements is heating up, one is desorbing, one is cooling down, and three are absorbing.  Heating is achieved by electrical resistance heaters.  Cooling is achieved by thermally connecting the compressor element to the so-called warm radiator, whose temperature is controlled (Sect.~2.4.2) by electrical heaters at a temperature in the range $272\pm10$\,\hbox{K}.  

The warm radiator covers three of the eight panels of the \hbox{SVM}.  The two compressor assemblies are mounted on the end panels of the three, each of which contains 16~straight heat pipes running parallel to the spacecraft spin axis and perpendicular to the compressor elements.  These heat pipes maintain a nearly isothermal condition across the panel, in particular distributing the heat of the compressor element that is in the cooldown cycle.  Eight long, bent, heat pipes run perpendicular to the others, connecting all three panels of the radiator together.  The external surfaces of all three panels are painted black.

Upon expansion through the JT valve, hydrogen forms liquid droplets whose evaporation provides the cooling power.  The liquid/vapour mixture then flows through two liquid/vapour heat exchangers (LVHX), the first  thermally and mechanically coupled to the HFI interface, where it provides precooling for the \HeJT\ cooler (Sect.~2.3.2) and the $^3$He-$^4$He dilution cooler (Sect.~2.3.3).  The second is coupled to the LFI interface, where it cools the LFI focal plane assembly to $\sim$20\,\hbox{K}. Any remaining liquid/vapour mixture flows through a third LVHX, which is maintained above the hydrogen saturated vapour temperature. This third LVHX serves to evaporate any excess liquid that reaches it, preventing flash boiling and thereby maintaining a nearly constant pressure in the low-pressure piping. Low-pressure gaseous hydrogen is re-circulated back to the cool sorbent beds for compression. 

Regulation of the system is done by simple heating and cooling; no active control of valves is necessary. The heaters for the compressors are controlled by a timed on-off heater system.  

The flight sorption cooler electronics and software were developed by the Laboratoire de Physique Subatomique et de Cosmologie (LPSC) in Grenoble. These electronics and their controlling software provide for the basic sequential operation of the compressor beds, temperature stabilization of the cold end, and monitoring of cooler performance parameters. In addition, they automatically detect failures and adapt operations accordingly.  Operational parameters can be adjusted in flight to maximise the lifetime and performance of the sorption coolers. 

The total input power to the sorption cooler at end of life (maximum average power) is 470\,\hbox{W}. Another 110\,W is available to operate the sorption cooler electronics.

%2.3.2

\subsubsection{\HeJT\ cooler}

\begin{figure}
\centering
\includegraphics[width=8.4cm]{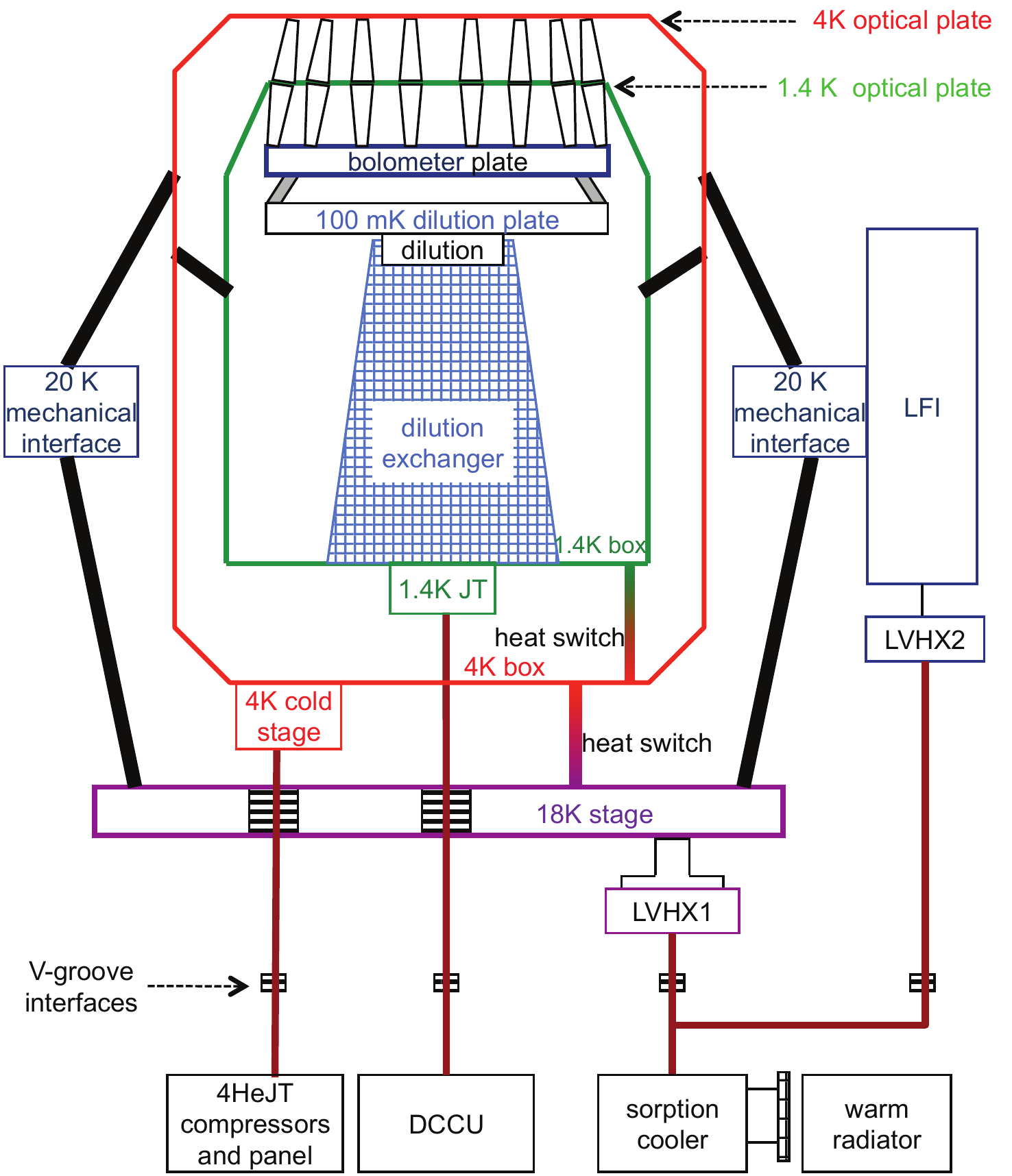}
\caption{Schematic of the HFI \hbox{FPU}.  The designations ``20\,K'', ``18\,K'', ``4\,K'', ``1.4\,K'', and ``100\,mK'' are nominal.  Actual operating temperatures are given in Sect.~4.}
\label{fig:ThermalInterfaces}
\end{figure}

Figure~\ref{fig:ThermalInterfaces} shows schematically the thermal interfaces of the HFI cooling system.  The compressors of the \HeJT\ cooler are mounted in opposition \citep{Lamarre2010} to cancel to first order momentum transfer to the spacecraft. Force transducers between the two compressors provide an error signal processed by the drive electronics servo system that controls the profile of the piston motions to minimise the first seven harmonics of the periodic vibration injected into the spacecraft.

\begin{table} 
\begingroup
\newdimen\tblskip \tblskip=5pt
\caption{Requirements and characteristics of the \HeJT\ cooler.} 
\label{table:4Kcooler}
\nointerlineskip
\vskip -6mm
\footnotesize
\setbox\tablebox=\vbox{
   \newdimen\digitwidth 
   \setbox0=\hbox{\rm 0} 
   \digitwidth=\wd0 
   \catcode`*=\active 
   \def*{\kern\digitwidth}
   \newdimen\signwidth 
   \setbox0=\hbox{+} 
   \signwidth=\wd0 
   \catcode`!=\active 
   \def!{\kern\signwidth}
\halign{\hbox to 2.6in{#\leaderfil}\tabskip=2em&
        #\hfil\tabskip=0pt\cr
\noalign{\vskip 3pt\hrule\vskip 5pt}
Working fluid &$^4$Helium\cr
\noalign{\vskip 4pt}
\omit Heat lift at 17.5\,K pre-cool temperature\hfil\cr
\hglue 2em Maximum&19.2\,mW\cr
\hglue 2em Required&13.3\,mW\cr
\noalign{\vskip 4pt}
\omit Pre-cool requirements\hfil\cr
\hglue 2em Third V-groove&$\leq 54$\,K\cr
\hglue 2em Sorption cooler LVHX1&17.5--19\,K\cr
Nominal operating temperature&4.5\,K\cr
\noalign{\vskip 4pt}
\omit Mass\hfil\cr
\hglue 2em Compressors, pipes, cold stage&27.7\,kg\cr
\hglue 2em Electronics and current regulator&8.6\,kg\cr
\noalign{\vskip 4pt}
Power  into current regulator&$\leq120$\,W\cr
\noalign{\vskip 5pt\hrule\vskip 3pt}}}
\endPlancktable
\endgroup
\end{table}

The 4\,K cold head is a small reservoir of liquid helium in a sintered material, located after the expansion of the gas through the JT orifice. This provides an important buffer with high heat capacity between the JT orifice and the rest of the \hbox{HFI}. It is attached to the bottom of the 4\,K box of the HFI FPU, as can be seen in Fig.~\ref{fig:ThermalInterfaces}.  It provides cooling for the 4\,K shield and also pre-cooling for the gas in the dilution cooler pipes described in the next section.

The cooling power and thermal properties of the \HeJT\ cooler,  measured by the RAL team at subsystem level and then in the system thermal vacuum tests, are summarised in the following relationships, which depend linearly on the adjustable parameters in the vicinity of the flight operating point:  
\begin{eqnarray}
{\HLMax} & = & 15.9\,{\rm mW} + 6.8(\strokeamp - 3.45\,{\rm mm}) \nonumber \\
         &   & \quad - 1.1(\Tpc - 17.3\,{\rm K}) + 0.6(\Pfill - 4.5\,\hbox{bar});\\
\noalign{\vskip 5pt}
\hbox{Heatload} & = & 10.6\,{\rm mW} + 0.5(\Tpc - 17.3\,{\rm K}) \nonumber \\
                &   &\quad + 0.065(\Tvg3 - 45\,{\rm K}) + \hbox{Heaters};\\
\noalign{\vskip 5pt}
{\TJT4K} & = & 4.4\,{\rm K} - 0.035 ({\HLMax} - \hbox{Heatload}). \end{eqnarray}
Here $\HLMax$ is the maximum heat lift, \Tpc\ is the pre-cooling temperature, \Tvg3\ is the temperature of V-groove 3, \strokeamp\ is the stroke half amplitude of the compressors, and \Pfill\ is the helium filling pressure.

The heat load on the 4\,K box was predicted using the thermal model and verified on the flight model during the CSL thermal balance/thermal vacuum test.  Performance in flight was unchanged (Sect.~4.3).

\begin{figure}
\centering
\includegraphics[width=9cm]{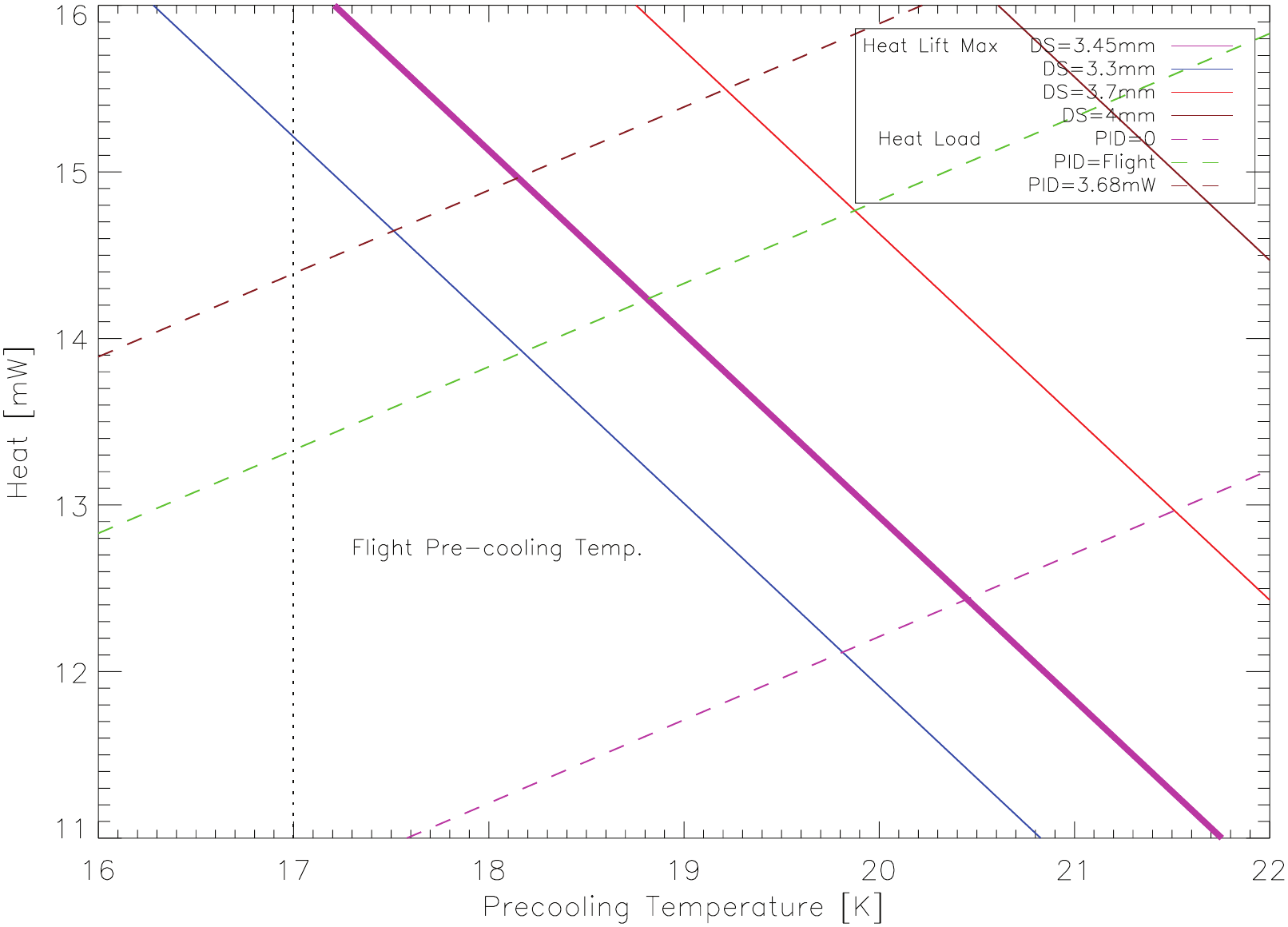}
\caption{Heat lift of the \HeJT\ cooler as a function of precool temperature, stroke half amplitude \strokeamp\ (``DS'' in the figure), and proportional, integral, differential (PID) control power.  For $\strokeamp=3.45$\,mm, a precool temperature of 19.5\,K gives the minimum required heat lift (Table~\ref{table:4Kcooler}) of 13.3\,mW.  The in-flight precool temperature of $\sim17.0$\,K (vertical dotted line, and Table\,\ref{table:SCS_performance}) allows the use of a low stroke amplitude, minimising stresses on the cooler, and provides a large margin in heat lift.}
\label{fig:heat_lift}
\end{figure}

The stroke amplitude, and to some degree the sorption cooler precool temperature, are adjustable in flight.  The interface with the sorption cooler, including the warm radiator temperature, is the most critical interface of the HFI cryogenic chain.  The \HeJT\ cooler heat load increases and its heat lift decreases as the sorption cooler precool temperature increases (see Fig.~\ref{fig:heat_lift}).  The \HeJT\ cooling power margin depends strongly on this temperature, itself driven mostly by the temperature of the warm radiator.  Warm radiator temperatures of 272\,K ($\pm10$\,K, Sect.~2.3.1) lead to sorption cooler temperatures between 16.5\,K and 17.5\,K \citep{Bersanelli2010}. It can be seen from Fig.~\ref{fig:heat_lift} that at a 20\,K precool temperature even the largest possible stroke amplitude leaves no margin.

The two mechanical compressors produce microvibrations and also induce electromagnetic interference, potentially affecting the science signals of bolometers. The risks associated with these effects were taken into account early in the design of the HFI by phase-locking the sample frequency of the data to a harmonic of the compressor frequency.

%2.3.3
\subsubsection{Dilution cooler}

The dilution cooler operates on an open circuit using a large quantity of $^4$He and $^3$He stored in four high pressure tanks.  The major components of the system are shown in Fig.~\ref{fig:100mK}, including a JT expansion valve producing cooling power for the ``1.4\,K stage'' of the FPU and pre-cooling for the dilution cooler.  The gas from the tanks (300\,bar at the start of the mission) is reduced to 19\,bar through two pressure regulators, and the flow through the dilution circuits is regulated by a set of discrete restrictions chosen by telecommand.  The flow rates for different configurations of the restrictions are given in Table~\ref{table:heliumflow} for the hot spacecraft case. The flow depends on the restriction temperature through changes of the helium viscosity.

\begin{table}  
\begingroup
\newdimen\tblskip \tblskip=5pt
\caption{Helium flow options for the dilution cooler.}
\label{table:heliumflow} 
\nointerlineskip
\vskip -3mm
\footnotesize
\setbox\tablebox=\vbox{
   \newdimen\digitwidth 
   \setbox0=\hbox{\rm 0} 
   \digitwidth=\wd0 
   \catcode`*=\active 
   \def*{\kern\digitwidth}
   \newdimen\signwidth 
   \setbox0=\hbox{+} 
   \signwidth=\wd0 
   \catcode`!=\active 
   \def!{\kern\signwidth}
\halign{\hbox to 1.0in{#\leaderfil}\tabskip=2em&
        \hfil#\hfil&
        \hfil#\hfil&
        \hfil#\hfil\tabskip=0pt\cr
\noalign{\doubleline}
\omit&$^4$He&$^4$He + $^3$He&$^3$He\cr
\omit\hfil Flow Level\hfil&[$\mu$mol\,s\mo]&[$\mu$mol\,s\mo]&[$\mu$mol\,s\mo]\cr
\noalign{\vskip 3pt\hrule\vskip 5pt}
Fmin2 & 14.5 & 19.8 & 5.4\cr
Fmin  & 16.6 & 22.9 & 6.3\cr
FNOM1 & 20.3 & 27.8 & 7.5\cr
FNOM2 & 22.6 & 30.8 & 8.2\cr
\noalign{\vskip 5pt\hrule\vskip 3pt}}}
\endPlancktable
\endgroup
\end{table}

The heat lift margin HL$_{\rm margin}$ (available for temperature regulation) is determined by:

\begin{itemize}

\item \HeFlow, the flow rate of the helium isotopes in \mumols\ given by
  the chosen restriction configuration --- for each restriction the flow is expressed  
  when the temperature of the dilution cooler control unit (DCCU) is at 273\,\hbox{K};

\item \TDCCU, the temperature of the DCCU in flight;

\item the heat loads from the bolometer plate, determined by the temperature 
      difference between \Tbolo\ and \Tdilu, and from the 1.4\,K stage at a 
      temperature of \T16; and

\item \Tdilu, the temperature of the dilution cold end.

\end{itemize}

This margin is given by: 
\begin{align}
  {\rm HL}_{\rm margin} \,[{\rm nW}] &= 3.2 \,10^{-3} \, \HeFlow [\mumols]
%                            \nonumber\\  
 %         &\qquad\qquad
           \Tdilu^2 \left({\TDCCU\over 273}\right)^{-1.5} \nonumber \\
   &\quad - 250\,(\T16 - 1.28) %\nonumber \\
   %&\quad 
   - 20\,(\Tbolo - \Tdilu) \nonumber \\
   &\quad - 490,
\end{align}
where the last two lines are, respectively, the heat loads from the 1.4\,K stage, the bolometer plate, and fixed conduction parasitics, all in nanowatts.

\begin{figure}
  \centering
\includegraphics[width=9cm]{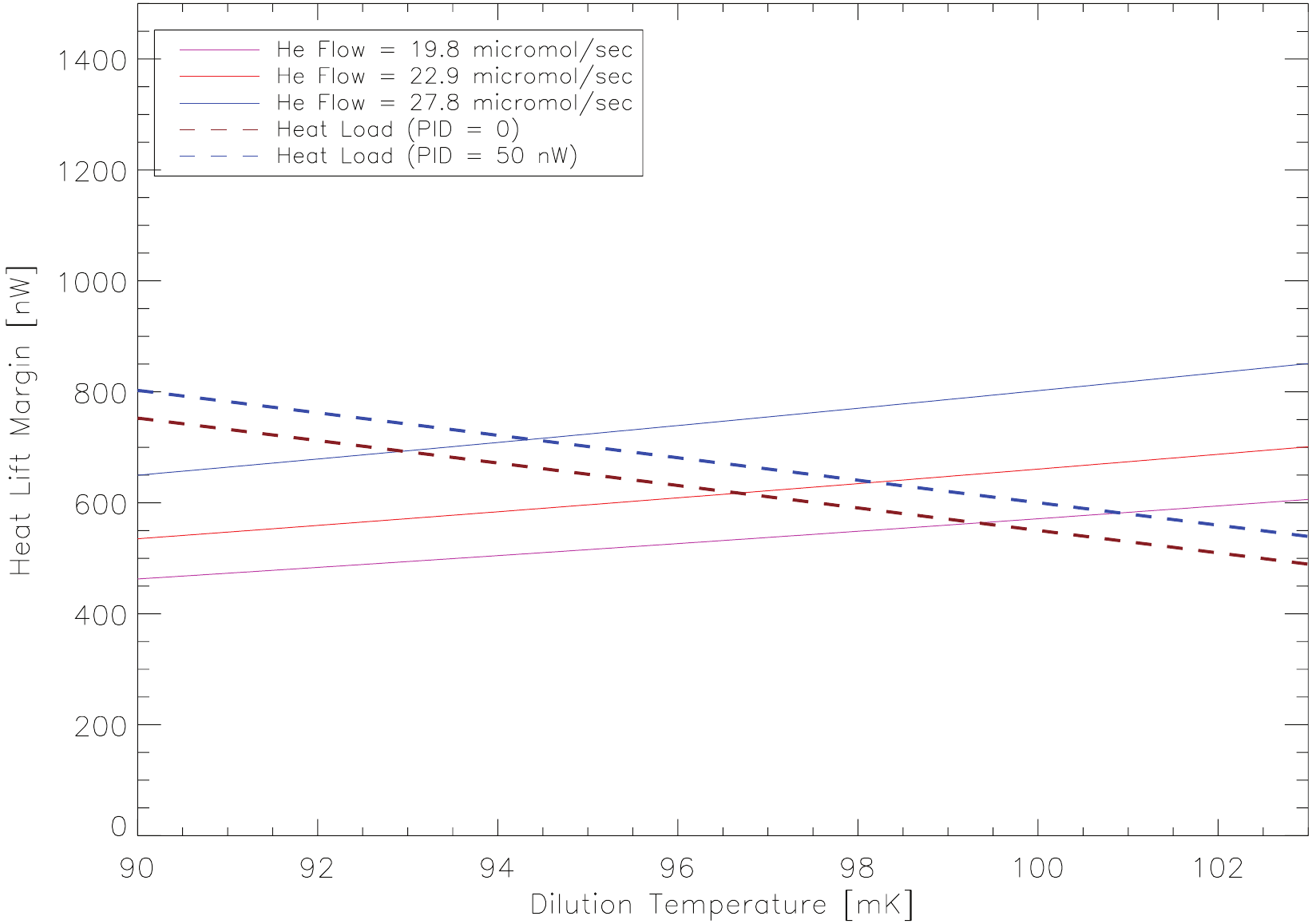}
  \caption{Heat lift margin of the dilution cooler as a function of dilution temperature and helium flow, with the dilution cooler control unit at 19\,\deg\hbox{C}.  For operation in flight at 101\,mK, 115\,nW is available to accommodate the heating from cosmic rays and for temperature regulation even at the lowest flow (Fmin2 = 19.8\,$\mu$mol\,\mo) used in flight (see Table~\ref{table:heliumflow} and Sect.~4.4).}
\label{fig:dilution_flow}
\end{figure}

As shown in Fig.~\ref{fig:dilution_flow}, even at the highest temperature of the dilution panel in the spacecraft (19\,\deg C, thus minimum flow for a given restriction due to higher viscosity of the helium), 101\,mK can be achieved in flight with the lowest flow (Fmin2), with 115\,nW of power available for regulation, and with the extra heat input in flight from cosmic rays and vibration.

%2.4
\subsection{Temperature control}

%2.4.1
\subsubsection{Service Vehicle Module}

The SVM thermal control system maintains all SVM components at their proper temperature, minimizes the heat flux to the payload module, and guarantees a stable thermal environment to the payload module.  No specific effort is made to control the temperature of the thrusters or their heating effect.  

The solar panels are nearly normal to the Sun and, as mentioned in Sect.~2.1, operate at 385\,\hbox{K}. To minimize their heat flux to the SVM they are covered by 20~layers of multilayer insulation (MLI) and mounted with low thermal conductivity titanium brackets.  The launch vehicle adaptor is always Sun-exposed; to minimize its flux to the SVM, it is covered with MLI to the maximum extent and, where not possible, with a proper ``cold'' thermo-optical coating.  

The SVM is octagonal.  Internal SVM components are distributed on the eight panels, each of which is provided with its own external radiator.  Three of the panels --- power, downlink transponder, and startracker/computer (STR/DPU) --- have temperature control of some sort.  The power panel is stable in dissipation and consequently in temperature. The other two panels experience temperature fluctuations for different reasons (see Sect.~5.1).

Another three panels are dedicated to the sorption coolers (Sect.~2.3.1).  The sorption cooler compressor elements have intermittent high dissipation.  To minimize the impact on the rest of SVM, the sorption cooler cavity is internally wrapped by \hbox{MLI}.  To maintain the sorption coolers above their minimum temperature limits (253\,K non-operating and 260\,K operating), several heaters have been installed on the eight horizontal heat pipes and grouped in seven heater lines working at different temperatures.  Temperature control of the warm radiator is described in Sect.~2.4.2 below.

The warm compressors of the \HeJT\ cooler are installed on another panel equipped with heaters to maintain a minimum temperature.  The LFI radiometer electronic backend units (REBAs) and the dilution cooler control unit (DCCU)  are installed on the last panel.  Heaters with PID control maintain the REBAs and the DCCU at a stable temperature at 2.75\, \deg\hbox{C}.  

The top surface of the SVM is covered with 20~layers of MLI to minimize radiation onto the payload.

%2.4.2
\subsubsection{Warm radiator}

The warm radiator is the means by which the heat generated inside the compressor elements during heat-up and desorption is rejected to cold space.  (Most of the heat lifted from the focal plane is rejected to space by the V-groove radiators.)  The temperature of the warm radiator determines the temperature of the hydride beds during the absorption part of the cooler cycle.  The lower the temperature, the lower the pressure of hydrogen on the low pressure side of the JT expansion, and therefore the lower the temperature of the thermal interfaces with the HFI and LFI (LVHX1 and LVHX2, respectively).  The strict requirement on the temperature of LVHX1 given in Table~\ref{table:SCS_requirements} translates into a requirement on the temperature of the warm radiator.

Temperature control of the warm radiator is achieved with seven independent heater lines.  The temperature of the warm radiator depends on the total heat input from the sorption cooler plus the heaters.  A listing of the heaters along with their control bands is given in Table~\ref{table:WRheaters}.  The average of three warm radiator thermistors, calculated once per minute for each loop, is used for control of all seven heaters.

\begin{table} 
\begingroup
\newdimen\tblskip \tblskip=5pt
\caption{Power levels and temperature ranges of the seven independent heater lines of the warm radiator temperature control system.  Heaters are always on, always off, or controlled on-off (``bang-bang'' control).  The sample configuration was typical during the first months of operation.}
\label{table:WRheaters} 
\nointerlineskip
\vskip -6mm
\footnotesize
\setbox\tablebox=\vbox{
   \newdimen\digitwidth 
   \setbox0=\hbox{\rm 0} 
   \digitwidth=\wd0 
   \catcode`*=\active 
   \def*{\kern\digitwidth}
   \newdimen\signwidth 
   \setbox0=\hbox{+} 
   \signwidth=\wd0 
   \catcode`!=\active 
   \def!{\kern\signwidth}
\halign{\hbox to 1.1in{#\leaderfil}\tabskip=2em&
        \hfil#\hfil&
        \hfil#\hfil&
        \hfil#\hfil\tabskip=0pt\cr
\noalign{\doubleline}
\omit&Power&Sample&$T$ Range\cr
\omit\hfil Loop Number\hfil&[W]&Configuration&[\deg C]\cr
\noalign{\vskip 3pt\hrule\vskip 5pt}
13&78&On&$-8$ to $-9$\cr
14&78&On&$-9$ to $-10$\cr
*8&91&On-Off&$-10$ to $-11$\cr
12&91&Off&$-11$ to $-12$\cr
32&91&Off&$-12$ to $-13$\cr
28&91&Off&$-13$ to $-14$\cr
27&91&Off&$-14$ to $-15$\cr
\noalign{\vskip 5pt\hrule\vskip 3pt}}}
\endPlancktable
\endgroup
\end{table}

%2.4.3
\subsubsection{20\,K stage}

LVHX1 provides a temperature below 18\,K, with fluctuations driven by the cooler (bed-to-bed variations, cycling, instabilities in the hydrogen liquid-gas flow after the JT, etc.).  Stabilization of the temperature of this interface with the HFI is not necessary, as temperature control of the subsequent colder stages is more efficient and very effective.

\begin{figure}
\begin{center}
\includegraphics[width=8.7cm]{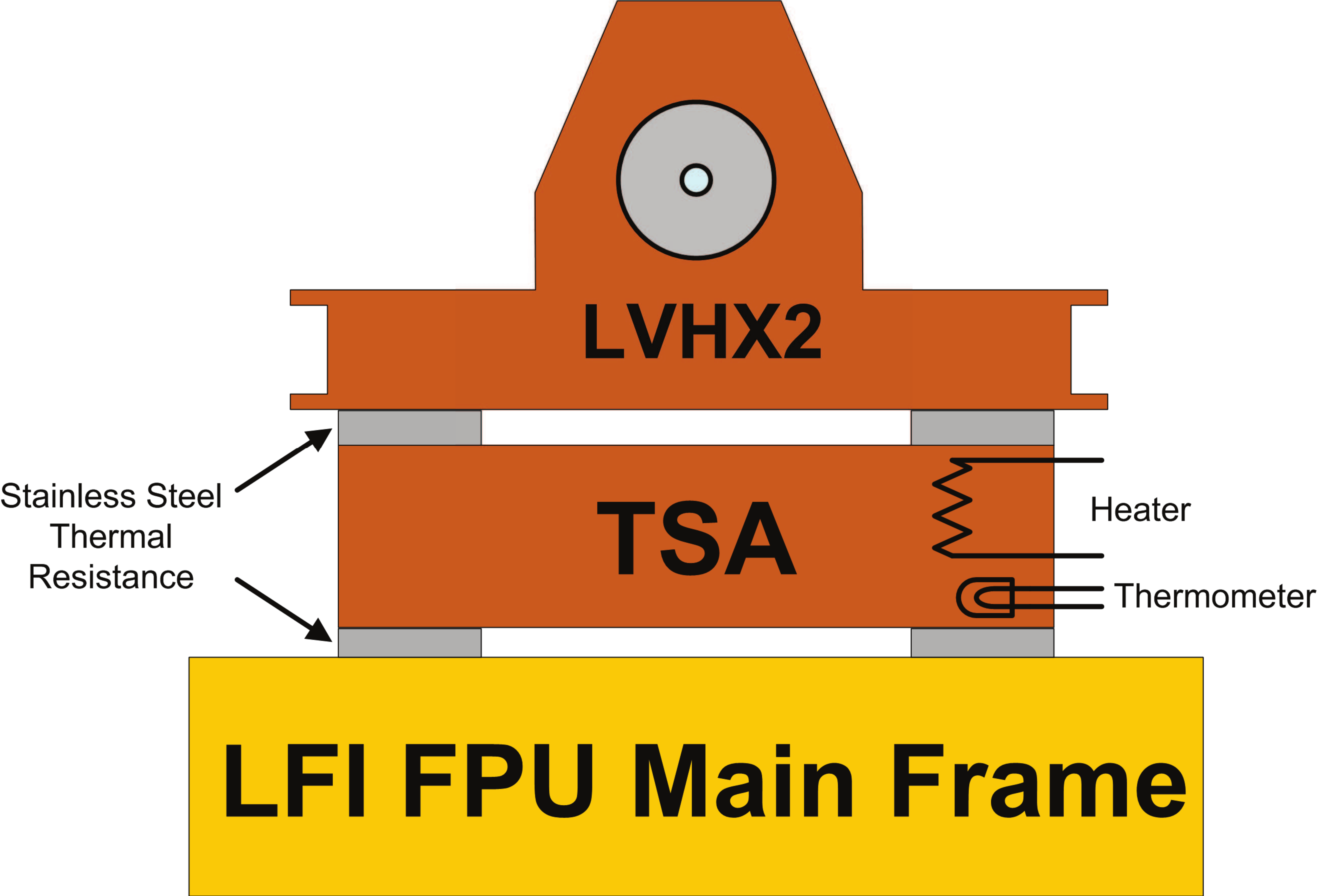}
\end{center}
\caption{Schematic of the temperature stabilization assembly (TSA).  The heater is controlled by a hybrid PID+predictive controller.  Stainless steel strips provide thermal resistance with a high heat capacity.    }
\label{fig:TSA}
\end{figure}

LVHX2 provides a temperature of about 18\,\hbox{K}.  To reduce cold end fluctuations transmitted to the radiometers, an intermediate stage, the temperature stabilization assembly (TSA, Fig.~\ref{fig:TSA}), is inserted between LVHX2 and the LFI \hbox{FPU}.   The TSA comprises a temperature sensor and heater controlled by a hybrid PID and predictive controller, plus a high-heat-capacity thermal resistance.   The set-point temperature of the TSA is an adjustable parameter of the sorption cooler system, chosen to provide dynamic range for control, but not to require more than 150\,mW of power from the heater.  As the hydride in the sorption cooler ages, the return gas pressure and thus the temperature of LVHX2 rise slowly.  The temperature of the warm radiator (Sect.~2.4.2) also affects the temperature of LVHX2.  Small adjustments of the set-point temperature are required now and then.  There is no other temperature regulation in the LFI focal plane.

The heater/thermometer is redundant. The thermal resistance between LVHX2 and the TSA causes a temperature difference proportional to the heat flux through it; thus the stage is stabilized to (or above) the highest temperature expected for LVHX2.  Thermal variation of the controlled stage is determined by several factors, including thermometer resolution, thermometer sampling/feedback cycle time, heater power resolution, and heater power slew rate.  Attached on the other side of  the controlled stage is another thermal resistance, through which flows all of the heat lifted from \hbox{LFI}. The transfer function of thermal noise from the controlled stage to the LFI FPU is determined by this resistance and by the effective heat capacity of LFI.

%2.4.4
\subsubsection{4\,K, 1.4\,K, and 0.1\,K stages} 
\label{4-1.4-0.1_stages}

The noise produced by thermal fluctuations of sources of stray radiation should be small relative to the photon noise if no correction is applied to the signal. This leads to a conservative requirement \citep{Lamarre2003} that the temperatures of the cryogenic stages that support optical elements must meet the stability requirements given in Table~\ref{table:HFIstability}.

Active thermal control of the 4\,K, 1.4\,K, and 100\,mK stages \citep{2003SPIE.4850..740P, 2000NIMPA.444..413P} is needed to meet these requirements.  Temperature is measured with sensitive thermometers made of optimised NTD Ge \citep{piat.JLTP.2001, 2002AIPC..605...79P} and read out by the same electronics as the bolometers.  Details of the temperature stability tests are given by \cite{Pajot2010}.

\begin{table} 
\begingroup
\newdimen\tblskip \tblskip=5pt
\caption{Temperature stability requirements on HFI components, over the frequency range 16\,mHz--100\,Hz.}
\label{table:HFIstability} 
\nointerlineskip
\vskip -6mm
\footnotesize
\setbox\tablebox=\vbox{
   \newdimen\digitwidth 
   \setbox0=\hbox{\rm 0} 
   \digitwidth=\wd0 
   \catcode`*=\active 
   \def*{\kern\digitwidth}§
   \newdimen\signwidth 
   \setbox0=\hbox{+} 
   \signwidth=\wd0 
   \catcode`!=\active 
   \def!{\kern\signwidth}
\halign{\hbox to 2.5in{#\leaderfil}\tabskip=1em&
        \hfil#\hfil\tabskip=0pt\cr
\noalign{\doubleline}
\omit\hfil Component\hfil&Requirement\cr
\noalign{\vskip 3pt\hrule\vskip 4pt}
4\,K horns and filters (30\% emissivity)&$\leq10$\muKHz\cr
\noalign{\vskip 2pt}
1.4\,K filters (20\% emissivity)&        $\leq28$\muKHz\cr
\noalign{\vskip 2pt}
0.1\,K bolometer plate&                  $\leq20$\,nK\,Hz$^{-1/2}$\cr
\noalign{\vskip 4pt\hrule\vskip 3pt}}}
\endPlancktable
\endgroup
\end{table}

Regulation of the stages is achieved by active control of low frequency fluctuations ($f \lsim 0.1$\,Hz) and passive filtering of high frequency fluctuations ($f\gsim 0.1$\,Hz). The active system uses the NTD Ge thermometers mentioned above and a PID control implemented in the on-board software.  Each heater is biased by a 24\,bit ADC (made of two 12\,bit ADCs). A passive electrical circuit connects the ADC to the heater to fix the maximum heat deposition and the frequency range.

\begin{figure*}
\centering
\leavevmode
\includegraphics[width=15cm]{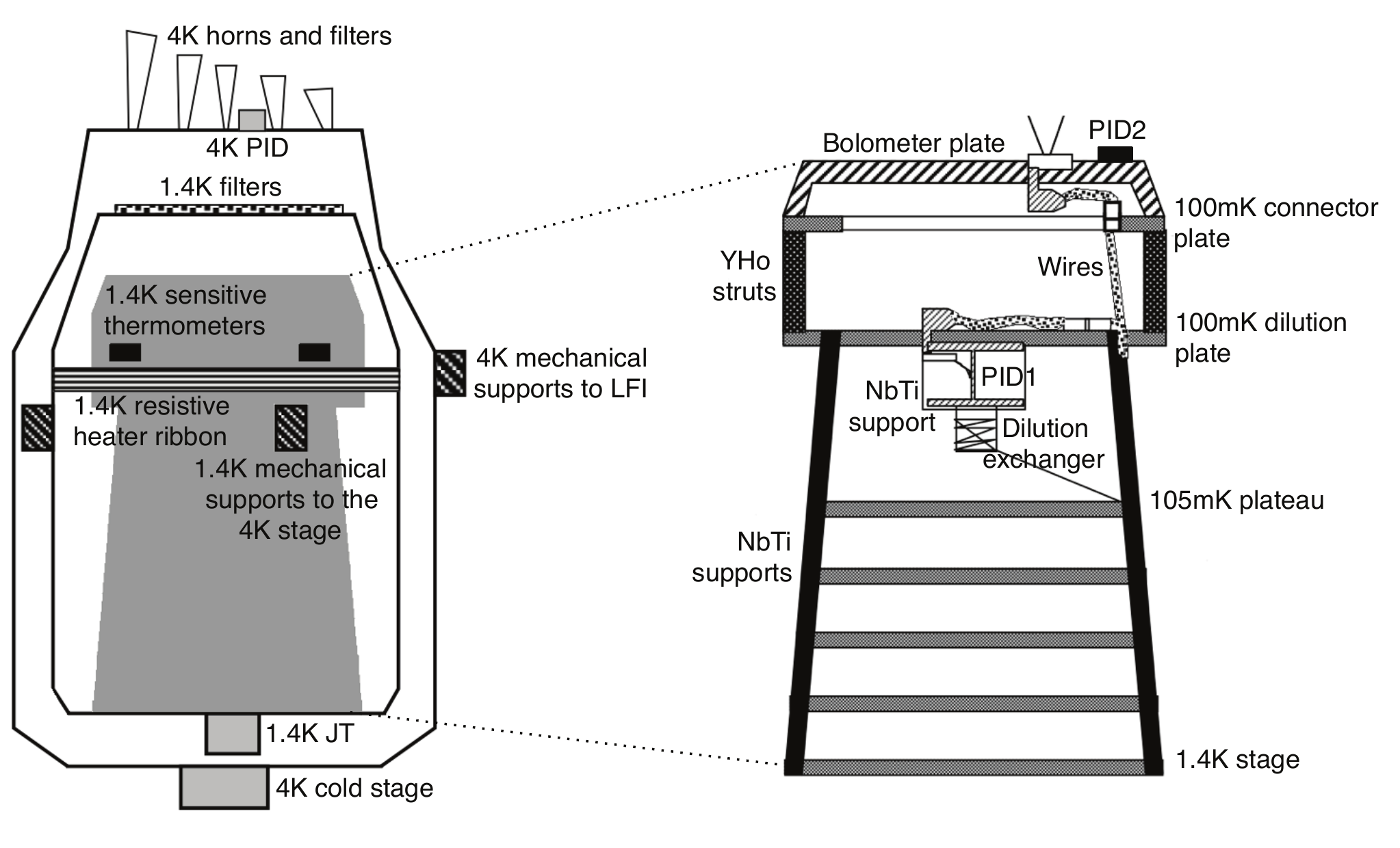}
\caption{Thermal control architecture of the 4\,K and 1.4\,K stages ({\it left\/}) and the 0.1\,K stage ({\it right\/}) .}
\label{fig:4K_16K_01K_control}
\end{figure*}

\smallskip\noindent
{\bf 4\,K} --- The main sources of thermal fluctuations on the 4\,K plate are the \HeJT\ cooler and the mechanical supports to LFI (Fig.~\ref{fig:4K_16K_01K_control}), which conduct thermal fluctuations introduced by the sorption cooler on the LFI chassis. The actively controlled heater is a ring located at the top of the cylindrical part of the 4\,K box (Fig.~\ref{fig:4K_16K_01K_control}).  Passive filtering is provided by the thermal path and the heat capacity of the 4\,K plate and box.

\smallskip\noindent
{\bf 1.4\,K} --- The 1.4\,K stage is a cylindrical box with a conical top (Fig.~\ref{fig:4K_16K_01K_control}).  Temperature fluctuations arise: (i)~at the bottom of the box where the JT expansion is located; and (ii)~on the side of the cylinder, at about 2/3 of its length, where mechanical supports attach at three points symmetrically located on the circumference.  To use the natural symmetry of this stage, the heater is a ribbon placed after the mechanical supports.  A sensitive thermometer on the 1.4\,K filter plate is used as the sensor in the regulation loop. Passive filtering is provided by the long thermal path between the sources of fluctuations and the filters, as well as by the heat capacity of the optical filter plate.

\smallskip\noindent
{\bf 0.1\,K} --- The principal sources of temperature fluctuations on the 100\,mK bolometer plate are the dilution cooler itself, the background radiation, and fluctuations in cosmic rays. This stage has been carefully optimised, since it is one of the main sources of noise on the bolometers themselves. The principles of the 100\,mK architecture are \citep{PhD.Piat}: (i)~control all thermal paths between sources of temperature fluctuations and parts that must be stable; (ii)~actively control the structure around the bolometer plate and  the bolometer plate itself to ensure long period stability; and (iii)~low-pass filter the structure to remove artifacts of the active system and to allow short-term stability.

Figure~\ref{fig:4K_16K_01K_control} shows a schematic of the 100\,mK architecture. The mechanical support of the heat exchanger consists of struts of Nb-Ti alloy and plates used to thermalise the heat exchanger tubes and the wiring.  A counterflow heat exchanger is thermally connected to all plates except the coldest one, the dilution plate. It goes directly from the next-to-last plate (at about 105\,mK) to the dilution exchanger. The dilution exchanger is a cylinder around which the dilution tubes are wound. The first regulation system (PID1) is directly attached to this cylinder and provides stability on long time scales. It is a hollow Nb-Ti alloy cylinder containing an I-shaped piece of copper with redundant PIDs (i.e., two heaters and two thermometers) in the thin part of the I \citep{2003SPIE.4850..740P}. Its function is to actively damp fluctuations induced by the dilution cooler.  The dilution plate supports the PID1 box,  wiring, and connectors. Yttrium-holmium (YHo) struts support the bolometer plate. They provide passive filtering with a thermal time constant of several hours thanks to a very large increase of heat capacity in YHo at low temperature \citep{PhD.Madet,PhD.Piat}. The bolometer plate is made of stainless steel covered with a 250\,$\mu$m film of copper, itself covered with a thin gold plating. This architecture was defined after thermal simulation of high energy particle interactions.  A second regulation stage (PID2) is placed directly on the bolometer plate. It ensures control of the absolute temperature of the plate and compensates for fluctuations induced by external sources such as cosmic rays or background radiation fluctuations.  Ground tests showed that the regulation systems would meet requirements as long as the in-flight fluctuations of the heat loads did not exceed predictions.

%2.5
\subsection{Dependencies}

\begin{figure}
\begin{center}
\includegraphics[width=9.4cm]{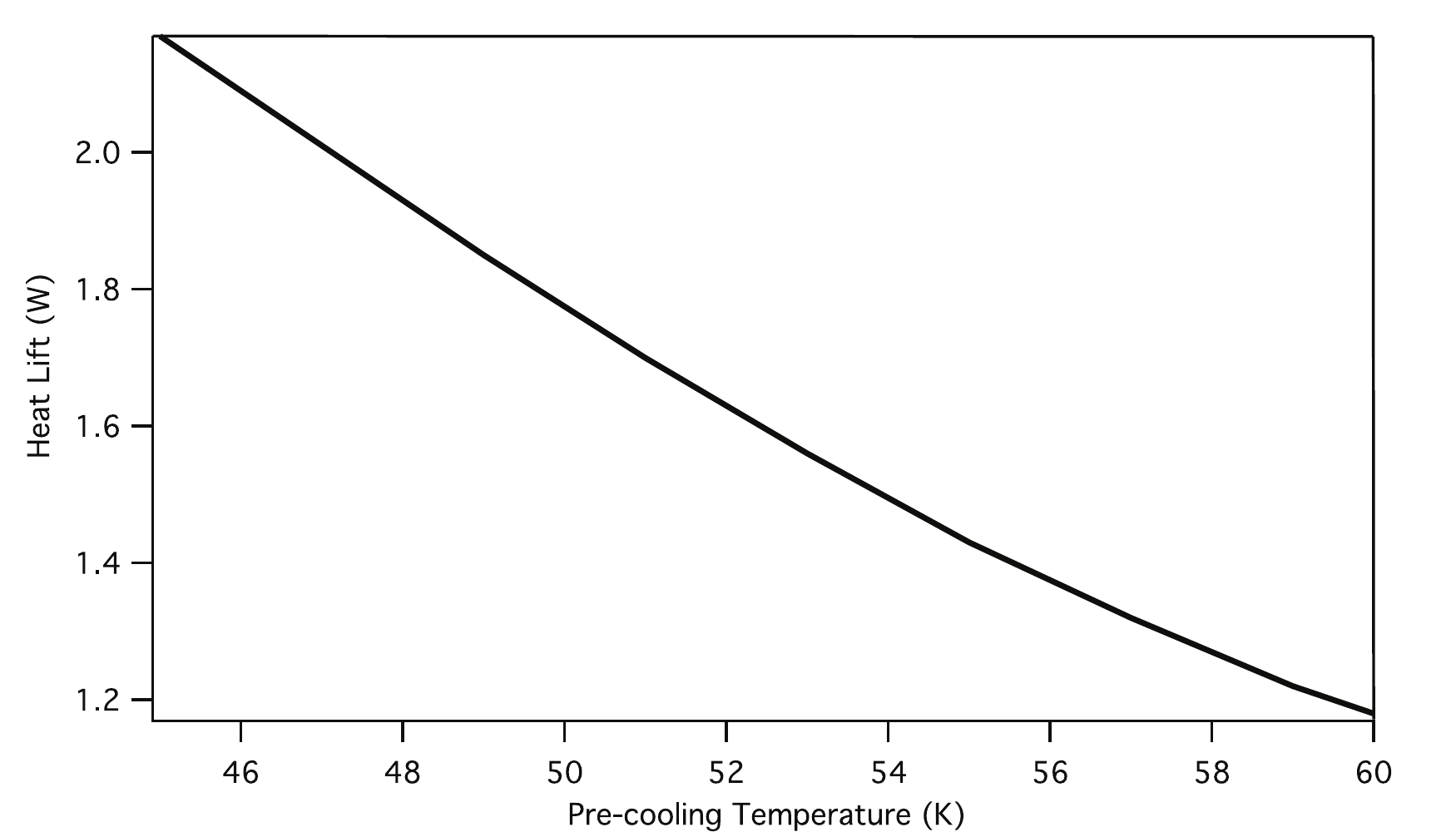}
\caption{Heat lift of the sorption cooler as a function of precool (V-groove 3) temperature.}
\label{fig:SCS_heatlift_vs_precool}
\end{center}
\end{figure}

\begin{figure}
\begin{center}
\includegraphics[width=9.0cm]{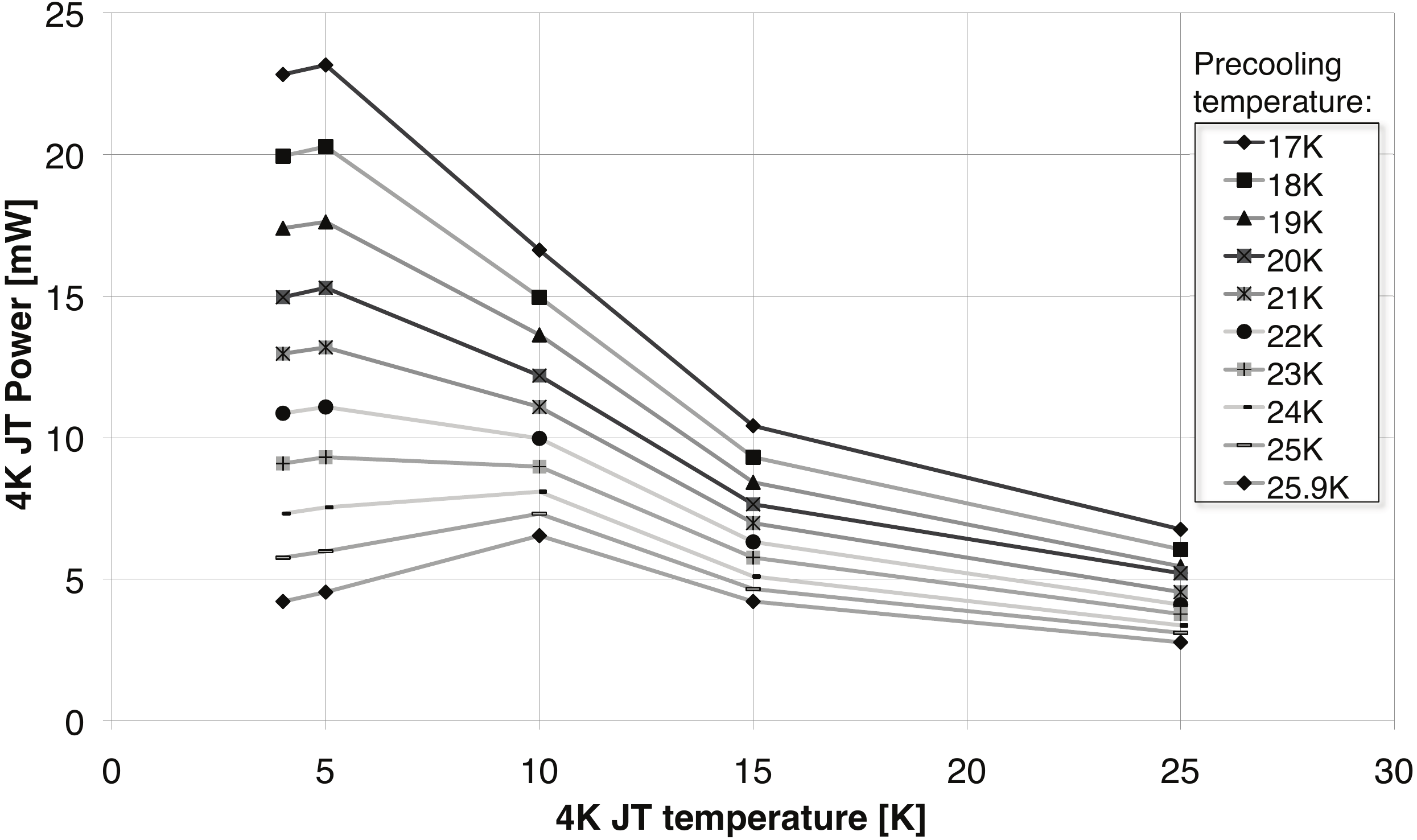}
\caption{Heat lift of the \HeJT\ cooler as a function of cold-end and precooling temperatures.}
\label{fig:4HeJTliftvsPCT}
\end{center}
\end{figure}

The \Planck\ cooling chain is complicated, with critical interfaces.  The most critical are: 

\begin{itemize}

\item the temperature of the passive cooling/sorption cooler interface at V-groove~3. The heat lift of the sorption cooler depends steeply on this precool temperature (Fig.~\ref{fig:SCS_heatlift_vs_precool}).

\item the \HeJT\ cooler helium precooling interface with the LVHX1 liquid reservoir of the sorption cooler.  The heat lift of the \HeJT\ cooler depends steeply on the cold-end temperature (Fig.~\ref{fig:4HeJTliftvsPCT}).  The precool temperature is also the dominant parameter for the heat load of the \HeJT\ cooler. 
\end{itemize}

Temperature fluctuations of the sorption cooler are driven by the variation of pressure with hydrogen concentration in the hydride beds during the desorption cycle, as well as by inhomogeneities in the hydride beds. This leads to fluctuations of rather large amplitude at both the bed-to-bed cycle frequency and the overall 6-bed cycle frequency. These are controlled at LVHX2 (Sect.~2.4.3), but not at LVHX1, the \HeJT\ cooler interface.  Temperature fluctuations in precooling the helium of the \HeJT\ cooler are  transferred to the cold head at 7--8\,mK\,K\mo.

Active temperature regulation of the 4\,K outer shell (Sect.~2.4.4) shields the lower temperature stages from temperature fluctuations coming from the warmer stages.

As long as the  \HeJT\ cooler precools the $^3$He and $^4$He gas below 4.7\,K, the dilution cooler operates in its nominal configuration.  It was found during system tests (Sect.~3) that it can operate with a precool as high as 5\,K, although this regime is expected to be rather unstable.  The heat lift of the dilution cooler (proportional to the flow rate of the helium gases and to the square of the dilution cold head temperature) is not a critical parameter of the cooling chain, as the temperature can adjust at the expense of a weak loss in sensitivity. The flow rate is adjustable (Sect.~2.3.3), and can be chosen to ensure the required mission lifetime, possibly trading off bolometer temperature (thus instantaneous  sensitivity) with lifetime.

The incoming $^3$He and $^4$He are precooled by the 4\,K stage before JT expansion to 1.4\,\hbox{K}.  They flow to the dilution capillary through a standard counterflow heat exchanger with the outgoing $^3$He-$^4$He mixture.  The passive and active cooling systems are highly interlinked and dependent on each other.  Table~\ref{table:dependencies} summarizes the principal dependencies.

\begin{table*}[tmb]
\begingroup
\newdimen\tblskip \tblskip=5pt
\caption{Dependencies in the \Planck\ cryosystem.}
\label{table:dependencies}
\nointerlineskip
\vskip -3mm
\footnotesize
\setbox\tablebox=\vbox{
   \newdimen\digitwidth 
   \setbox0=\hbox{\rm 0} 
   \digitwidth=\wd0 
   \catcode`*=\active 
   \def*{\kern\digitwidth}
   \newdimen\signwidth 
   \setbox0=\hbox{+} 
   \signwidth=\wd0 
   \catcode`!=\active 
   \def!{\kern\signwidth}
\halign{\hbox to 1.8in{#\leaderfil}\tabskip=1.0em&
    #\hfil&
    \vtop{\hsize=2.03in\noindent\tolerance=1600\hbadness=1600\hangindent=1em\hangafter=1\strut#
          \strut\par}&
    \vtop{\hsize=1.33in\noindent\tolerance=1600\hbadness=1600\hangindent=1em\hangafter=1\strut#
          \strut\par}\tabskip=0pt\cr
\noalign{\doubleline}
\omit\hfil Component\hfil&\omit\hfil Affected By\hfil&\omit\hfil Affects\hfil&\omit\hfil Trade-off\hfil\cr
\noalign{\vskip 3pt\hrule\vskip 5pt}
{\bf V-groove 3 temp.}&Sorption cooler mass flow&Efficiency of the sorption cooler system; sorption cooler required mass flow; sorption cooler power \& cycle time.\cr
\noalign{\vskip 8pt}
{\bf Warm Radiator temp.}&Sorption cooler power&$T_{\rm LVHX1,2}$\cr
\noalign{\vskip 2pt}
\omit&                         Warm radiator heater power&\HeJT\ cooler efficiency\cr
\noalign{\vskip 2pt}
\omit&                                                   &\HeJT\ cooler heat load\cr
\noalign{\vskip 8pt}
{\bf Warm Radiator stability}&  Warm radiator heater control&$T_{\rm LVHX1,2}$ stability\cr
\noalign{\vskip 8pt}
{\bf Sorption Cooler System}&V-groove 3 temperature&Sorption cooler heat lift\cr
\noalign{\vskip 2pt}
\omit\hglue 2em\vtop{\hbox{Power}\hbox{Cycle time}\hbox{TSA power}\hbox{LPSB}}&
                               Warm radiator temperature&Sorption cooler temperature; \HeJT\   
                               cooler efficiency; 1.4\,K efficiency\cr
\noalign{\vskip 2pt}
\omit&                         Warm radiator stability&\omit
            \vtop{\hbox{Sorption cooler stability}
                  \hbox{Sorption cooler lifetime}
                  \hbox{Temperature stability of 4\,K box}
                  \hbox{Temperature stability of 4\,K loads}
                  \hbox{Temperature and stability of LFI}}\cr   
\noalign{\vskip 8pt}
{\bf \HeJT\ Cooler}&\omit\vtop{\hbox{LFI temperature}\hbox{$T_{\rm LVHX1}$}\hbox{V-groove 3 temperature}}&&$T$, heat lift\cr  
\noalign{\vskip 2pt}
\hglue 2em Fill pressure&&                   \HeJT\ cooler efficiency and temperature\cr
\noalign{\vskip 2pt}
\hglue 2em Stroke frequency&&                Could affect microphonics if vibration control system off.  Small effect on cooling power.\cr
\noalign{\vskip 2pt}
\hglue 2em Stroke amplitude&&                Affects gas flow&Heat lift, lifetime\cr
\noalign{\vskip 2pt}
\hglue 2em PID power&&                       Temperature and temperature stability; margin on heat lift (found to be large in thermal balance/thermal vacuum tests).&$T$, $T$ stability\cr
\noalign{\vskip 2pt}
\hglue 2em Vibration control system&&        Dilution (via microvibrations if vibration control system off)\cr
\noalign{\vskip 5pt}
{\bf Dilution Cooler}&4\,K and 1.4\,K precool temp&Instability of 4\,K and 1.4\,K if 1.4\,K too cold (unstable evaporation)&Heat lift, $T$, lifetime\cr
\noalign{\vskip 2pt}
\omit\hglue 2em\vtop{\hbox{Flow rate}\hbox{1.4 K PID power}\hbox{Dilution plate PID}\hbox{Bolometer plate PID}}&                           SVM temperature&Changes the isotope flow for a given choice of restrictions&$T$, $T$ stability, margin (long timescale)\cr
\noalign{\vskip -10pt}
\omit&&                           &$T$, $T$ stability, ~margin (short timescale)\cr
\noalign{\vskip 5pt\hrule\vskip 3pt}}}
\endPlancktable
\endgroup
\end{table*}

\begin{figure*}[htbp]
\centering
\includegraphics[width=13cm]{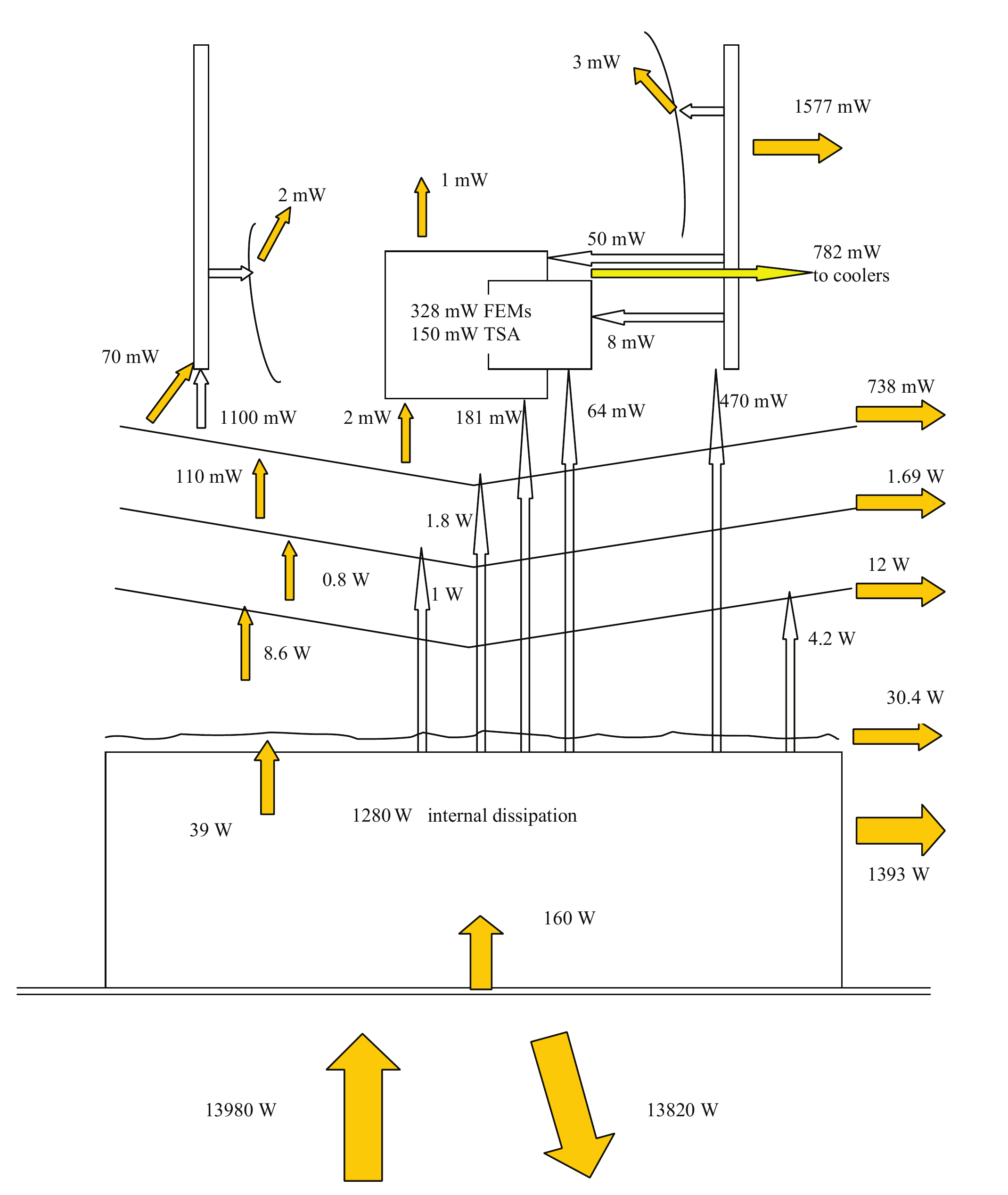}
\caption{Principal heatflows for \Planck.  Nearly 14\,kW is incident on the solar panel, but less than 1\,W reaches the \hbox{FPU}.  Passive radiative cooling and thermal isolation are the keys to the overall thermal performance.  Yellow arrows indicate radiation.  White arrows indicate conduction.}
\label{fig:Planck_heatflows}
\end{figure*}

\begin{figure*}[htbp]
\centering
\includegraphics[width=15.6cm]{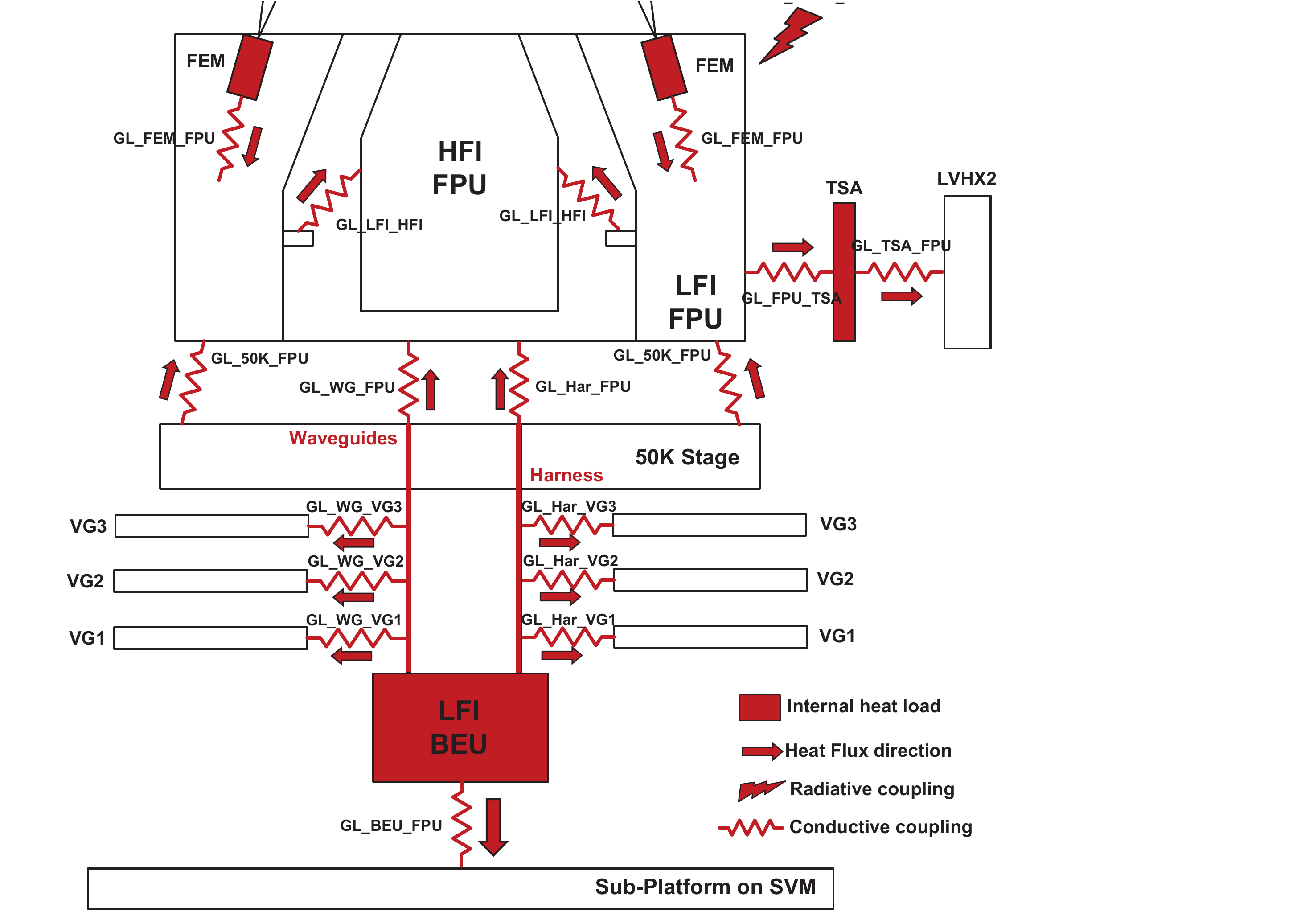}
\caption{Principal heatflows for the LFI.}
\label{fig:LFI_Q_scheme}
\end{figure*}

\begin{figure*}[htbp]
\centering
\leavevmode
\includegraphics[width=8.4cm]{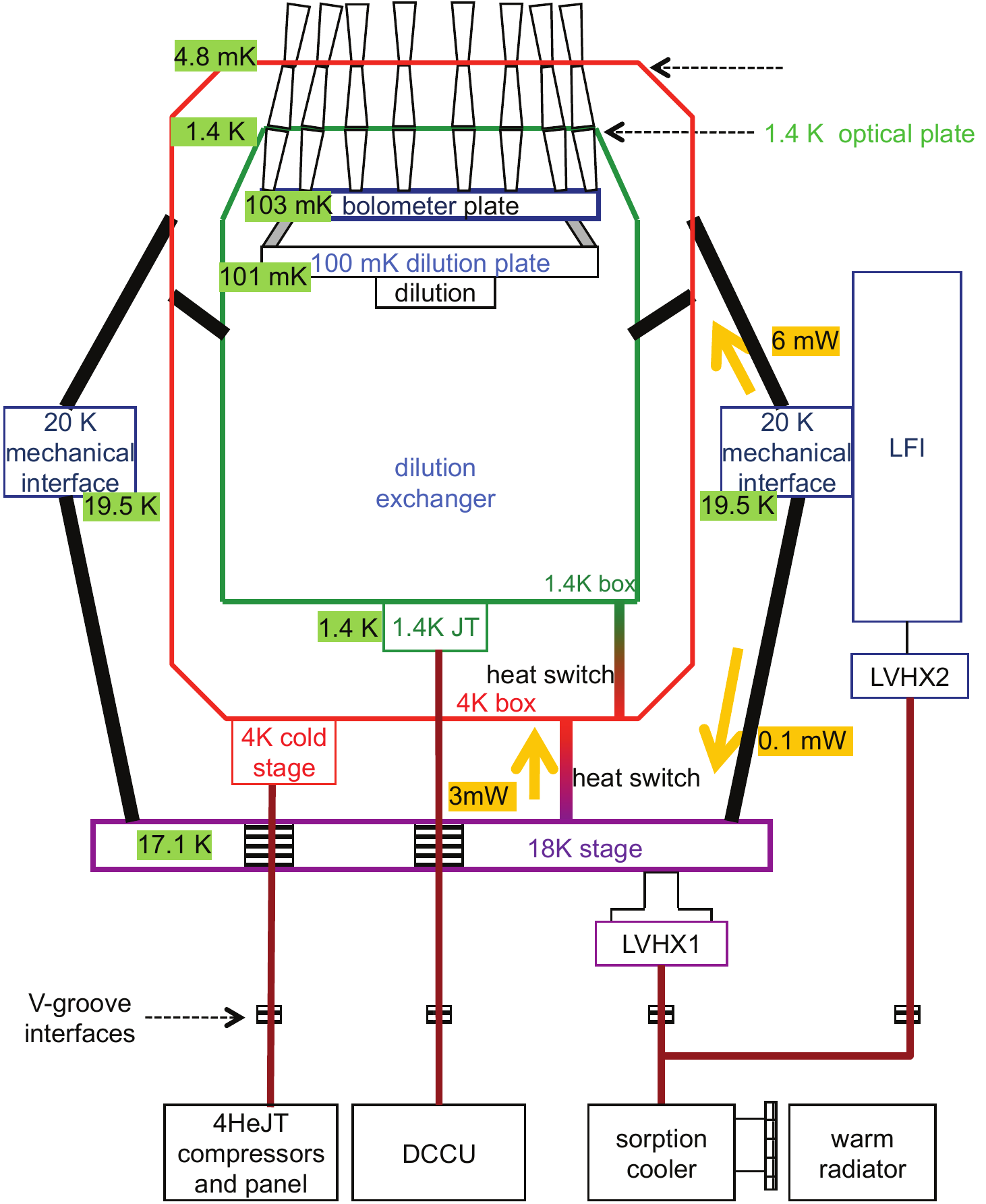}\hfill
\includegraphics[width=8.4cm]{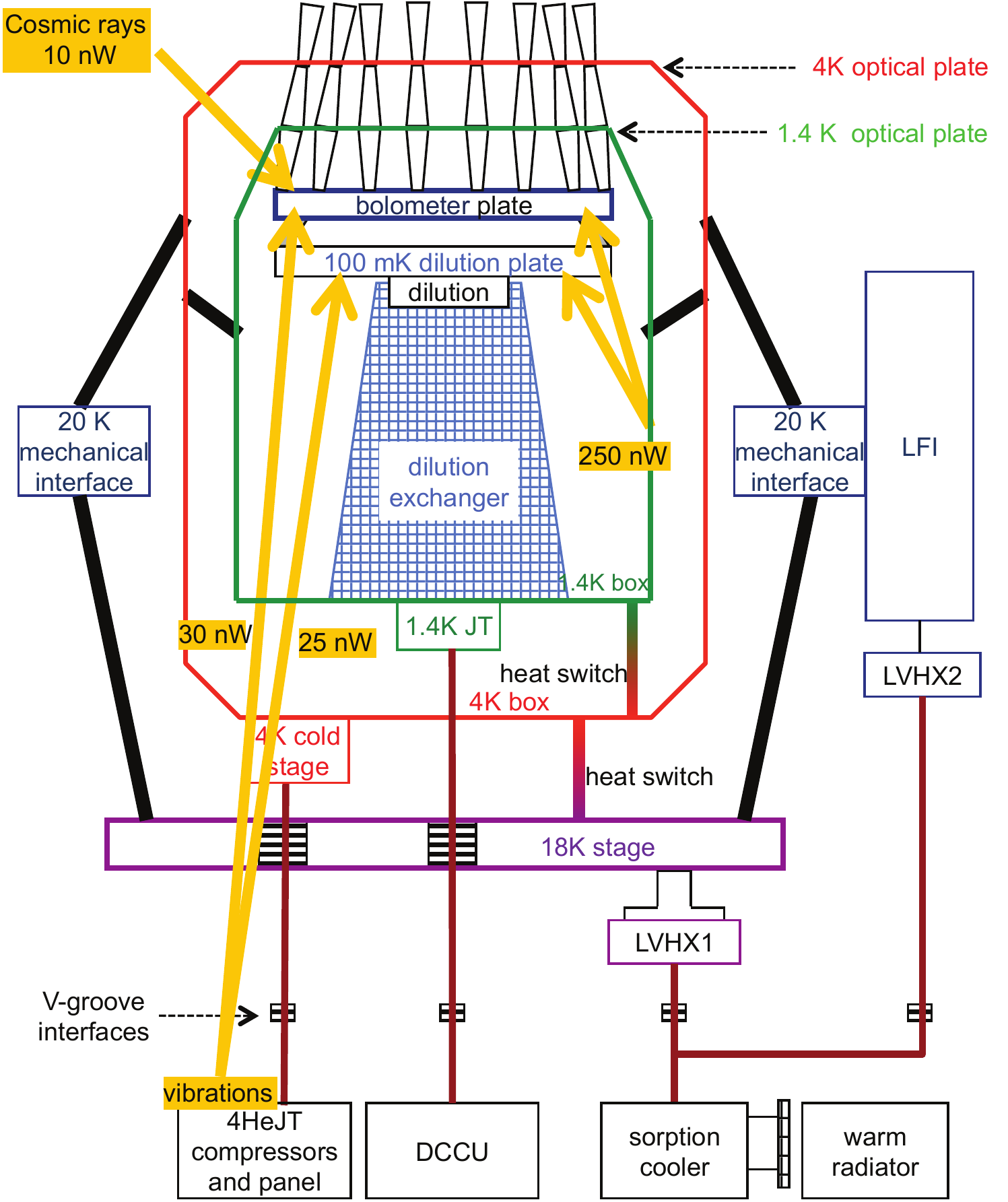}
\caption{Principal heatflows for the HFI ``4\,K'' system (\textit{left}\/) and dilution system (\textit{right}\/).}
\label{fig:HFI_Q_scheme}
\end{figure*}

The principal passive radiative and conductive heatflows and interfaces for \Planck\ are shown in Fig.~\ref{fig:Planck_heatflows}. The heat lifts and loads from the coolers are too small to include in that figure.  Expanded diagrams for LFI and HFI are shown in Figs.~\ref{fig:LFI_Q_scheme} and \ref{fig:HFI_Q_scheme}. Table~\ref{table:LFIheatflows} summarizes LFI heatflows.

\begin{table*}[tmb]                    
\begingroup
\newdimen\tblskip \tblskip=5pt
\caption{Principal heat flows for the \hbox{LFI}.  Figure~\ref{fig:LFI_Q_scheme} defines and locates the thermal couplings.}  
\label{table:LFIheatflows}
\nointerlineskip
\vskip -3mm
\footnotesize
\setbox\tablebox=\vbox{
   \newdimen\digitwidth 
   \setbox0=\hbox{\rm 0} 
   \digitwidth=\wd0 
   \catcode`*=\active 
   \def*{\kern\digitwidth}
   \newdimen\signwidth 
   \setbox0=\hbox{+} 
   \signwidth=\wd0 
   \catcode`!=\active 
   \def!{\kern\signwidth}
\halign{\hbox to 1.3in{#\leaderfil}\tabskip=3em&
        #\hfil&
        #\hfil&
        \hfil$#$\hfil\tabskip=4pt\cr
\noalign{\doubleline}
\omit&&&\omit\hfil Power\hfil\cr
\omit\hfil Unit\hfil&\omit\hfil Load From\hfil&\omit\hfil Thermal Coupling\hfil&\omit\hfil[mW]\hfil\cr
\noalign{\vskip 3pt\hrule\vskip 5pt}
{\bf FPU}&FEM total&              GL\_FEM\_FPU&   345\cr
\omit&Waveguides&                 GL\_WG\_FPU&    142\cr
\omit&LFI Harness&                GL\_Har\_FPU&   *26\cr
\omit&Bipods&                     GL\_50K\_FPU&   *39\cr
\omit&Shielding (rad)&            GR\_Shield\_FPU&!**2.1\cr
\omit&Telescope (rad)&            GR\_Tel\_FPU&   !**1.9\cr
\omit&Baffle (rad)&               GR\_Baffle\_FPU&!**1.4\cr
\omit&HFI I/F support ring (cond)&GL\_LFI\_HFI&   **-5.7\cr
\omit&HFI (rad)&                  GR\_LFI\_HFI&   **-0.1\cr
\omit&Space (rad)&                GR\_Space\_FPU& **-1.4\cr
\noalign{\vskip 5pt}
\omit&&\omit\hfil{\bf Total FPU}&                 {\bf 550}\cr
\noalign{\vskip 10pt}
{\bf SCS LVHX2}&FPU&              GL\_FPU\_TSA&   550\cr
\omit&TSA&                        GL\_TSA\_LVHX2& 100\cr
\noalign{\vskip 5pt}
\omit&&\omit\hfil{\bf Total LVHX2}&               {\bf 650}\cr
\noalign{\vskip 5pt\hrule\vskip 3pt}}}
\endPlancktable
\endgroup
\end{table*}

%SECTION 3. 

\section{Integration and test }
\label{sec:IandT}

\subsection{Introduction, philosophy, and system-level tests}

Considering the tight dependance of the instruments' active cooler chain on the spacecraft passive cooling (V-grooves and warm radiator), testing and verification could not be separated in the usual way between the instruments and the spacecraft plus isolated payload elements.  That meant that a full test of the complete flight cooling chain, the most difficult component of \Planck, had to be performed following integration of the flight system.  Interface temperatures and heat loads were designed with margins at each stage.  To keep requirements on each cooler reasonable, the worst case, for which each element of the cooling chain was at its poor-performance limit,  had a rather small overall margin.  This meant that earlier tests at the component, sub-system, and sub-assembly level, had to be very carefully designed, and required§ ´§  special ground support equipment.

Early, pre-delivery testing of the individual coolers proceeded as follows.  The sorption cooler was tested alone with water cooling substituting for the warm radiator interface.  The 4\,K cooler was tested using a laboratory 20\,K cooler for precooling.  The dilution system was tested together with an engineering model of the 4\,K cooler in an early demonstration test.  A qualification model (QM) of the  HFI focal plane unit was built and tested together with a qualification model of the {\HeJT\ cooler}.  This test was done using the large helium test tank Saturne at IAS-Orsay.  In this test, the 4\,K and 18\,K interfaces were provided by regulated interfaces linked to the helium plate between 2.7 and 4\,\hbox{K}.  In all of these tests, the non-flight interfaces could reproduce the required temperatures, but could not reproduce the dynamical interactions between coolers. 

A test with a representative full cooling chain was done in September 2005 using:

\begin{itemize}

\item a thermally representative mockup of the spacecraft, including three V-grooves;

\item an engineering model of the 20\,K sorption cooler piping and cold-end fed from hydrogen bottles regulated to the required pressure, and using a vacuum pump and two pressure controllers to simulate the high and low presure manifolds;
 
\item a thermal mockup of the LFI instrument, using heaters;

\item the QM \HeJT\ cooler; and 
 
\item  the QM HFI focal plane unit and dilution cooler.
  
\end {itemize}

This test, in the largest cryogenic test chamber at the Centre Spatial de Li\`ege, required a 4\,K liquid helium shroud to test the critical pasive cooling of the V-grooves and warm radiator.  

In March 2006, a test in a representative configuration, with the protoflight model of the spacecraft and sorption cooler warm radiator, FM1 flight sorption cooler, and LFI instrument together with the flight model of the \HeJT\ cooler, dilution cooler, and HFI focal plane unit, was performed in the same test chamber.  Only the first V-groove was mounted; the important interface with the third V-groove was simulated thermally.  This long and difficult test was not only a system test, but also the first realistic test of the  full HFI cooling chain and its influence on the instrument performance.  To shorten the cooldown phase of this test, a ``precooling loop'' of circulating helium gas  was introduced into the HFI design.  (The difficulty of adequately evacuating this loop after the end of cooldown was a problem in several tests.)  This was the first time the margin on the full cryo chain could be measured directly.  Fortunately, it was close to the maximum expected.  Most elements performed well above their worst-case limits.  It also confirmed the performance of the instruments, which was consistent with subsystems tests.  Finally, it verified the thermal model, which was ultimately able to predict the inflight cooling time within a couple of days.

The final test of the full flight cryo system, in July and August 2008, finally included the flight spacecraft and sorption cooler warm radiator, and used the FM2 sorption cooler.  It confirmed the performance of the entire system.  Details of these and earlier tests performed on the individual coolers are given in the following sections.

%3.2
\subsection{Sorption cooler}

An engineering model of the sorption cooler was built at JPL and tested with an engineering model of the cooler electronics built at the Laboratoire de Physique Subatomique et de Cosmologie in Grenoble \citep{pearson2007}.  A temperature-controlled mounting plate substituted for the warm radiator.  All requirements were verified \citep{morgante2009}.

The flight coolers were shipped fully assembled to avoid welding on the spacecraft or the use of demountable field joints to join the compressor with the cold-end piping, and possible associated contamination.  This had a substantial impact on the integration.

The thermal test program for the sorption coolers is summarized in Table~\ref{table:SCS_test_program}.  Subsystem testing was done by thermally simulating the main spacecraft interfaces and testing the flight-allowable range for both interfaces.  Requirements were met except for that on temperature fluctuations.  Although it was recognized that the fluctuation requirement itself was a poor representation of the temperature stability required for good scientific performance (Sect.~2.3.1), a ``tiger team'' was formed to investigate the source of the fluctuations.  The conclusion was that the excess fluctuations were gravitationally induced and were likely to be smaller in the micro-gravity space environment.

\begin{table*}[tmb]                    
\begingroup
\newdimen\tblskip \tblskip=5pt
\caption{Sorption cooler test program.}  
\label{table:SCS_test_program}
\nointerlineskip
\vskip -3mm
\footnotesize
\setbox\tablebox=\vbox{
   \newdimen\digitwidth 
   \setbox0=\hbox{\rm 0} 
   \digitwidth=\wd0 
   \catcode`*=\active 
   \def*{\kern\digitwidth}
   \newdimen\signwidth 
   \setbox0=\hbox{+} 
   \signwidth=\wd0 
   \catcode`!=\active 
   \def!{\kern\signwidth}
\halign{\hbox to 1.5in{#\leaderfil}\tabskip=1.5em&
     \vtop{\hsize=1.7in\hangindent=1em\hangafter=1\strut #\strut\par}&
     \vtop{\hsize=1.7in\hangindent=1em\hangafter=1\strut #\strut\par}&
     \vtop{\hsize=1.7in\hangindent=1em\hangafter=1\strut #\strut\par}\tabskip=0pt\cr
\noalign{\doubleline}
\omit&\multispan2\hfil C{\sc onfiguration}\hfil\cr
\noalign{\vskip -4pt}
\omit&\multispan2\hrulefill\cr
\omit\hfil T{\sc est}\hfil&\omit\hfil Cooler\hfil&\omit\hfil Interfaces\hfil&\omit\hfil V{\sc erification}\hfil\cr
\noalign{\vskip5pt\hrule\vskip 5pt}
Subsystem acceptance&Actual FM1 and FM2 coolers, PACE, and compressors joined in non-flight configuration.&
  Thermal simulation of the two interfaces.&
  Verified performance over flight-allowable temperature range for warm radiator and 
  final pre-cooling stage.\cr
\noalign{\vskip 5pt}
FM1 CQM test&Engineering model of PACE, compressor simulated by hydrogen gas system.&
  Engineering model of V-groove system (no compressor).&
  Heat load from sorption cooler to V-grooves.\cr
\noalign{\vskip 5pt}
FM1 spacecraft bus&FM1 flight cooler.&Flight warm radiator, V-groove 1, pre-cooler 
  thermally simulated.&
  Interaction of sorption cooler and warm radiator, minimum and maximum power.\cr
\noalign{\vskip 5pt}
FM2 spacecraft&FM2 flight cooler.&Actual spacecraft interfaces.&Requirements verification 
  with full spacecraft and payload.\cr
\noalign{\vskip 5pt\hrule\vskip 3pt}}}
\endPlancktable
\endgroup
\end{table*}

The interaction between the warm radiator and the compressor, its impact on temperature fluctuations, and the performance of the V-grooves and sorption cooler piping could not be tested at subsystem level.  These, and the interactions of all three coolers and passive components, were verified in the three high-level tests performed at CSL and described in Sect.~3.1.  All requirements on the sorption cooler system were verified in those tests. 

%3.3
\subsection{\HeJT\ and dilution coolers}

An engineering model of the \HeJT\ cooler was built at RAL using development models of the compressors.  After testing at RAL, it was transported to IAS (Orsay), integrated with an engineering model of the dilution cooler system, and used to demonstrate the ``4\,K to 0.1\,K'' cryogenic concept.  The performance of the two-cooler system and its dependence on the interface temperatures between the coolers was measured and was a key element in the definition of the cryogenic interface requirements between the JPL, RAL, and IAS teams.

A qualification model of the \HeJT\ cooler was built and tested at \hbox{RAL}.  In parallel, a qualification model of the HFI FPU was built, including a fully representative dilution cooler (including launch locks on the focal plane) and eight flight-like bolometers.  This qualification model of the FPU was tested in the Saturne 4\,K test tank at IAS for function as well as for the performance of the bolometers, filters, and feed horns.  The 18\,K and 4\,K interfaces of the FPU were provided by the 4\,K Saturne cryostat. The radiation environment of the FPU was nearly a blackbody, the temperature of which could be changed from 2.7\,K to 5\,\hbox{K}.

The September 2005 system test at CSL is described in Sect.~3.1.  The flight models were tested in Saturne in December 2007, and at CSL in the satellite level thermal test in July and August 2008.  The CSL test was the first  of the full \Planck\ cryogenic chain.  These tests showed that the sorption cooler precooling temperature was close to 17.5\,K, that the \HeJT\ cooling power margin at zero PID power was about 5\,mW for a stroke amplitude \strokeamp\ of 3.5\,mm (well below the maximum value of 4.4\,mm), and that the cold end temperature was about 4.4\,K, well below the maximum of 4.7\,K required for the operation of the dilution cooler with reasonable margins.  This showed that the system as built had significant margins, leaving a range of $\pm 2$\,mW for temperature regulation of the 4\,K stage. 

During the CSL test instabilities developed when the 1.4\,K stage was too cold.  Liquid helium was overproduced and filled the pipes between the 1.4\,K stage and the 4\,K stage.  Unstable evaporation then generated large temperature fluctuations.  

Adjustment of the coolers for maximum cooling performance gives the best margins in operation of the cooling chain; however, because of such instabilities associated with excess liquid, the highest cooler performance does not necessarily lead to the optimum overall configuration.  It is essential that the PIDs have enough power to warm the 1.4 and 4\,K stages to the optimal operating point, far away from the unstable evaporation region.

Vibration from the compressors could affect the HFI data in several ways.  No microphonic noise was seen in system tests when the vibration control system was activated in the drive electronics of the compressors;  however, electromagnetic interference was seen in the qualification and flight model system tests at several beat frequencies of the compressor frequency and sampling frequency.  In the FPU instrument tests with no \HeJT\ cooler, the average amount of heat dissipated in the bolometer plate was around 10\,n\hbox{W}.  In the system tests at CSL, it was between 40 and 50\,nW, a good indication that microvibrations were heating the 100\,mK stage, although it was not possible to separate the contributions of the CSL test tank from those of the \HeJT\ cooler. These heat imputs had negligible fluctuations and affected only the heat loads in the Planck cryo system.  The heat inputs on the 100\,mK stage were measured in flight and  are discussed in Sect.~4.4.

The system tests also tested the cooldown, and defined the full cooldown procedure in flight.  This is very important.  One could design a cooling chain with an operating point with margins, but no way to reach it.  The system tests had to demonstrate not only the final configuration and its margins, but also the margins along the entire cooling path.  We will return to this point in~Sect.~7.

%SECTION 4.
\section{Performance: Temperature}
\label{sec:performance}

\begin{figure*}[htbp]
\centering
\includegraphics[width=15.5cm]{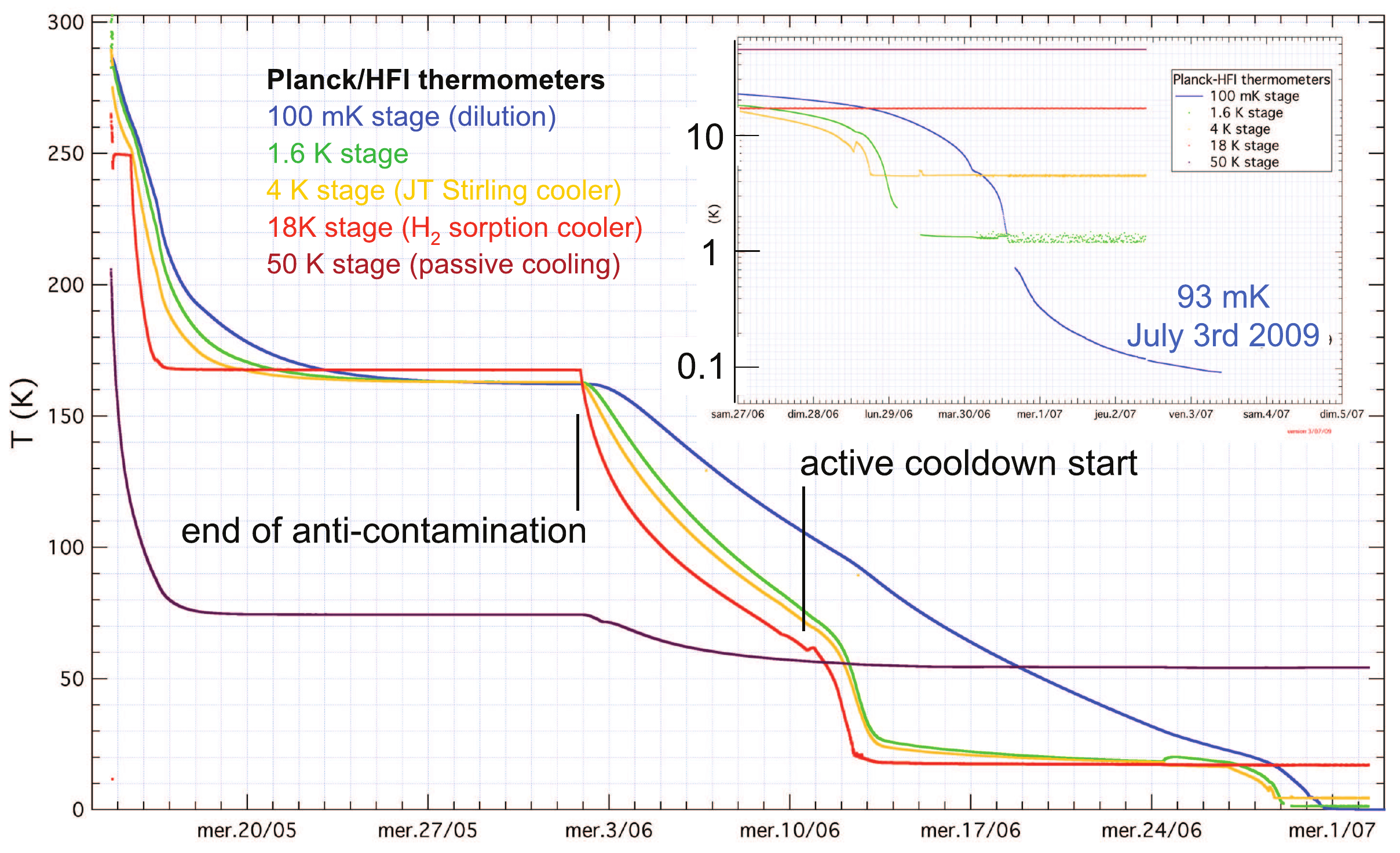}
\includegraphics[width=13cm]{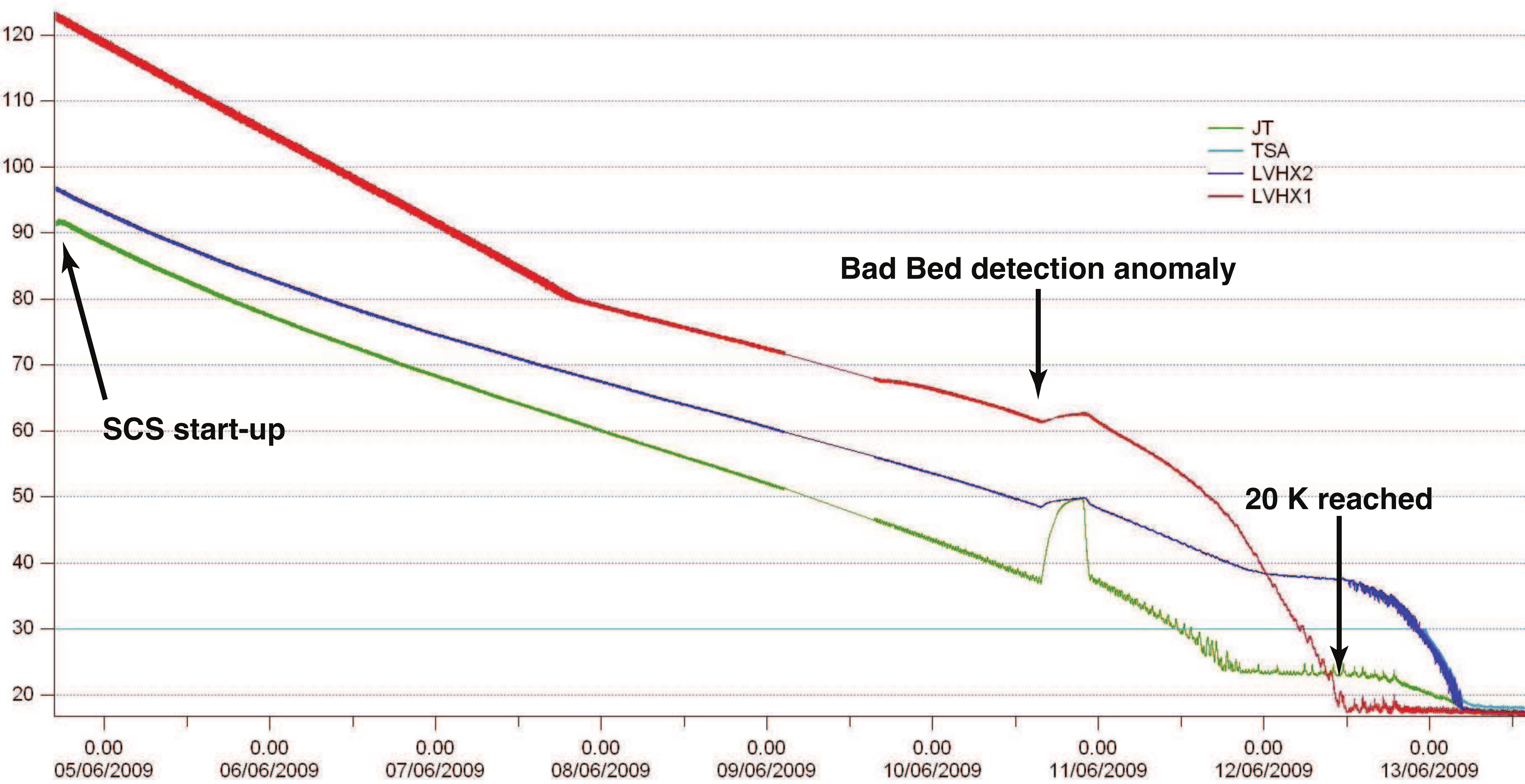}
\includegraphics[width=13cm]{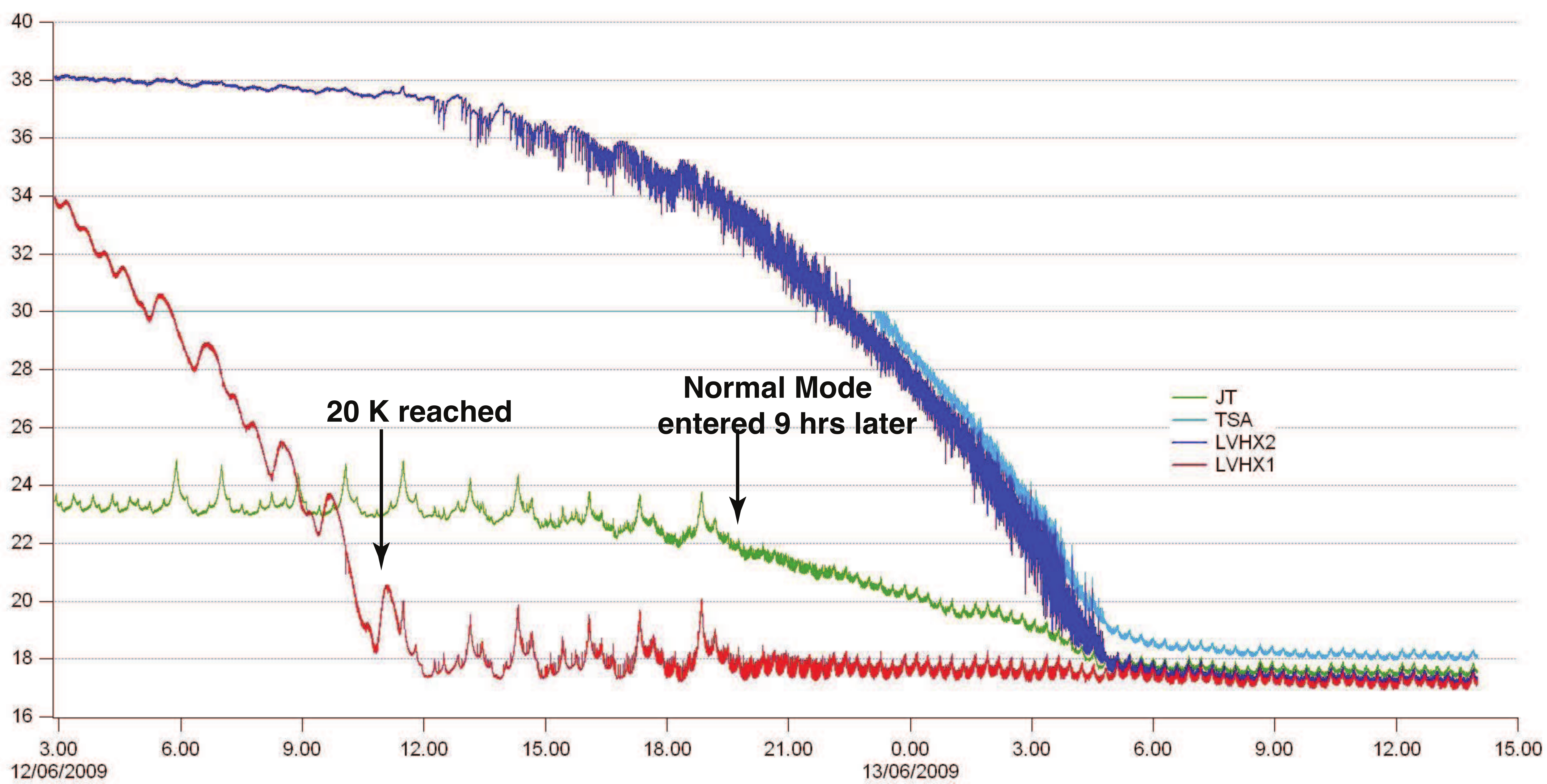}
\caption{{\it Top:\/} Initial cooldown of \Planck\/, starting from launch on 14~May~2009 and concluding on 3~July~2009, when the bolometer plate reached 93\,m\hbox{K}.  The inset above shows the last week of the HFI cooldown.  {\it Middle:\/} Expanded view of the LFI cooldown from sorption cooler start-up on 5~June 2009 until 13~June~2009.  {\it Bottom\/} Last day of the sorption cooler cooldown.}
\label{fig:cooldown}
\end{figure*}

Radiative cooling began immediately after launch on 14 May 2009.  Figure~\ref{fig:cooldown} shows the temperatures of the key elements as a function of time.   Operating temperatures were reached on 3 July 2009 after a cooldown in agreement with the ground tests and simulations.  Cooling from 20\,K to 4\,K was deliberately slowed down to allow calibration of the \hbox{LFI}.  Table~\ref{table:Planck_temperatures} lists the temperatures achieved in flight for key components.

\begin{table}                    
\begingroup
\newdimen\tblskip \tblskip=5pt
\caption{Temperatures of key components in flight.}  
\label{table:Planck_temperatures}
\nointerlineskip
\vskip -3mm
\footnotesize
\setbox\tablebox=\vbox{
   \newdimen\digitwidth 
   \setbox0=\hbox{\rm 0} 
   \digitwidth=\wd0 
   \catcode`*=\active 
   \def*{\kern\digitwidth}
   \newdimen\signwidth 
   \setbox0=\hbox{+} 
   \signwidth=\wd0 
   \catcode`!=\active 
   \def!{\kern\signwidth}
\halign{\hbox to 1.3in{#\leaderfil}\tabskip=0.6em&
        \hfil#\hfil&
        \hfil#\hfil&
        \hfil#\hfil\tabskip=0pt\cr 
\noalign{\doubleline}
\omit&&\multispan2\hfil T{\sc emperature}\hfil\cr
\noalign{\vskip -3pt}
\omit&&\multispan2\hrulefill\cr
\omit&\omit\hfil Number of\hfil&\omit\hfil Range\hfil&\omit\hfil Mean\hfil\cr
\omit\hfil Component\hfil&\omit\hfil Sensors\hfil&\omit\hfil [K]\hfil&\omit\hfil [K]\hfil\cr
\noalign{\vskip 3pt\hrule\vskip 5pt}
Solar panels&    6&377--398&384\cr
SVM&             3&273--294&286\cr
Star trackers&   2&262--272&267\cr
Helium tanks&    3&287--290&289\cr
V-groove 1&      2&135.6--140.2&137.9\cr
V-groove 2&      2&*91.7--*91.9&*91.8\cr
V-groove 3&      5&*45.2--*47.7&*46.1\cr
Telescope baffle&4&*42.1--*42.7&*42.3\cr
Primary mirror&  2&*35.9--*35.9&*35.9\cr
Secondary mirror&3&*39.6--*39.6&*39.6\cr
LFI focal plane& 3&*19.8--*23.2&*22.0\cr
HFI FPU&         3&*18.2--*18.2&*18.2\cr
HFI ``4\,K stage''&  2&**4.7--**4.8&**4.75\cr
HFI ``1.4\,K stage''&1&\dots&***1.393\cr
HFI ``100\,mK stage''&1&0.1028&****0.1028\cr
\noalign{\vskip 5pt\hrule\vskip 3pt}}}
\endPlancktable
\endgroup
\end{table}

%4.1
\subsection{V-grooves, baffles, telescope}

The primary and secondary telescope mirrors cooled after launch to 36.3\,K and 39.6\,K, respectively.  The temperatures are quite stable, with a small annual variation driven by the distance of \Planck\ from the Sun (Fig.~\ref{fig:seasonal_variations}).

The background (steady state) optical loading on the bolometers, $P_{\rm opt}$, was measured during the CPV phase in July 2009.   From this, thermal emission in the HFI bands can be estimated.  Several contributors to $P_{\rm opt}$ are well known from ground and flight calibration, including the in-band power $P_{\rm CMB}$ from the CMB, the thermal emission $P_{\rm 4K}$ from the filter and horns at 4\,K, and the thermal emission $P_{\rm 1.4K}$ from the filters at 1.4\,K \citep{planck2011-1.5}.  $P_{\rm opt}$ is measured by the difference in apparent base temperature of the bolometers $T_{\rm b}$ without electrical bias ($VI_{\rm b}=0$), the measured base temperature of the 0.1\,K plate, and the thermal conductance $G = g_0T^{\beta}$:
\begin{equation}
P_{\rm opt} = \frac{g_0}{\beta + 1} \left(T_{\rm b}^{\beta + 1} + T_0^{\beta + 1} \right)
\label{opt_load}
\end{equation}

\begin{figure}
\begin{center}
\includegraphics[width=9cm]{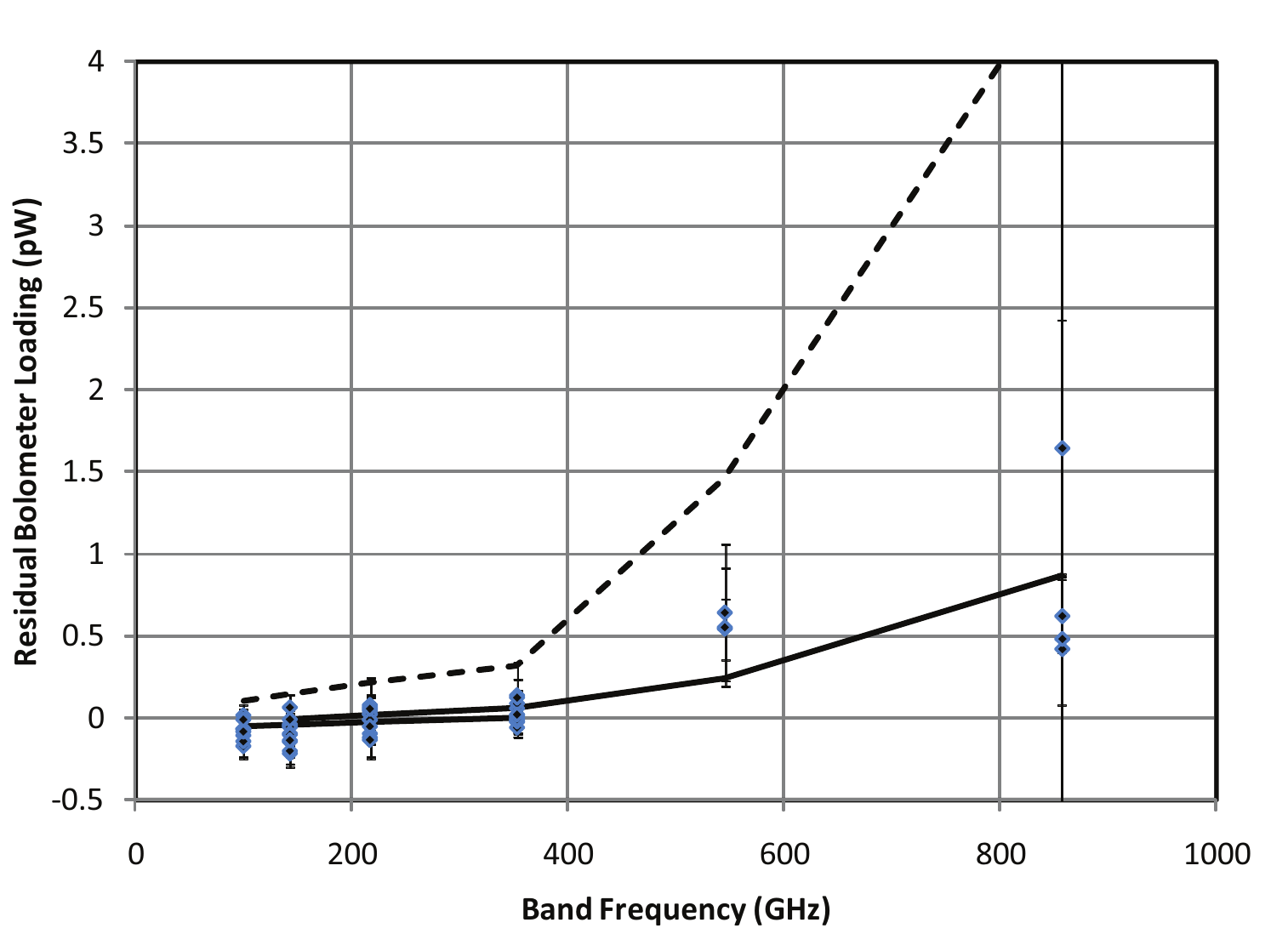}
\caption{Measured background power in the bolometer bands from the primary and secondary mirrors of the \Planck\ telescope. The solid line through the data is a best fit of a constant emissivity, $\epsilon = 0.07\%$ times the computed blackbody power from the primary mirror at $\sim 36.2$\,K and the secondary mirror at $\sim 39.4$\,K in the measured bandpass for each detector. The dashed line is the best case performance estimated before flight, which corresponds to a mirror with $\epsilon = 0.6\%$ at a temperature of 40\,\hbox{K}. The scatter in the measured power is largely due to uncertainty in gradients at the level $<1$\,mK in the 100\,mK plate.}
\label{fig:telescope_emissivity}
\end{center}
\end{figure}

The value of $P_{\rm residual} = P_{\rm opt} - P_{\rm CMB} - P_{\rm 4K} - P_{\rm 1.4K}$ is shown in Fig.~\ref{fig:telescope_emissivity} for the HFI bands.  There is a clear trend of increasing $P_{\rm residual}$ with band center frequency.  In the two highest frequency bands, the dependence of $P_{\rm residual}$ on frequency is stronger, as expected, since these bands are multimoded.  In addition, the 857\,GHz band has a contribution of sky signal from the Galaxy that at the time of the tests was at least comparable to the error in $P_{\rm residual}$.  The dominant sources of error in $P_{\rm residual}$ are a constant temperature offset in the 0.1\,K plate, estimated to be $\pm 1$mK, due either to a calibration error in the thermometry or a gradient between the thermometer and bolometer locations, and an error in the value of $g_0$ as determined from ground based measurements. For most bolometers, these errors are a few percent. However, for some bolometers, especially at 545 and 857\,GHz, the optical loading produces a temperature rise in the detector of $T_{\rm b} - T_0$, comparable to the thermometer accuracy. Thus, error bars on the these points are $>100\%$ as shown in Fig.~\ref{fig:telescope_emissivity}.  The line through the data points is a least squares fit of the computed in-band power from the two mirrors multiplied by an emissivity, $\epsilon$, independent of frequency. The best fit value of $\epsilon < 0.07\%$ is lower than the requirement for the mirror of $0.6\%$, shown by the dotted line in Fig.~\ref{fig:telescope_emissivity}.

%4.2
\subsection{Sorption cooler}

The FM2 sorption cooler was activated on 2 June 2009 with the JT expander at a temperature of 90\,\hbox{K}.  Liquid hydrogen formed after 187\,h, and the cooler transitioned to normal operating mode 9\,h later.  Initial tuning parameters were set close to those used during ground testing, except that the TSA set-point was lowered to minimize the heat load.   Heat lift and temperature were similar to those in ground tests; however, temperature fluctuations were higher than those on the ground.  Table~\ref{table:SCS_performance} summarizes the results.  As will be discussed later, the temperature fluctuations have small impact on the science data.

\begin{table*}                    
\begingroup
\newdimen\tblskip \tblskip=5pt
\caption{Summary of sorption cooler system flight performance.}  
\label{table:SCS_performance}
\nointerlineskip
\vskip -3mm
\footnotesize
\setbox\tablebox=\vbox{
   \newdimen\digitwidth 
   \setbox0=\hbox{\rm 0} 
   \digitwidth=\wd0 
   \catcode`*=\active 
   \def*{\kern\digitwidth}
   \newdimen\signwidth 
   \setbox0=\hbox{+} 
   \signwidth=\wd0 
   \catcode`!=\active 
   \def!{\kern\signwidth}
\halign{\hbox to 1.3in{#\leaderfil}\tabskip=2em&
        #\hfil&
        #\hfil&
        \hfil#\hfil\tabskip=0pt\cr
        \noalign{\doubleline}
\omit\hfil Characteristic\hfil&\omit\hfil CPV$^{\rm a}$ Flight Result\hfil&\omit\hfil Ground Test Result\hfil&\omit\hfil Requirement\hfil\cr
\noalign{\vskip 3pt\hrule\vskip 5pt}
Cold end $T$&17.0\,K at LVHX1&17.1\,K at LVHX1&LVHX1 $<$ 19.02\,K\cr
\omit&       18.5\,K at TSA&  18.7\,K at TSA&  LVHX2 $<$ 22.50\,K\cr
\noalign{\vskip 8pt}
Cooling power&$1125\pm75$\,mW&$1125\pm75$\,mW&at LVHX1 $>$ 190\,mW\cr
\omit&&&                                     at LVHX2 $>$ 646\,mW\cr
\noalign{\vskip 8pt}
Input power&  301\,W&304\,W&$<426$\,W at BOL\cr
\noalign{\vskip 3pt}
Cold end $\Delta T$&580\,mK at LVHX1&550\,mK at LVHX1&at LVHX1 $<$ 450\,mK\cr
\omit&              140\,mK at TSA&  120\,mK at TSA&  at LVHX2 $<$ 100\,mK\cr
\noalign{\vskip 5pt\hrule\vskip 3pt}}}
\endPlancktablewide
\tablenote a Calibration and performance verification phase.\par 
\endgroup
\end{table*}

\begin{figure}
\centering
\includegraphics[width=9cm]{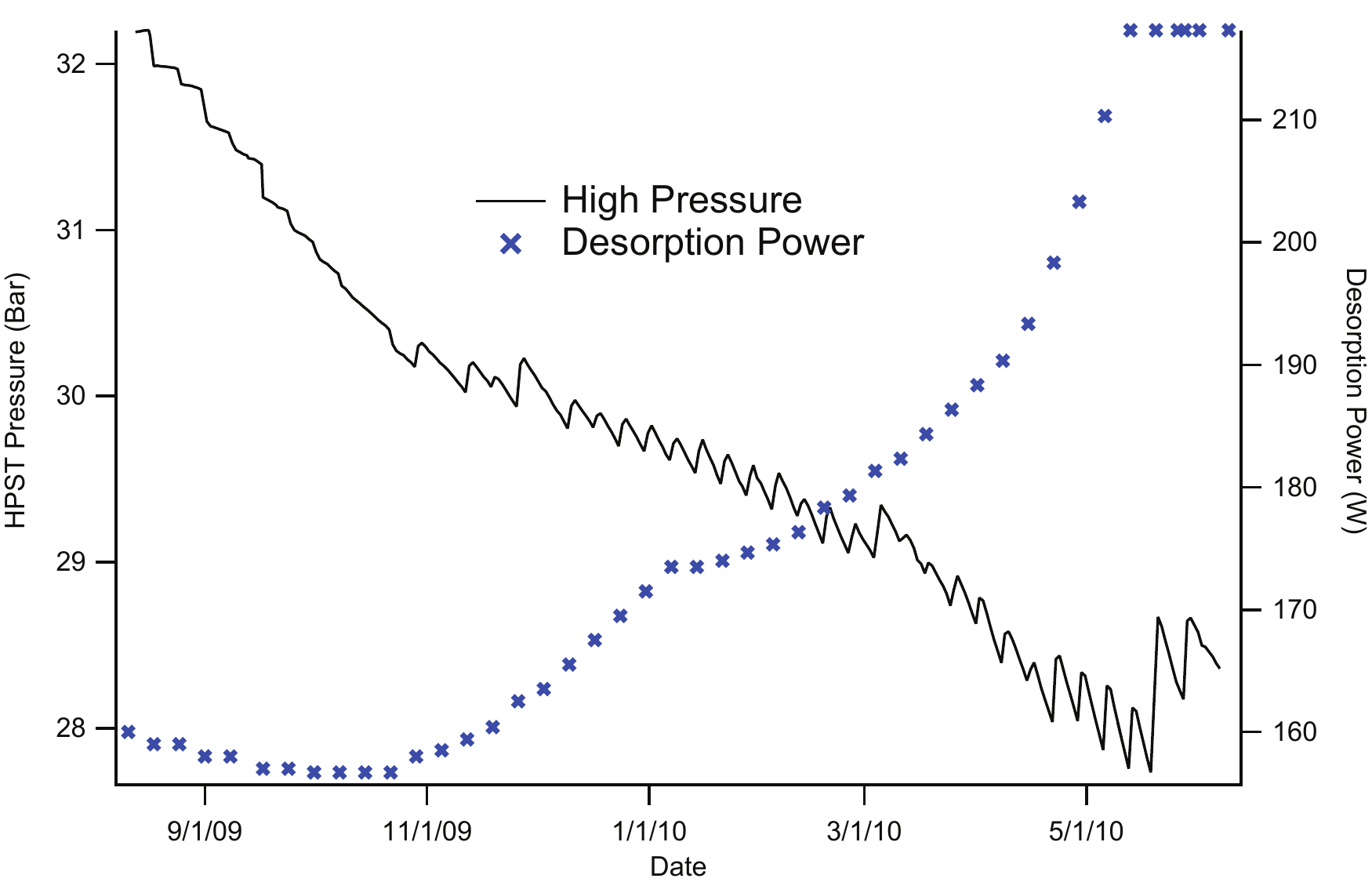}
\caption{Sorption cooler high pressure and desorption power as a function of time.}
\label{fig:SCS_pressure_power}
\end{figure}

Due to degradation of the hydride with operation, the input power required to maintain the required heat lift increases with time.  Figure~\ref{fig:SCS_pressure_power} shows the high pressure and the desorption power as a function of time.

As discussed in Sect.~2.4.2, the temperature of the warm radiator, the age of the hydride, and the temperatures of LVHX1 and LVHX2 are related.  For the LFI, the set-point for the TSA at LVHX2 is set to a minimum to reduce the heat load on the sorption cooler.  For the HFI, the temperature of LVHX1 must have margin to accommodate temperature drift.

\begin{figure}
\centering
\includegraphics[width=9cm]{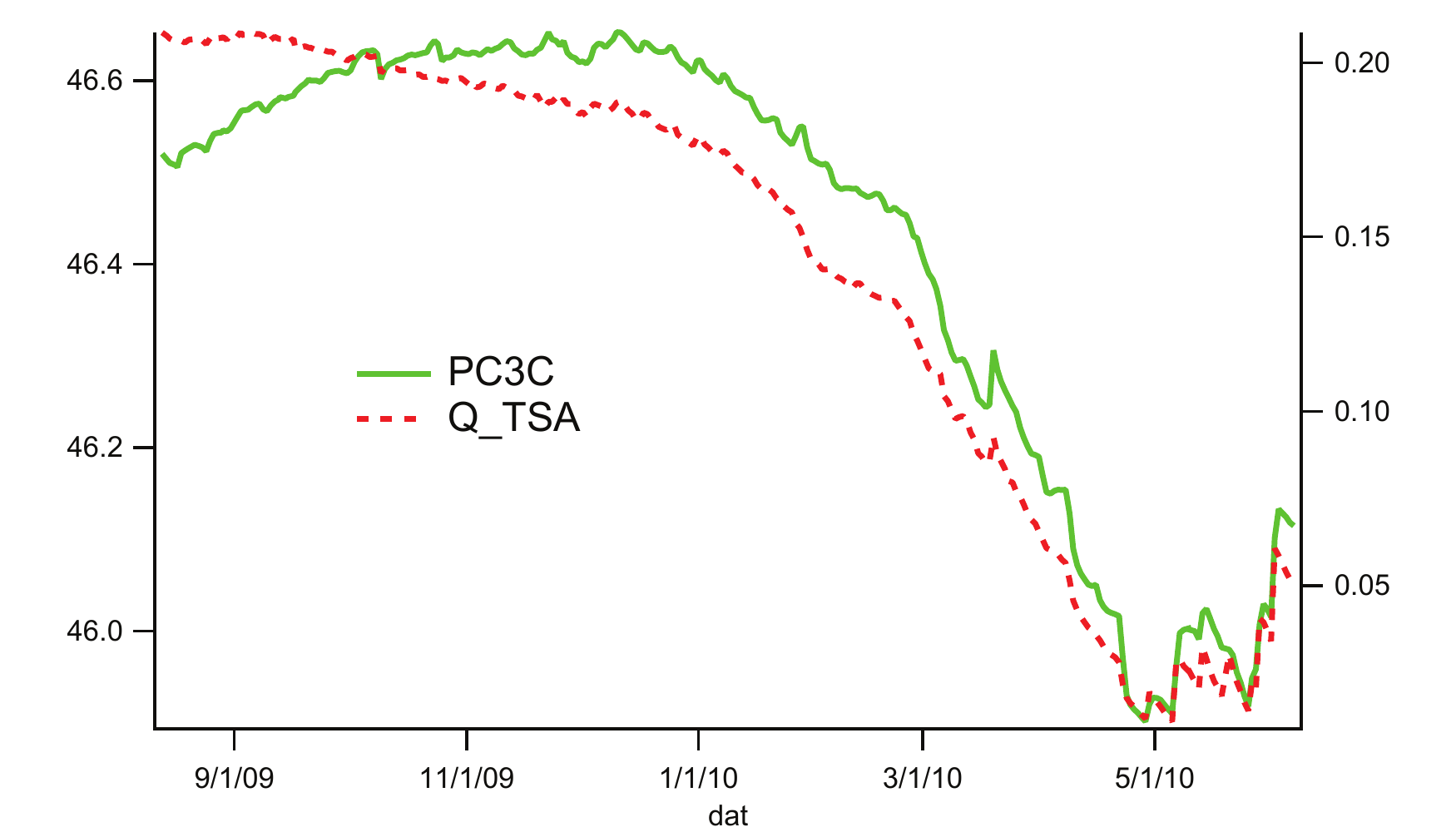}
\caption{Temperature of the sorption cooler heat exchanger on V-groove 3 (PC3C), and the temperature stabilization assembly (TSA) power, as functions of time. The sorption cooler including the TSA is the dominant heat load on V-groove 3.  As TSA power is increased, the heat lifted by the sorption cooler increases, the load on V-groove 3 increases, and its temperature rises.  A smaller seasonal effect of the distance of \Planck\ from the Sun is superimposed on this general trend.  The closest approach to the Sun occurred at the beginning of January. }
\label{fig:PC3C_TSA_power}
\end{figure}

The sorption cooler also affects the three V-groove temperatures, as the heat lifted from the focal plane is mostly radiated to space by the V-grooves.  As TSA power is increased, the heat lifted by the sorption cooler increases, the load on V-groove 3 increases, and its temperature rises.  Figure~\ref{fig:PC3C_TSA_power} shows the temperature of the heat exchanger on V-groove 3 (PC3C) and the power dissipated by the TSA as a function of time.  The relationship is clear.  In addition, there is a seasonal variation in the temperature of V-groove 3 driven by changes in \Planck's distance from the Sun.  The closest approach to the Sun occurred about 1 January (see also Fig.~\ref{fig:seasonal_variations}).

%4.3
\subsection{\HeJT\ cooler}

As seen in Sect.~2.3.2, the performance of the \HeJT\ cooler is characterised by the temperature \TJT4K\ of the JT and the maximum heat lift \HLMax\ that the cooler can generate.  While \HLMax\ depends on both the high and  low pressures of helium in the cooler, \TJT4K\ depends only on the low pressure. The operating characteristics of the \HeJT\ cooler depend on three adjustable parameters and two environmental conditions:

\begin{itemize}

\item{The stroke amplitude \strokeamp.}

\item{The compressor frequency \fcomp.  This can be chosen in the range 35--45\,Hz.  The choice is driven by four factors.  First, the lowest resonance frequency of the spacecraft panel on which the compressors are mounted is 72\,Hz, driving the choice above 37\,Hz.  Second, the operating efficiency shows a broad maximum around 40\,Hz.   Third, we want to minimize electromagnetic contamination and microphonic lines in the data.  Fourth, in order to minimize mechanical stress, the frequency should be below 45\,Hz.  We chose 40.08\,Hz as the nominal frequency.}

\item {The filling pressure \Pfill.  This is adjustable only before flight. The filling pressure was 4.5\,bar.}

\item{The pre-cool temperature \Tpc\ provided by the sorption cooler.  This temperature is at most 0.1\,K higher than the LVHX1 cold head temperature and was found to be 17.3\,K, providing a large system margin}

\item{The temperature \Tcomp\ of the base plate on which the compressors are mounted. This temperature is set by the spacecraft architecture and is not easily tuned. The temperature in flight was 7\,\deg C, and did not change by more than 1\,\deg C during the nominal mission.  This leads to negligible changes of performance during the mission.}

\end{itemize}

We end up finally with one parameter adjustable in flight, the stroke amplitude \strokeamp, and one environment parameter, the pre-cooling temperature provided by the sorption cooler.
The performance of the \HeJT\ cooler as measured in flight is consistent with the ground based tests given in Sect.~2.3.2.

No extra heat load on the 4\,K box was identified with respect to the CSL thermal balance/thermal vacuum test, showing that the launch had not affected the cryogenic configuration.

In flight, unstable evaporation of helium between 1.4\,K and 4\,K  stages was seen, induced by the  4\,K  stage being too cold. The 1.4\,K--4\,K temperature plane was mapped to find the unstable zones, which turned  out to be  $T_{1.4\,{\rm K}} < 1.34\,{\rm K}$ and $T_{4\,{\rm K}} < 4.64\,{\rm K}$.  The best operating region is $1.36\,{\rm K} < T_{1.4\,{\rm K}} < 1.4\,{\rm K}$ and $4.7\,{\rm K} < T_{4\,{\rm K}} < 4.75\,{\rm K}$. The operating point for flight was chosen in this region. 

During CPV, \strokeamp\ was reduced to 3.45\,mm to decrease the power of the \HeJT\ cooler and optimise its operating conditions, keeping the temperature above the unstable evaporation range. The vibration control system converged quickly to a new stable configuration, demonstrating that adjustment of this important subsystem would be easy should it be necessary. No further adjustment has been required.

The 4\,K PID flight configuration has fixed power of 840\muW\ on the redundant PID, and regulating power of around 900\muW\ on the nominal \hbox{PID}.  The in-flight operating temperature of the 4\,K cold tip is 4.37\,\hbox{K}. The PID for the 4\,K feed horn plate was set at 4.81\,K. 

%4.4
\subsection{Dilution cooler}

The isotope flow was set to FNOM1 (Table~\ref{table:heliumflow}) for the cooldown.  The bolometer plate reached 93\,mK and was still decreasing on 3 July 2009 when the flow was changed to Fmin2, the lowest available. The 1.4\,K stage reached 1.23\,\hbox{K}. The flow for a given restriction is slightly higher than it was during ground tests because the pressure regulators on the high pressure tanks had to be changed after a leak was found in the system test. The replacement regulators provide 19\,bar rather than the original 18\,bar.

The dilution stage is stabilized by a PID control with a power initially around 31\,n\hbox{W}, providing a temperature near 101\,mK.  This power decreased slowly to a minimum of 24\,nW in November 2010, and then began to rise again.  This trend was expected, and is due to a small nonlinearity of the pressure regulators.  The regulated pressure decreases slightly as the tank pressure drops to about half its initial value, then increases again as the tank pressure goes lower.  The amplitude of the effect is fully consistent with the heat lift given in Sect.~2.3.3.

The bolometer plate is stabilized at 102.8\,mK with a PID power around 5\,n\hbox{W}. The PID parameters give a time constant of about 0.5\,h.  This ensures that glitches induced by cosmic rays on the PID thermometer do not induce temperature fluctuations on the bolometer plate.  Both PIDs show fluctuations on time scales of days to weeks, discussed in Sect.~5.4.

The heat inputs on the bolometer plate are: (i)~the input from the bias current of the bolometers (less than 1\,nW); (ii)~the microwave radiation reaching the bolometers (0.12\,nW); (iii)~cosmic rays that penetrate the FPU box and deposit energy in the bolometer plate (variable); (iv)~the heat dissipated by microvibrations in the bolometer plate; and (v)~heating from PID2.   

\begin{figure}
\begin{center}
\includegraphics[width=9cm]{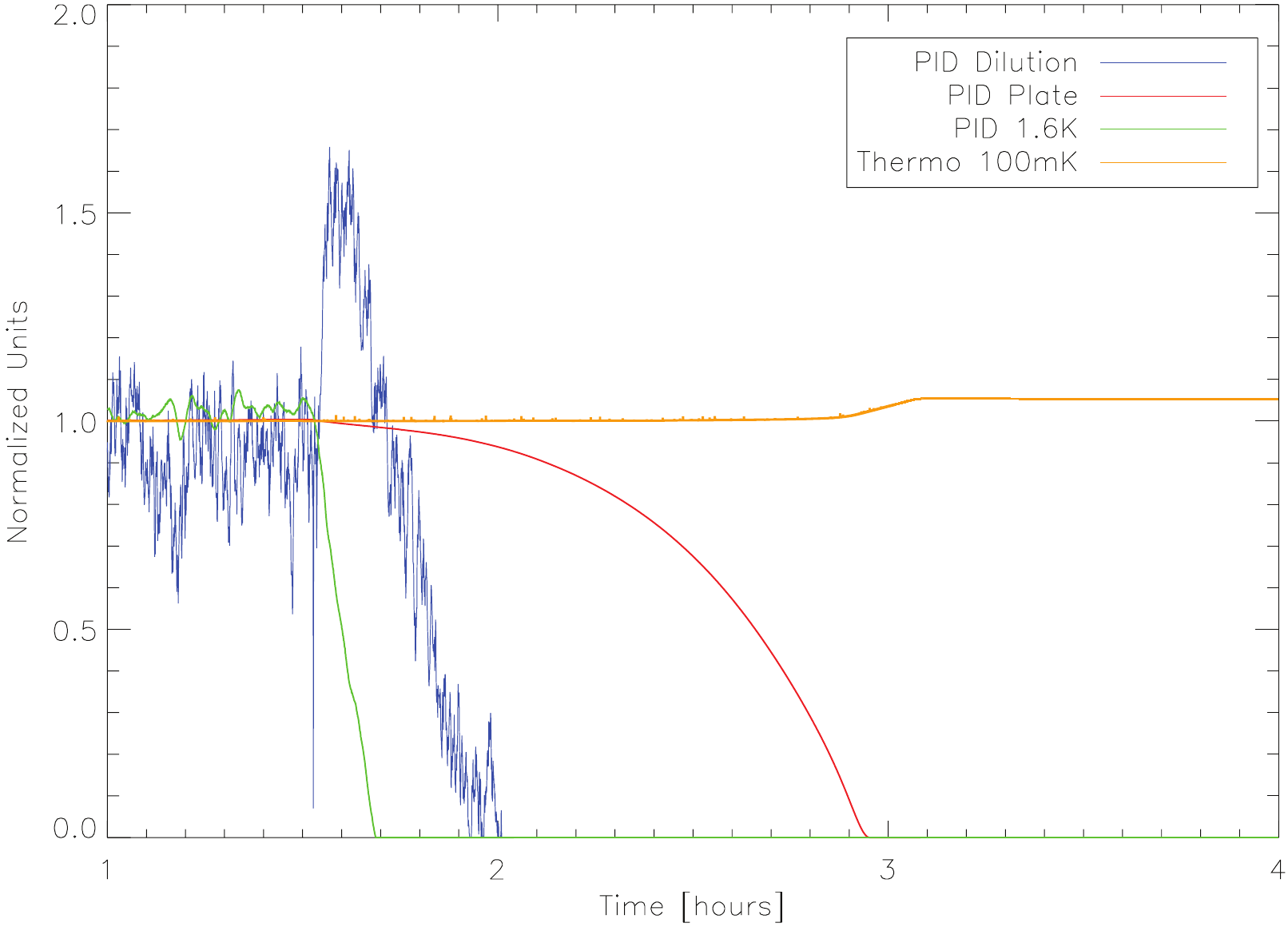}
\caption{Thermal behaviour of the 100\,mK stage during the shutdown of the \HeJT\ cooler on 6 August 2009. Vibrational power from the 4\,K stage propagated to the dilution stage is measured by the step in the power dissipated by the PID regulating the temperature of the dilution stage (``PID Dilution'', in blue) after the rapid drop of the 4\,\hbox{K} stage. It is equal to 26\,n\hbox{W}.  The power dissipated by the PID regulating the temperature of the bolometer plate (``PID Plate'', in red) shows the effect of the long (0.5\,h) time constant of this control.}
\label{figure:4ksd}
\end{center}
\end{figure}

The heat dissipated in the dilution and bolometer plates by micro-vibrations from the 4\,K compressors was not measured precisely during ground tests, but was of order 30\,nW on the bolometer plate.  On one occasion during the in-orbit checkout period the \HeJT\ cooler was turned off by the charge regulator that controls the current call from the drive electronics to the spacecraft.  This single event allowed us to measure precisely the heat input from compressor micro-vibrations on the dilution plate (see Fig.~\ref{figure:4ksd}). It was 26\,nW. There was evidence for comparable heating of the bolometer plate, but the long time constant of the PID meant that it could not be assessed accurately.  The heat input from micro-vibrations is essentially constant.

Correlation of the bolometer plate temperature with the Space Radiation Environment Monitor (SREM) data and with the glitch rate on the bolometers over periods of days to weeks (Sect.~5.4) shows that cosmic rays also heat the bolometer plate.  The variable part of the power input from the cosmic rays can be calibrated using the PID response on long periods, allowing estimation of the total power input from cosmic rays.  We find about 12\,nW on the bolometer plate and 8\,nW on the dilution plate for the first 100 days of the survey.  These numbers were checked using the small solar flare on 5--7 April 2010, and within uncertainties the powers agree.  Starting in January 2010, an increase in Solar activity brought these numbers down slowly by 1.5 and 1\,nW respectively.  There is thus a consistent picture in which the Galactic cosmic rays detected by the SREM  account for both the temperature fluctuations discussed in Sect.~5.4 and the extra heating seen in flight on the bolometer and dilution plate.

In summary, the bolometer plate receives 10--12\,nW (depending on the period) from cosmic rays and about 30\,nW from micro-vibrations from the \HeJT\ cooler, in good agreement with the 40\,nW obtained empirically from the gradients within the 100\,mK stage.  The dilution plate receives 26\,nW from micro-vibration from the \HeJT\ cooler and 7--8\,nW from the cosmic rays, in good agreement with the 35\,nW obtained from the temperature gradients.
 
Heating by cosmic rays is higher than expected before flight by almost an order of magnitude.  The period following the Planck launch was a period of exceptionally low solar activity, resulting in very weak solar modulation of cosmic rays at 1\,AU (\citealt{Mewaldt2010} and references therein).  Tthe flux of low energy (200\,MeV\,nucleon\mo) nuclei from carbon to iron was four times higher than in the period 2001--2003, and 20\% higher than in previous solar minima over the last 40~years.  Nevertheless, the total flux of Galactic cosmic ray protons, which dominates and peaks around 200\,MeV, gives a total proton flux of 3--$4 \times10^3$\,particles\,m$^{-2}$\,sr\mo\,s\mo\,MeV\mo.  This accounts only for half the heating of the bolometer plate. The SREM total particle flux above 10\,MeV is 7--$10\times10^3$\,particles\,m$^{-2}$\,sr\mo\,s\mo\,MeV\mo, depending on the effective acceptance solid angle of the \hbox{SREM}. This could indicate that the particles in the range 30--100\,MeV, which can enter the FPU and have a high energy loss rate, contribute also to the bolometer plate heating.  Note that this is the energy range of the so-called anomalous cosmic rays (neutral Galactic atoms ionized and accelerated in the heliosphere and also strongly affected by the solar modulation). 

In summary, the correlation between SREM data and PID power when applied to the total SREM count rate accounts well for the heating of the bolometer plate.  Attempts to reconcile the SREM count rate with the Galactic cosmic rays, and modeling the heating of the plate from the Galactic cosmic rays, both fall short by nearly a factor of 2, indicating a possible significant contribution from the anomalous cosmic rays.

%SECTION 5.
\section{Performance: Temperature fluctuations}
\label{sec:fluctuations}

%5.1
\subsection{Warm elements}

The downlink transponder in the SVM dissipates 70\,W when on.  A heater on the panel is not close enough to compensate effectively for the varying dissipation of the transponder.  During initial checkout of the flight system, downlink was continuous and the transponder did not produce temperature fluctuations in the \hbox{SVM}.  On 6 July 2009, 53~days after launch, however, downlinks were reduced to 6\,h\,day\mo. On 8 August 2009, 100~days after launch, they were reduced further to the normal operating time of 3\,h\,day\mo.  During those hours, the power dissipated by the transponder caused a rise in the temperature of the $^3$He and $^4$He tanks, increasing their pressure and thus affecting the flow of gas in the dilution cooler.  The temperature of the LFI radiometer backends, mounted on the \hbox{SVM}, also varied.  Both variations led to fluctuations in the detector outputs that were easy to see.  Because the timescale was long, the spin of the spacecraft and normal processing removed the effects in the data essentially completely.  Nevertheless, it was decided to leave the transponder on all the time.  This was done on 25 January 2010 (day number 259), leading to a reduction of the daily fluctuations of the relevant temperatures in the SVM by up to an order of magnitude (see Fig.~\ref{fig:bemmod}).

\begin{figure*}
\centering
\includegraphics[width=9cm]{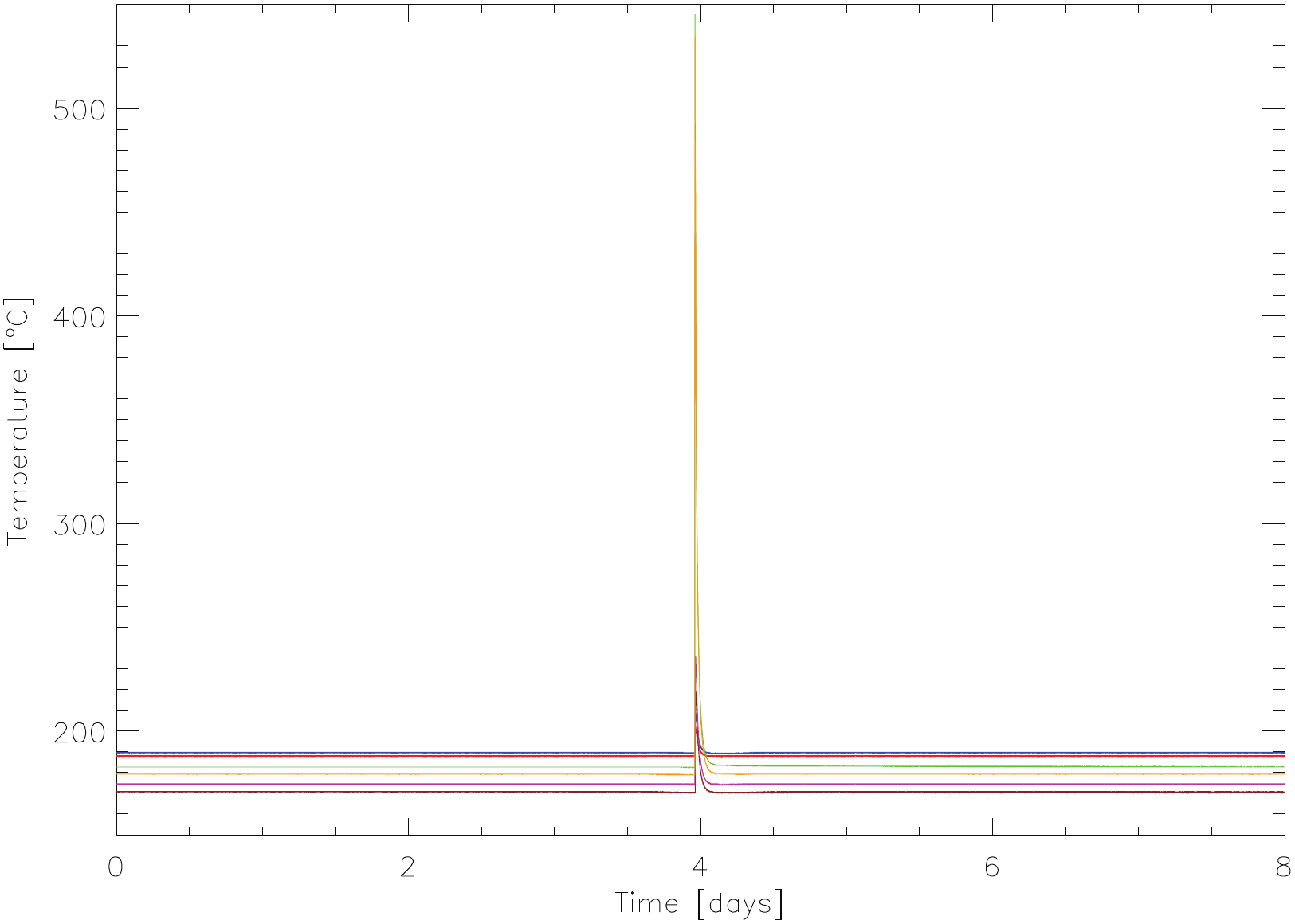}
\includegraphics[width=9cm]{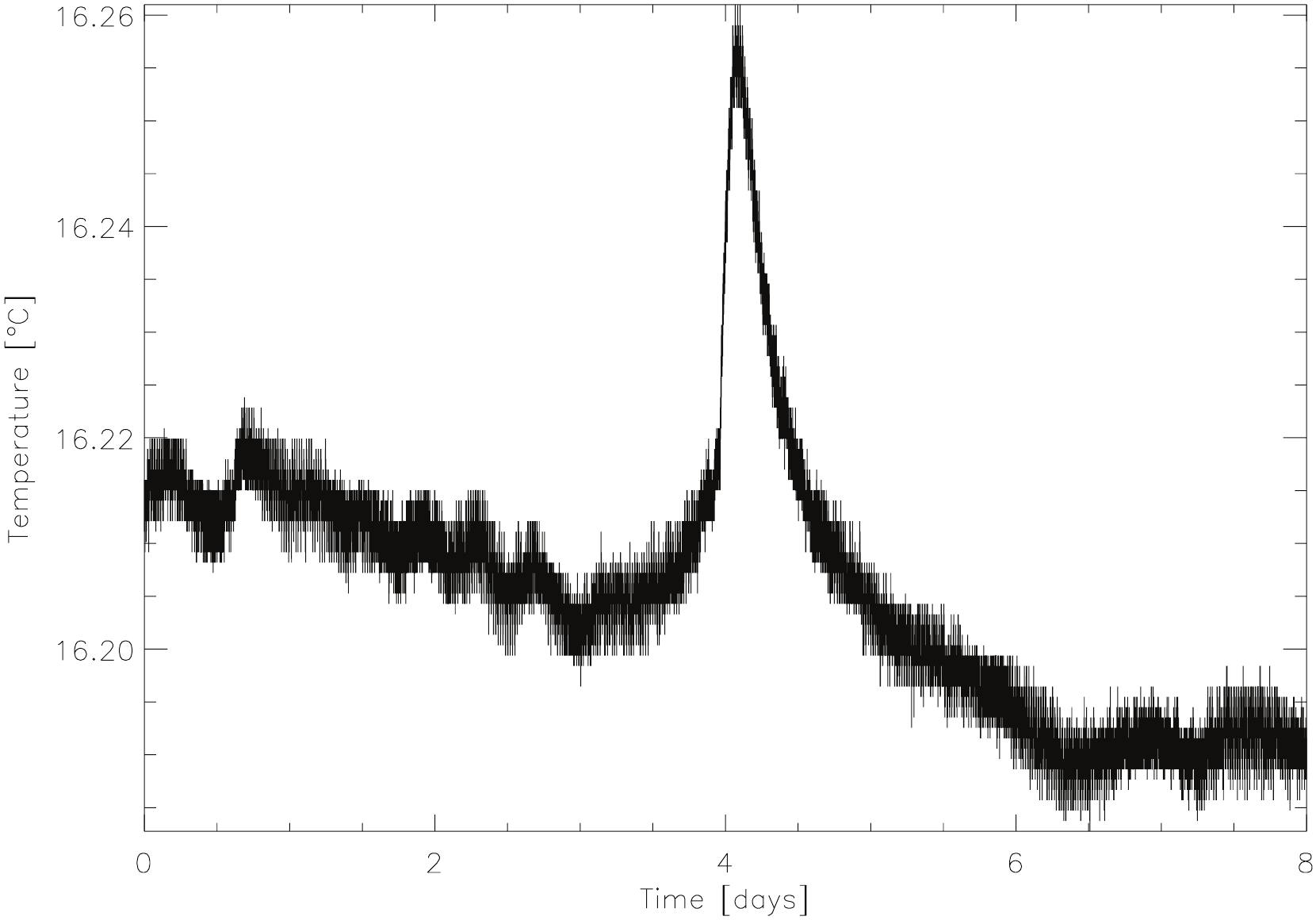}
\caption{Temperatures of the 20\,N thrusters (\textit{left}) and the $^4$He tank (\textit{right}) over eight days centered on orbit correction manoeuvre number 5 on 26 February 2010.}
\label{fig:20Nthruster-tank}
\end{figure*}

Fig.~\ref{fig:20Nthruster-tank} shows the temperature of the 20\,N thrusters at the time of orbit correction manoeuvre number 5 on 26 February 2010. Fig.~\ref{fig:20Nthruster-tank} shows the effect of the thruster firings on the temperature of one of the $^4$He tanks.   The manoeuvre delta-V began at UT 23:01:50 and ended at 23:09:14, with a commanded size of 4.260966\,cm\,s$^{-1}$.  The corresponding pressure increase in the helium tank leads to an increased flow rate for a while, but the downstream controls on temperature adjust so there is no detectable thermal effect on the detectors.

During the first 18 months of the mission, three unexpected star tracker switchovers occurred, two of them during science operations.  The longest lasted 10\,hr.  It is expected that swapping electrical power from one star tracker to the other changes the relative alignment of the star tracker platform with the focal plane due to thermal effects.   Alignment stability during switchover was measured precisely by reobserving the same sky immediately after a scheduled switchover.  The effect can be removed from the data if the redundant star tracker is used.

%5.2
\subsection{17--20\,K elements}

\begin{figure*}
\begin{center}
\includegraphics[width=9.0cm]{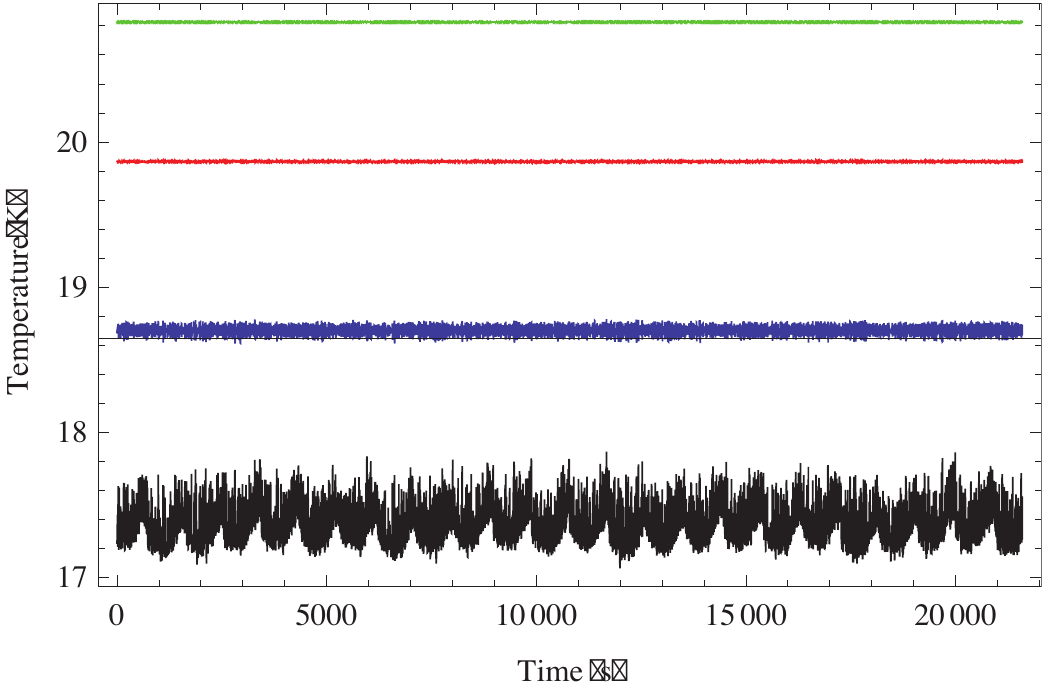}
\includegraphics[width=9.0cm]{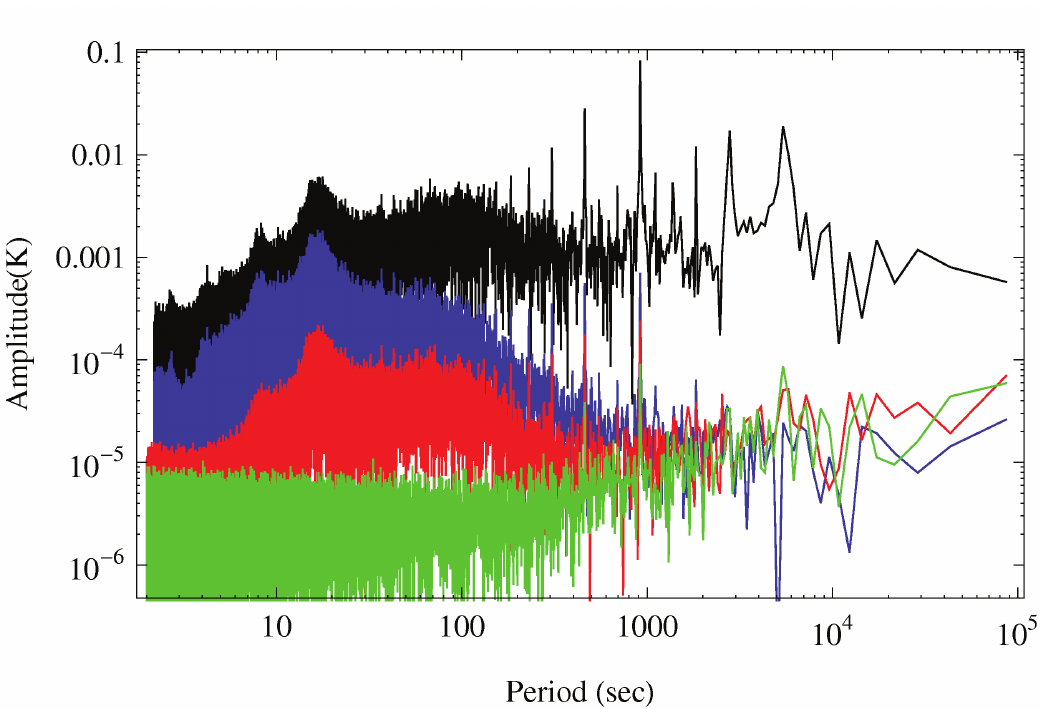}
\end{center}
\caption{\textit{Left --- }LFI focal plane temperatures in October 2009.  \textit{Right --- }corresponding Fourier transform aplitudes;  Black --- LVHX2; blue --- TSA; red --- TSR1 (focal plane sensor closest to LVHX2); green --- TSL6 (focal plane sensor farthest from LVHX2).}
\label{Tsum1}
\end{figure*}

\begin{figure*}
\begin{center}
\includegraphics[width=9.0cm]{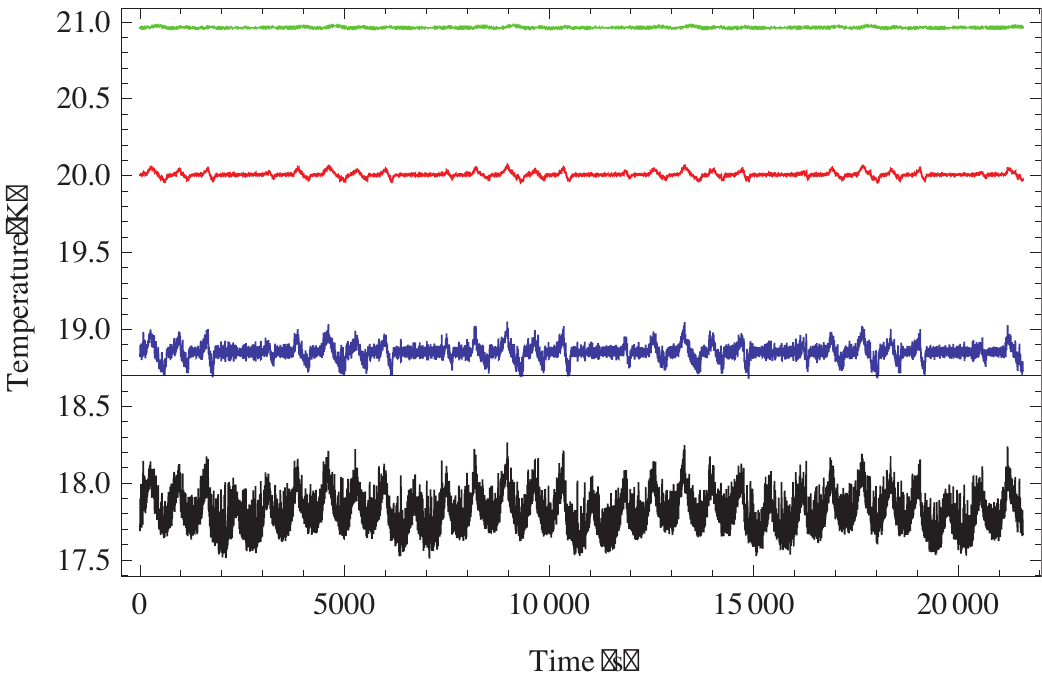}
\includegraphics[width=9.0cm]{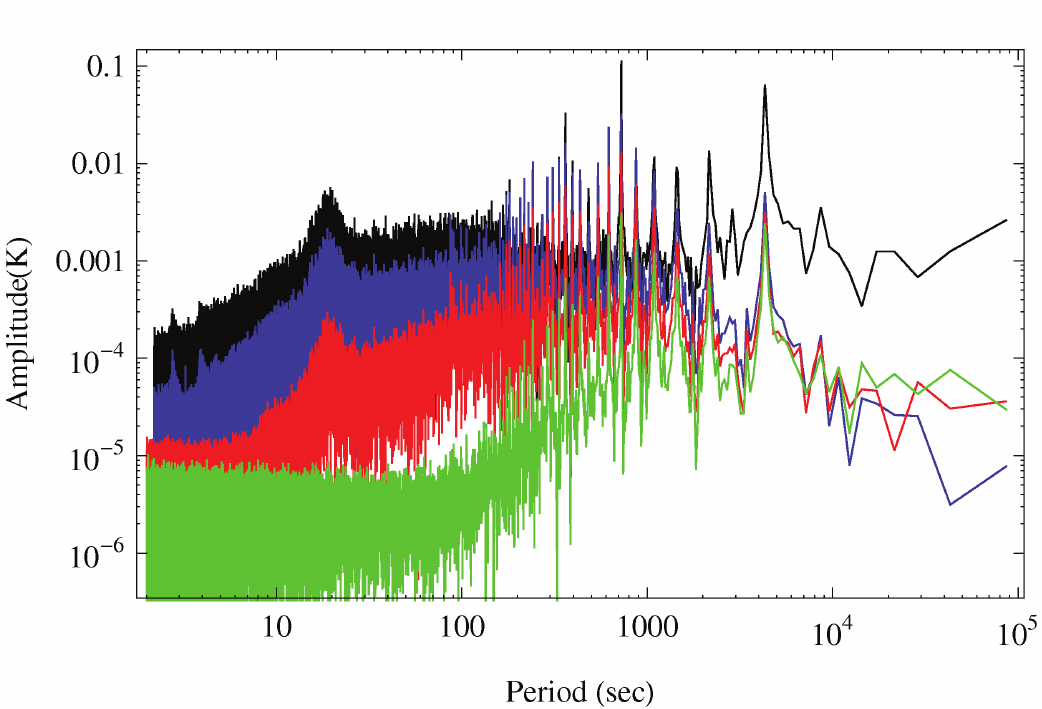}
\end{center}
\caption{\textit{Left --- }LFI focal plane temperatures in May 2010.  \textit{Right --- }Corresponding Fourier transform amplitude.  Black --- LVHX2; blue --- TSA; red --- TSR1 (focal plane sensor closest to LVHX2); green --- TSL6.}
\label{Tsum2}
\end{figure*}

Figures~\ref{Tsum1} and \ref{Tsum2} show temperature as a function of time and the corresponding Fourier transforms for several LFI temperature sensors: LVHX2, the sorption cooler cold end interface to the LFI where fluctuations arise; the TSA, where temperature is actively controlled; and the closest and farthest focal plane sensors from LVHX2.  Fig.~\ref{Tsum1} is for October 2009; Fig.~\ref{Tsum2} is for May 2010.   Thermo-mechanical damping from the source of temperature fluctuations to the LFI radiometers is evident in the decrease in fluctuations with distance from LVHX2 and with frequency.  A maximum fluctuation amplitude of a few millikelvin is obtained throughout the period.  Note the expected (see Sect.~2.4.3) slow upward drift in cold end temperature over the nine month period spanned by Figs.~\ref{Tsum1} and \ref{Tsum2}.  As the cold end temperature approaches the TSA set-point, the margin of power available for control decreases, leading to an increase of residual fluctuations.

\begin{figure*}
\centering
\includegraphics[width=13cm]{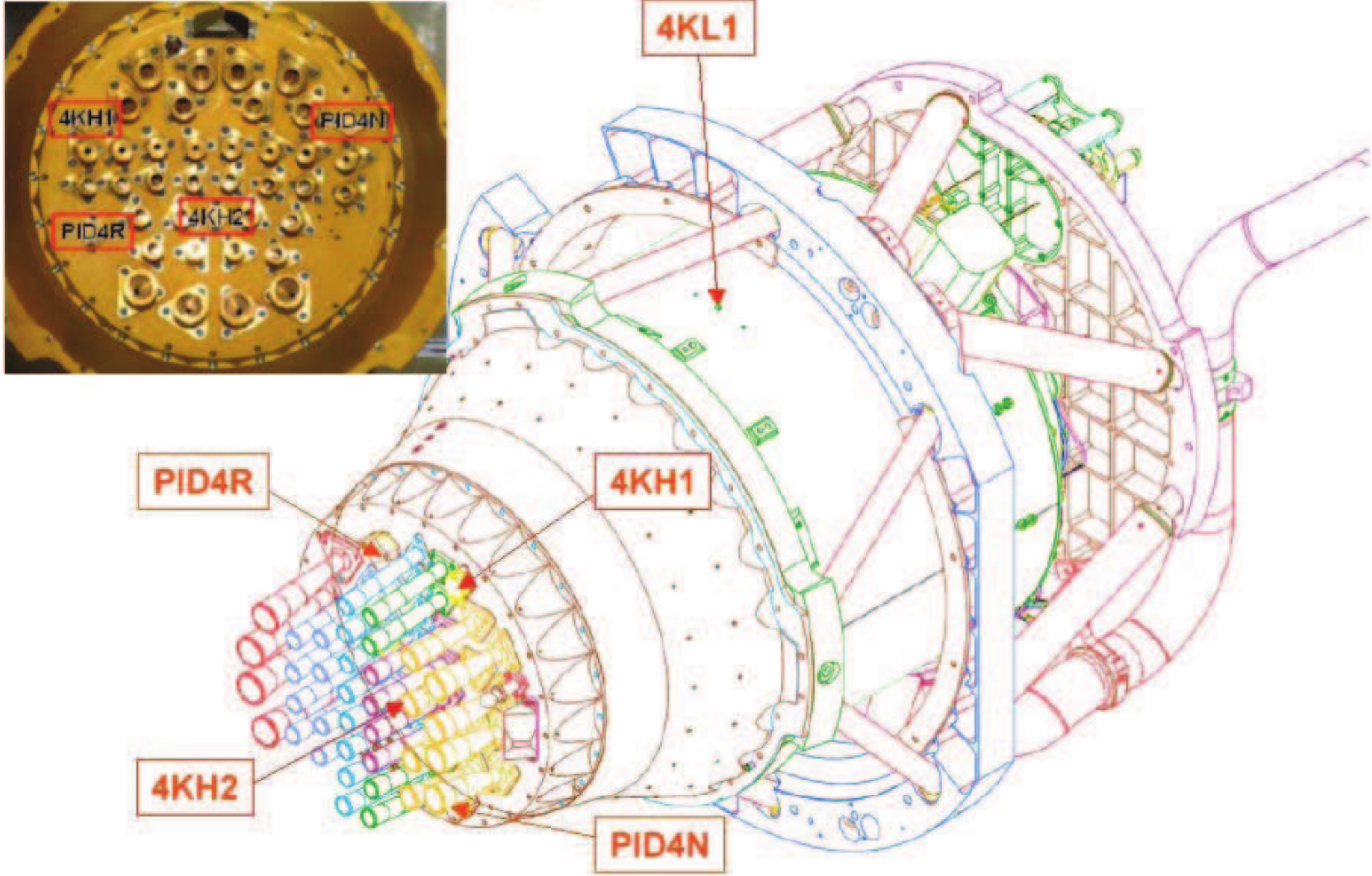}
\caption{Position of thermometers near the LFI reference loads.}
\label{fig:LFI_loads}
\end{figure*}

\begin{figure*}
\begin{center}
\includegraphics[width=9.0cm]{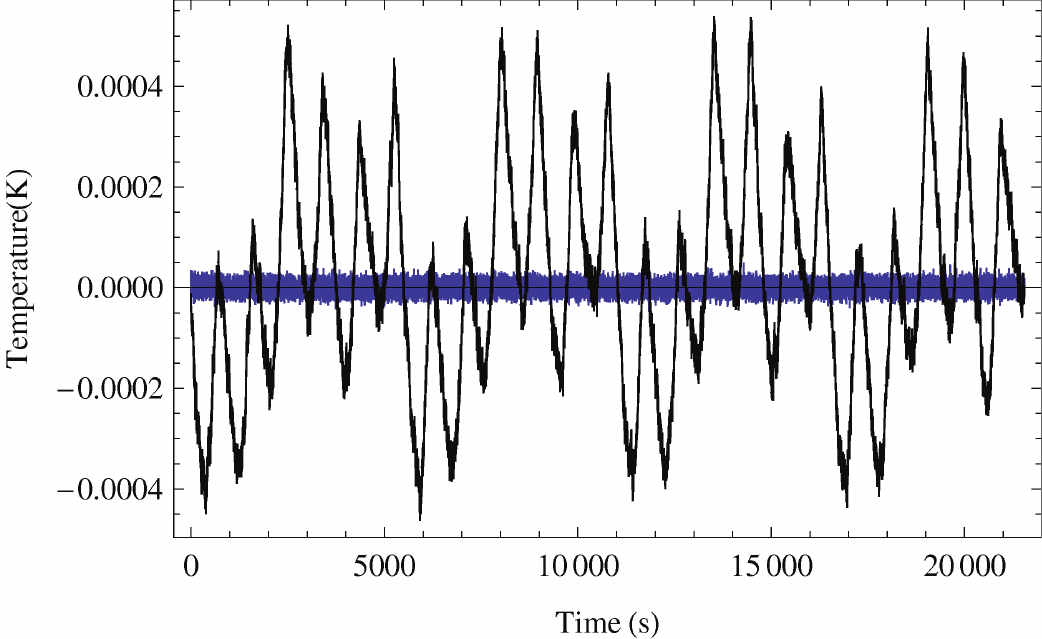}
\includegraphics[width=9.0cm]{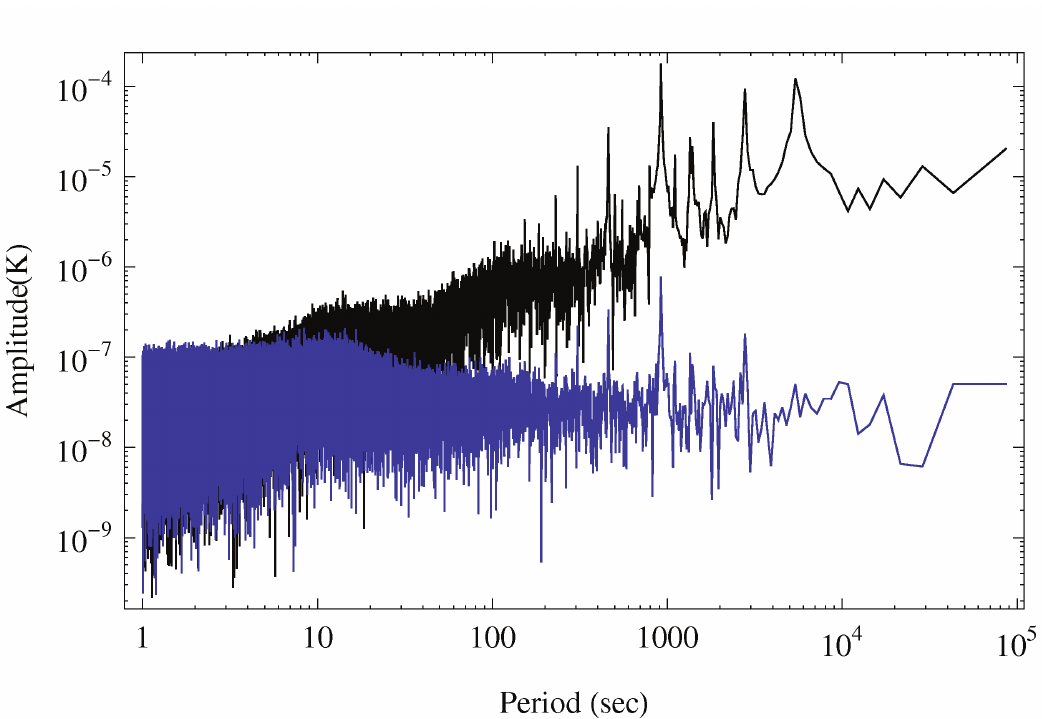}
\end{center}
\caption{\textit{Left --- }Daily behaviour of temperatures in the HFI outer shield in October 2009.  \textit{Right --- }corresponding Fourier transform amplitude.  Black --- 4KL1 sensor (close to 30 and 44 GHz loads); blue --- PIDN sensor (close to 70\,GHz loads).}
\label{t4k1}
\end{figure*}

\begin{figure*}
\begin{center}
\includegraphics[width=9.0cm]{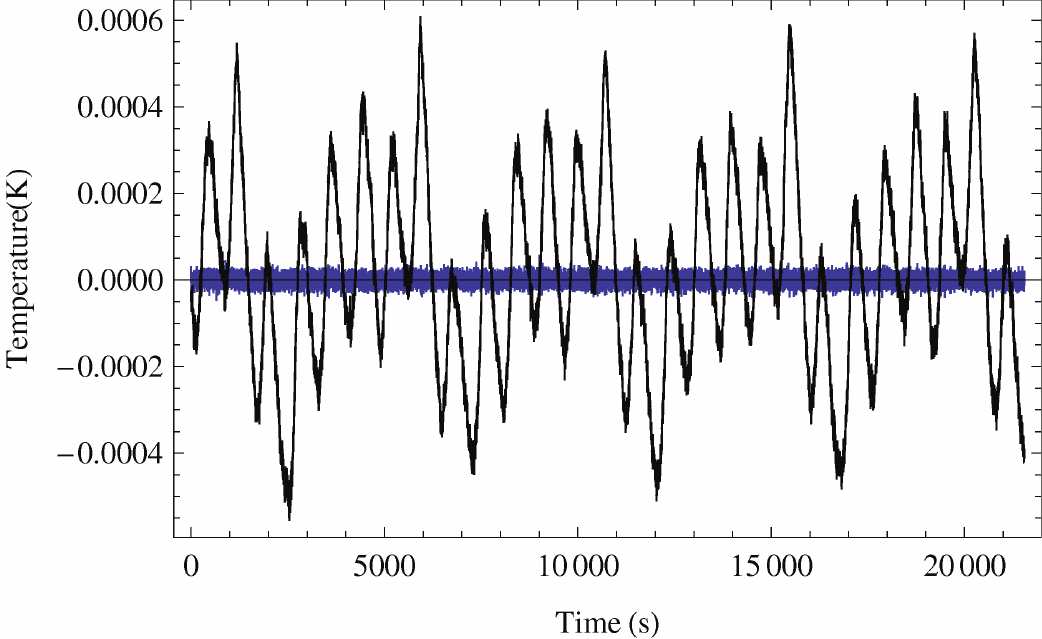}
\includegraphics[width=9.0cm]{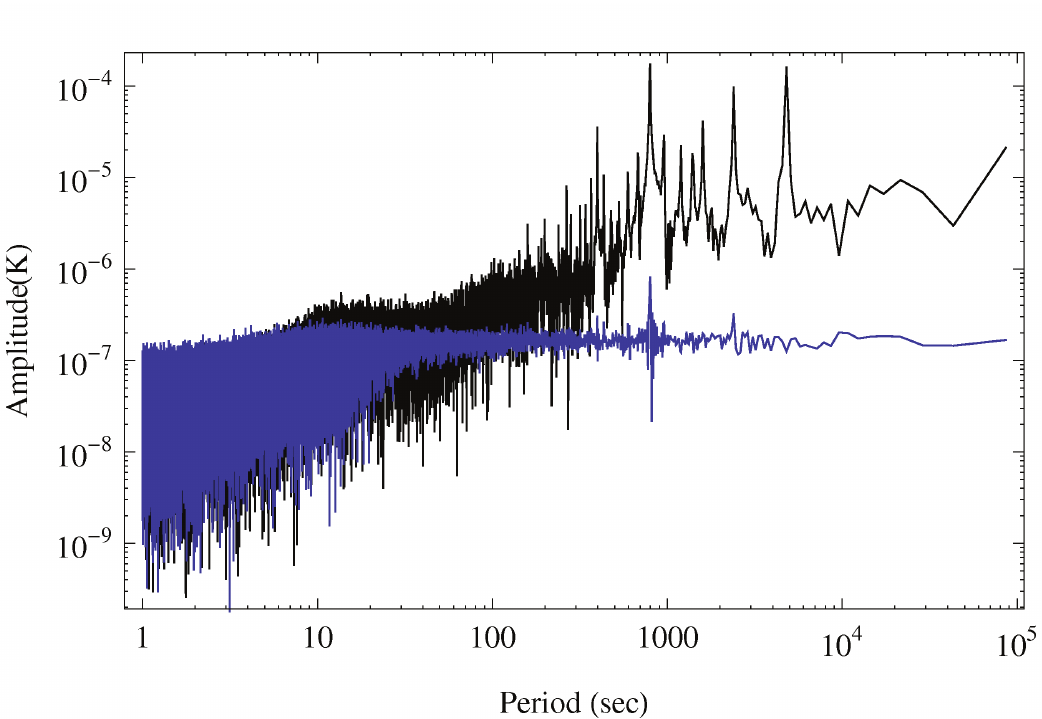}
\end{center}
\caption{\textit{Left --- }Daily behaviour of temperatures in the HFI outer shield in May 2010.  \textit{Right --- }corresponding Fourier transform amplitude.  Black --- 4KL1 sensor (close to 30 and 44 GHz loads); blue --- PIDN sensor (close to 70\,GHz loads).}
\label{t4k2}
\end{figure*}

Temperatures of the HFI outer shield and LFI reference loads (Fig.~\ref{fig:LFI_loads}) are relatively   homogeneous and uniform throughout the year.  Figs. \ref{t4k1} and \ref{t4k2} show temperature as a function of time and the corresponding Fourier transform amplitudes for two sensors from October 2009 and May 2010, respectively.   Peaks at the sorption cooler frequencies are always below 1\,mK amplitude, and at the microkelvin level in the region close to the 70\,GHz reference loads.  

\begin{figure*}
\begin{center}
\includegraphics[width=9.0cm]{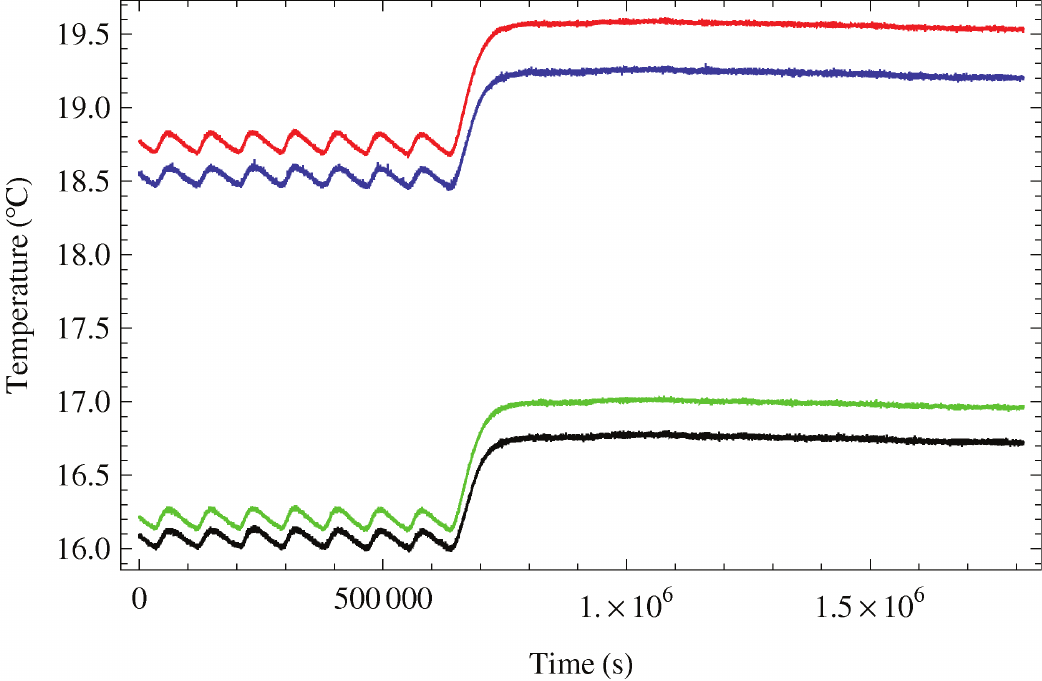}
\includegraphics[width=9.0cm]{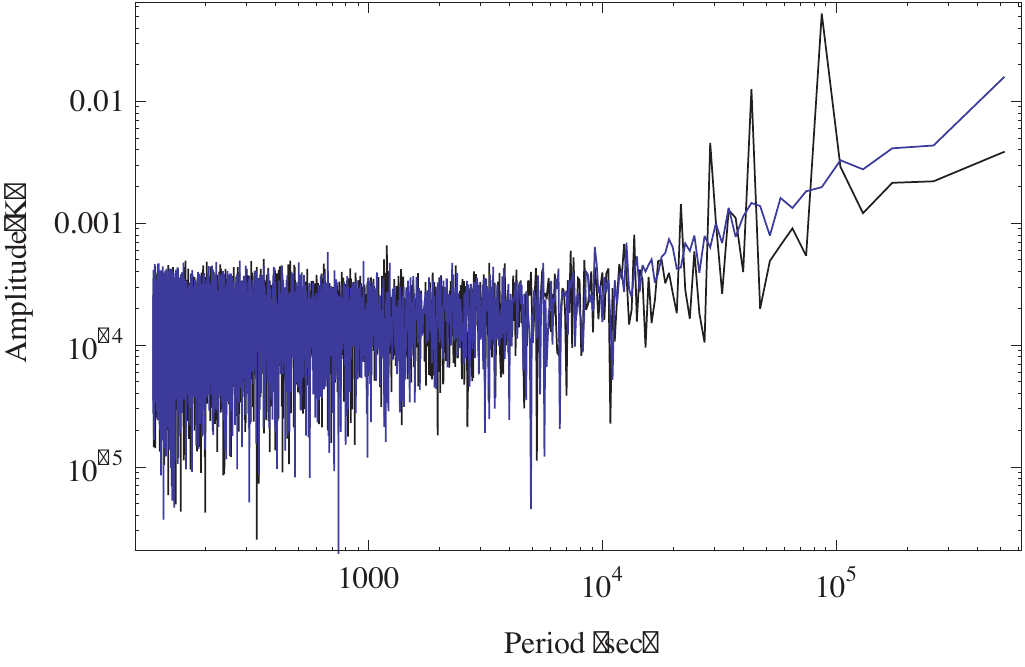}
\end{center}
\caption{\textit{Left --- }LFI backend temperatures (black --- left BEM1 sensor; blue --- left BEM2 sensor; red --- right BEM1 sensor; green --- right BEM2 sensor) around day 259 of flight operations, when the transition to an always-on condition for the transponder was implemented. \textit{Right --- }comparison of the Fourier transform amplitude of the Left BEM1 sensor before (black) and after (blue) transition to stable, showing the daily time scale peaks disappearing.}
\label{fig:bemmod}
\end{figure*}

The temperature of the LFI backend unit (BEU, see Fig.~\ref{fig:LFI_Q_scheme}) showed a 0.15\,K daily modulation in the first part of the mission, driven by the transponder on-off cycle discussed in Sect.~5.1.  After the transponder was turned on continuously (day 259), the temperature variation decreased dramatically (Fig. \ref{fig:bemmod}).

Table \ref{LFI_summary} gives typical daily mean temperatures and temperature variations of the LFI focal plane and backend during the first year of operations.

\begin{table*}
\begingroup
\newdimen\tblskip \tblskip=5pt
\caption{Mean values and daily stability of the main LFI temperature sensors in two cases representative of the best and worst case. The early phase was the best case for the focal plane and the worst case for the backend; the late phase was the opposite.}
\label{LFI_summary}
\nointerlineskip
\vskip -3mm
\footnotesize
\setbox\tablebox=\vbox{
   \newdimen\digitwidth 
   \setbox0=\hbox{\rm 0} 
   \digitwidth=\wd0 
   \catcode`*=\active 
   \def*{\kern\digitwidth}
   \newdimen\signwidth 
   \setbox0=\hbox{+} 
   \signwidth=\wd0 
   \catcode`!=\active 
   \def!{\kern\signwidth}
\halign{\hbox to 1.5in{#\leaderfil}\tabskip=1.5em&
   \hfil#\hfil&
   \hfil#\hfil&
   \hfil#\hfil\tabskip=3em&
   \hfil#\hfil\tabskip=1.5em&
   \hfil#\hfil&
   \hfil#\hfil\tabskip=0pt\cr
\noalign{\doubleline}
\omit&\multispan3\hfil E{\sc arly} P{\sc hase}$^{\rm a}$\hfil&\multispan3\hfil L{\sc ate} P{\sc hase}$^{\rm b}$\hfil\cr
\noalign{\vskip -3pt}
\omit&\multispan3\hrulefill&\multispan3\hrulefill\cr
\omit&$T$&$\sigma_T$&$\Delta T_{\rm p-p}$&$T$&$\sigma_T$&$\Delta T_{\rm p-p}$\cr
\omit\hfil S{\sc ensor}\hfil&[K]&[K]&[K]&[K]&[K]&[K]\cr
\noalign{\vskip 5pt\hrule\vskip 5pt}
TS1R&  *19.865& 0.0036&0.026& *20.007&0.0169& 0.127\cr
TS2R&  *20.506& 0.0008&0.007& *20.639&0.0075& 0.050\cr
TS3R&  *20.399& 0.0008&0.007& *20.532&0.0077& 0.052\cr
TS4R&  *20.037& 0.0014&0.011& *20.176&0.0118& 0.085\cr
TS5R&  *20.388& 0.0005&0.005& *20.521&0.0078& 0.053\cr
TS6R&  *20.587& 0.0005&0.004& *20.719&0.0053& 0.031\cr
TS1L&  *20.635& 0.0008&0.006& *20.771&0.0038& 0.022\cr
TS2L&  *20.480& 0.0008&0.007& *20.614&0.0051& 0.032\cr
TS3L&  *20.090& 0.0012&0.009& *20.227&0.0098& 0.072\cr
TS4L&  *20.755& 0.0008&0.006& *20.891&0.0038& 0.022\cr
TS5L&  *20.569& 0.0005&0.004& *20.705&0.0044& 0.025\cr
TS6L&  *20.826& 0.0005&0.004& *20.964&0.0039& 0.021\cr
\noalign{\vskip 5pt}
L-BEM1&289.232& 0.040*&0.177& 289.899&0.0156& 0.095\cr
L-BEM2&291.694& 0.041*&0.198& 292.380&0.0158& 0.100\cr
R-BEM1&291.923& 0.045*&0.179& 292.709&0.0141& 0.080\cr
R-BEM2&289.366& 0.044*&0.181& 290.137&0.0144& 0.085\cr
\noalign{\vskip 5pt\hrule\vskip 3pt}}}
\endPlancktablewide
\tablenote a ``Early Phase'' means sorption cooler FM2 operating, transponder cycling.  ``Late Phase'' means sorption cooler FM1 operating, transponder always on.\par
\endgroup
\end{table*}

%5.3
\subsection{1--5\,K elements}

\begin{figure*}
\begin{center}
\includegraphics[width=9cm]{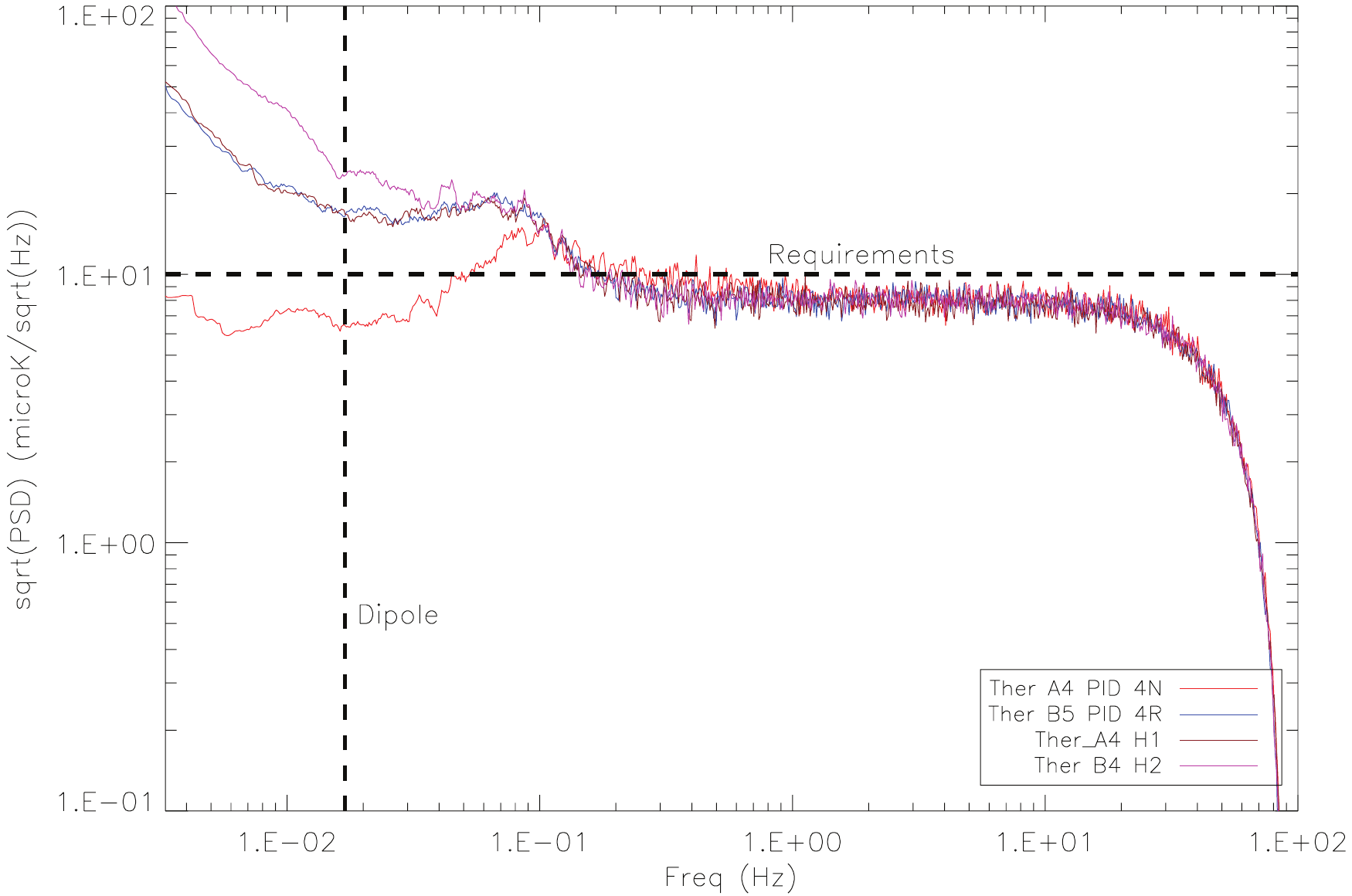}
\includegraphics[width=9cm]{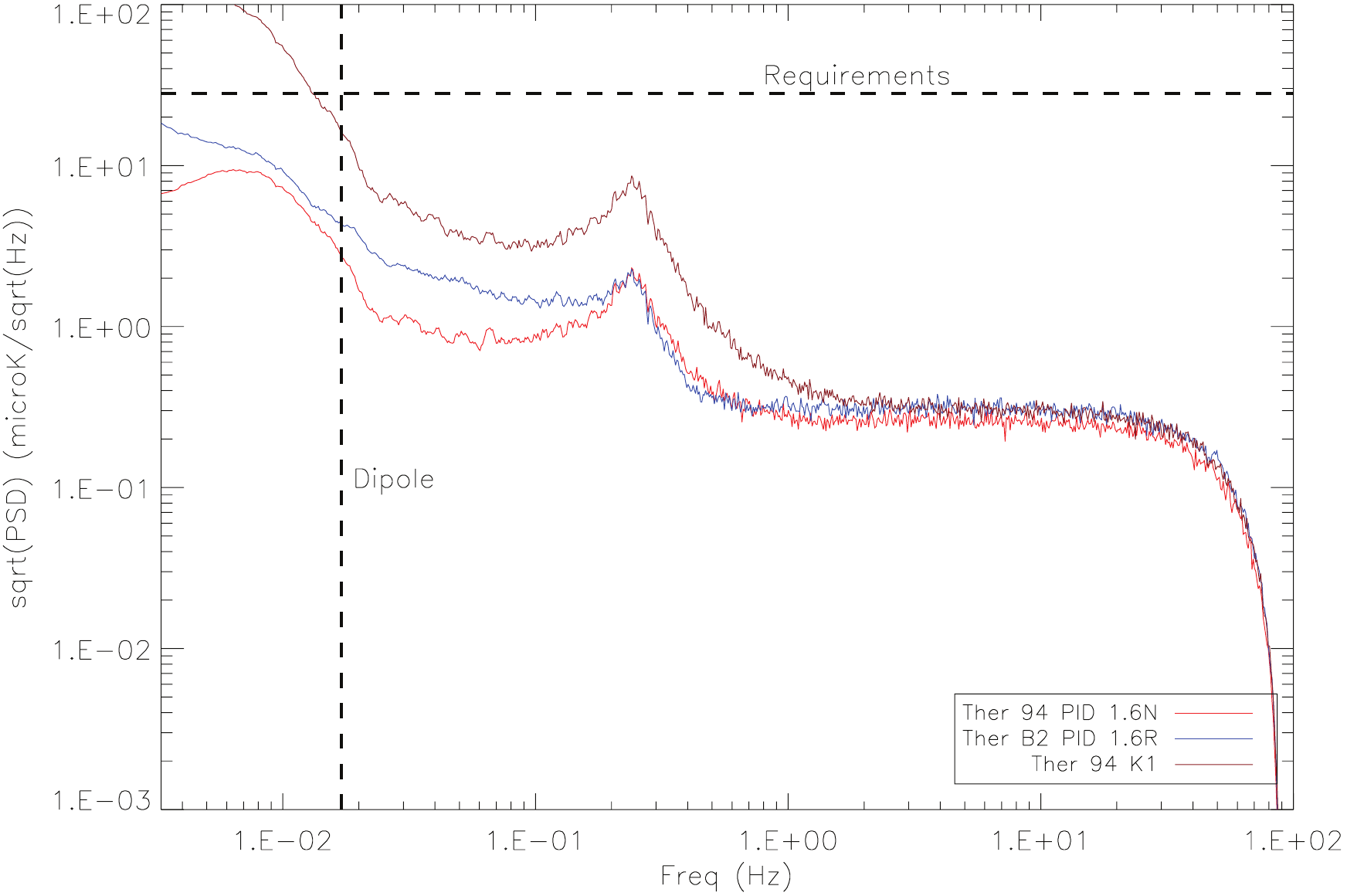}
\caption{\textit{Left --- }Power spectrum of thermal fluctuations measured at the feedhorns that couple to the telescope.  \textit{Right --- }Power spectrum of thermal fluctuations measured at the 1.4\,K filter plate.}
\label{fig:4K-1.4K-fluctuations.PS}
\end{center}
\end{figure*}

As described in Sect.~\ref{4-1.4-0.1_stages}, the temperature of the 4\,K box is regulated by a PID servo system with a heating belt on the 4\,K box.  The feedhorns that couple to the telescope are all at nearly the same temperature, which is controlled by four high sensitivity thermometers sampled at the same rate as the bolometers. Figure~\ref{fig:4K-1.4K-fluctuations.PS} left  shows the power spectrum of each of these thermometers over a broad range of frequencies. It can be seen that in the frequency range above 16\,mHz containing the scientific signal, the thermometer used for regulation (PID4N) is basically within the very strict requirement set initially. The three others show fluctuations at about twice that level, which introduce additional noise of about 20\% of the full detector chain noise. It also shows that the gradients on the 4\,K plates are very low, and comparable to the temperature fluctuations observed in ground tests  \citep{2008SPIE.7017E..32L}.  The reference loads for the LFI 70\,GHz radiometers are mounted above the heater ring, and benefit from the same temperature stability.  The 30 and 44\,GHz reference loads are mounted below the ring. They are also protected from the fluctuations induced by the sorption cooler on the \HeJT\ cooler, but not as well as the 70\,GHz loads.  

An equivalent PID servo system controls the stability of the 1.4\,K screen of the FPU using one of two redundant thermometers on the 1.4\,K stage structure. Two thermometers on the 1.4\,K filter plate monitor its temperature with the same sampling as the detectors.  Figure~\ref{fig:4K-1.4K-fluctuations.PS} right shows the power spectrum of each of these thermometers over the same range of frequencies. They stay well below the requirement of 28\muKHz\ in the 0.016--100\,Hz frequency range containing the scientific signal.  The additional noise is typically smaller than 2\% of the detector chain noise, in line with ground measurements  \citep{Pajot2010}.

In summary, these two stages behave exactly as designed and as tested on the ground. They do not add any significant noise to the measurements.

%5.4
\subsection{0.1\,K elements}

In ground tests the natural fluctuations of the dilution cold head were essentially eliminated by PID control at low frequencies and passive filtering at high frequencies.  The PIDs on the dilution plate and bolometer plate inject variable but known heat inputs with average values of 28 and 5\,nW, respectively.  Fluctuations were basically within requirements during the ground tests \citep{Pajot2010}, but perturbations from the test setting such as vibrations associated with helium transfer limited our ability to test this requirement precisely.

In flight, the microwave radiation reaching the bolometers is variable but so small (0.12\,nW) that it does not affect thermal behaviour.   On the other hand, cosmic rays above about 30\,MeV penetrate the FPU box, causing glitches in the PID thermometers and depositing energy in the bolometer and dilution plates as discussed in Sect.~4.4.  The bolometer plate PID had to be set with long enough time constants that it did not generate temperature fluctuations from the glitch signals.  As discussed in Sects.~4.1.3 and 5.3.1, the flux of Galactic cosmic rays is modulated by the solar wind.  This is a significant variable heat source not present in ground tests.

\begin{figure*}
\begin{center}
\hbox{
\includegraphics[width=9cm]{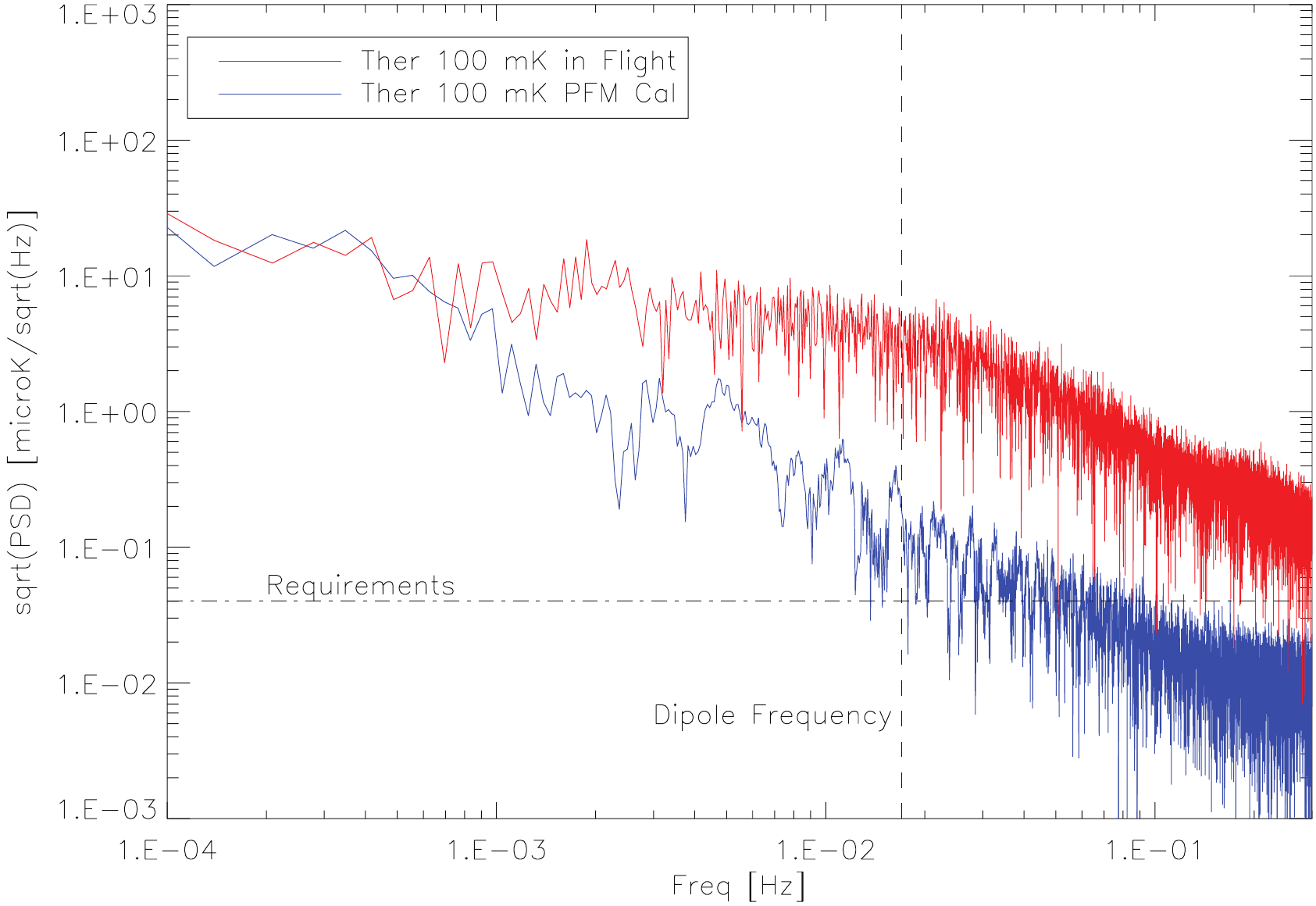}
\includegraphics[width=9.3cm]{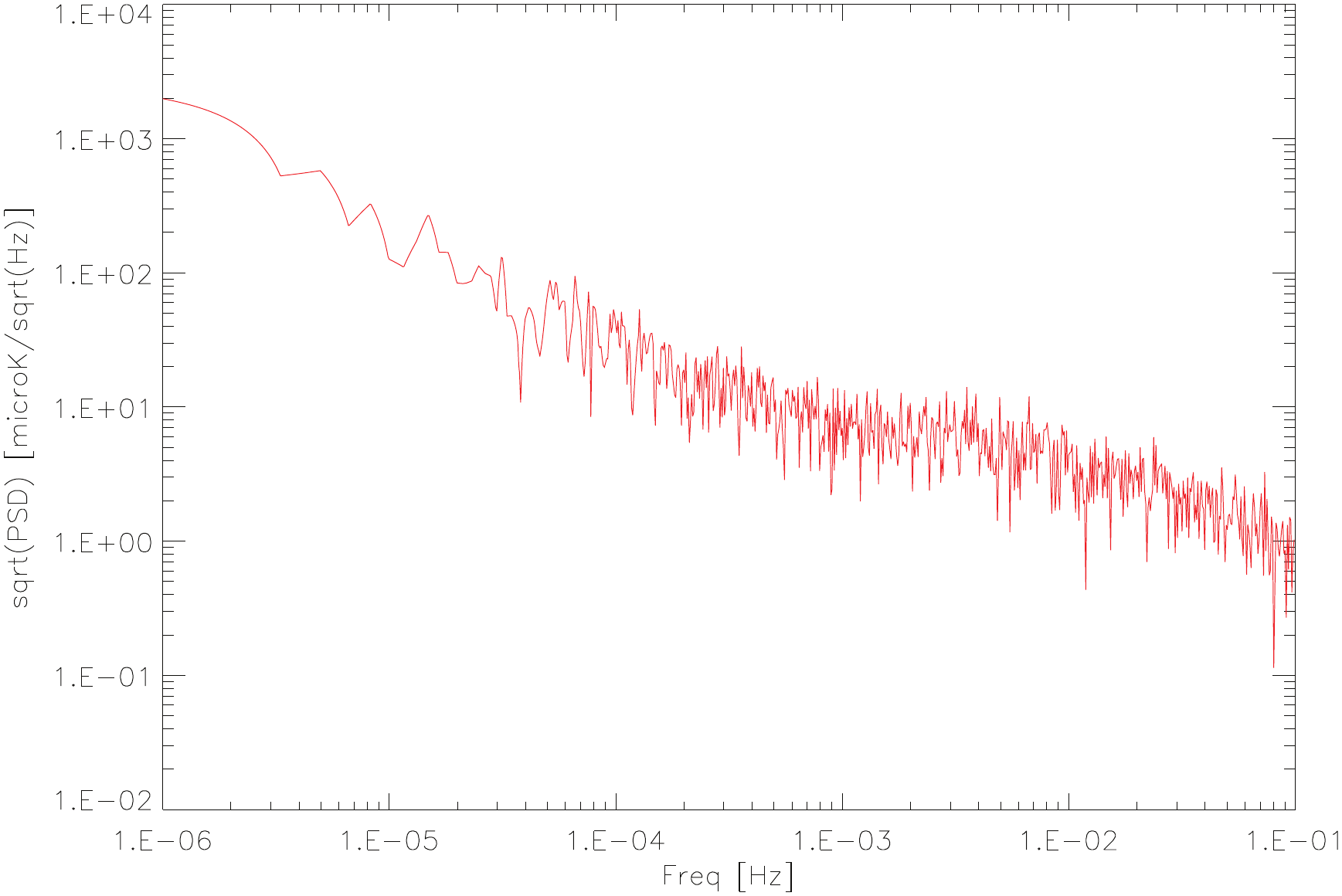}
}
\caption{\textit{Left --- }Frequency spectrum of the temperature of the bolometer plate, measured in flight (red) and on the ground (blue).    \textit{Right --- }Spectrum of the flight measurements over a wider frequency range.  The shoulder on the low frequency side is due to the temperature fluctuations described in Fig.~\ref{figure:correction}.  The bump in the $10^{-2}$ to $10^{-3}$\,Hz range seen, also seen in the left panel but only in the flight curve, is probably associated with the effect of cosmic rays in the bolometer structures.}
\label{figure:thermo}
\end{center}
\end{figure*}

Figure~\ref{figure:thermo} shows the bolometer plate temperature and the corresponding power spectrum for ground tests and in flight.  The temperature data have been ``deglitched'' (cosmic ray hits producing a signal significantly larger than the noise have been detected and removed).  It is clear that in the frequency range below 10\,mHz there is an excess of fluctuations in flight with respect to ground testing.  These fluctuations are removed by the scanning strategy and map-making, and have no residual effect on the science data. There is a smaller excess  above the requirement in the 16--40\,mHz range and affecting the data. This noise component is not fully correlated across the bolometer plate.

%5.5
\subsection{Bolometer Plate --- Particle Contribution}

Big solar flares --- typically 10 to 20 per solar cycle or at most three during the mission --- are expected to heat up the bolometer plate from 100\,mK to 500\,\hbox{mK}. This will create gaps in the survey of a few days.  None has been seen up to the end of 2010.

\begin{figure}
\begin{center}
\includegraphics[width=9cm]{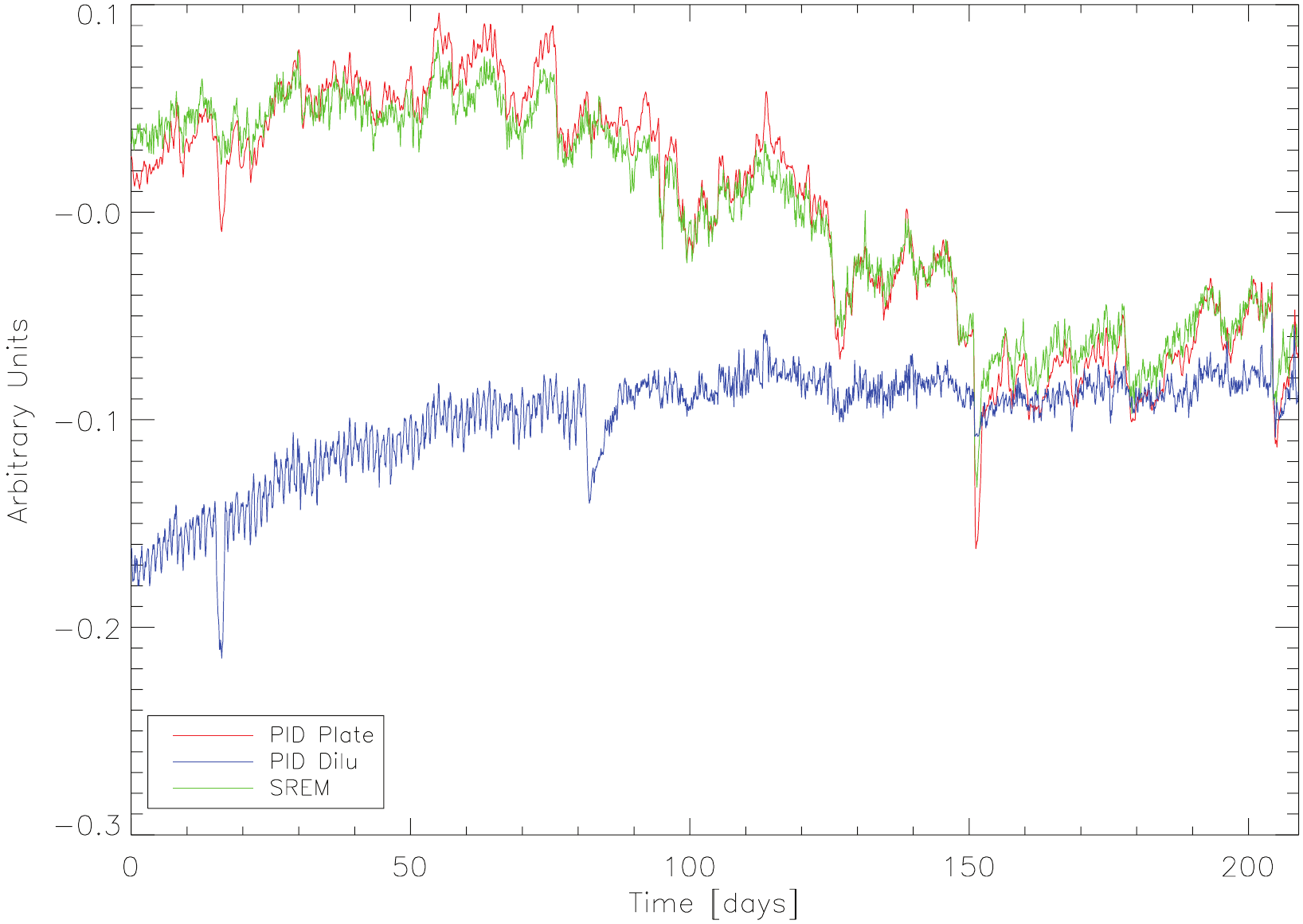}
\caption{Correlation between the signal of the SREM (red) and the signal of the active regulation of the temperature of the bolometer plate (green) for the frequency range $10^{-7}$--$10^{-5}$\,Hz. The equivalent signal of the active regulation of the dilution plate is shown in blue.}
\label{figure:PIDvsSREM}
\end{center}
\end{figure}

\begin{figure*}
\begin{center}
\hbox{
\includegraphics[width=8.8cm]{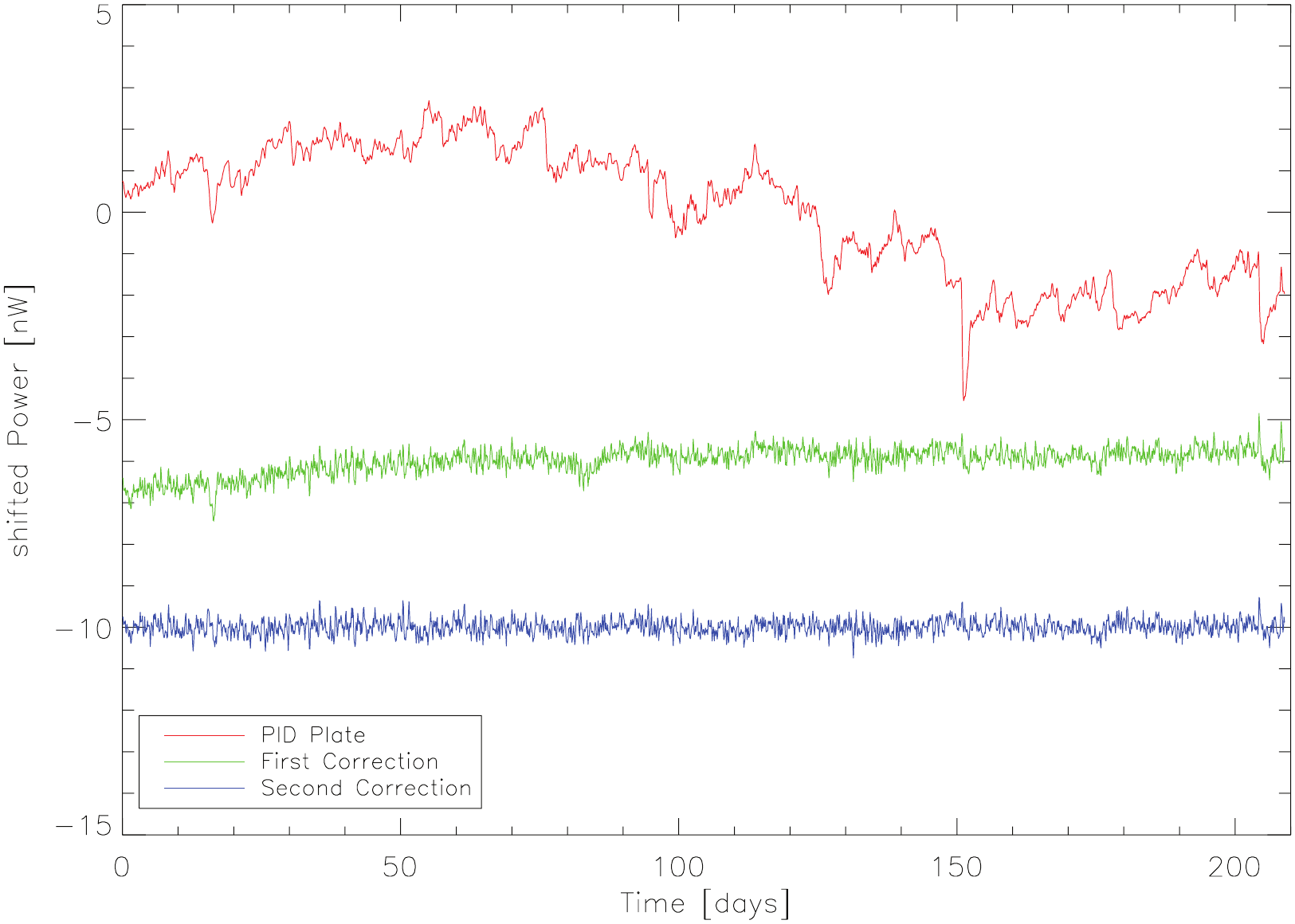}
\includegraphics[width=9.5cm]{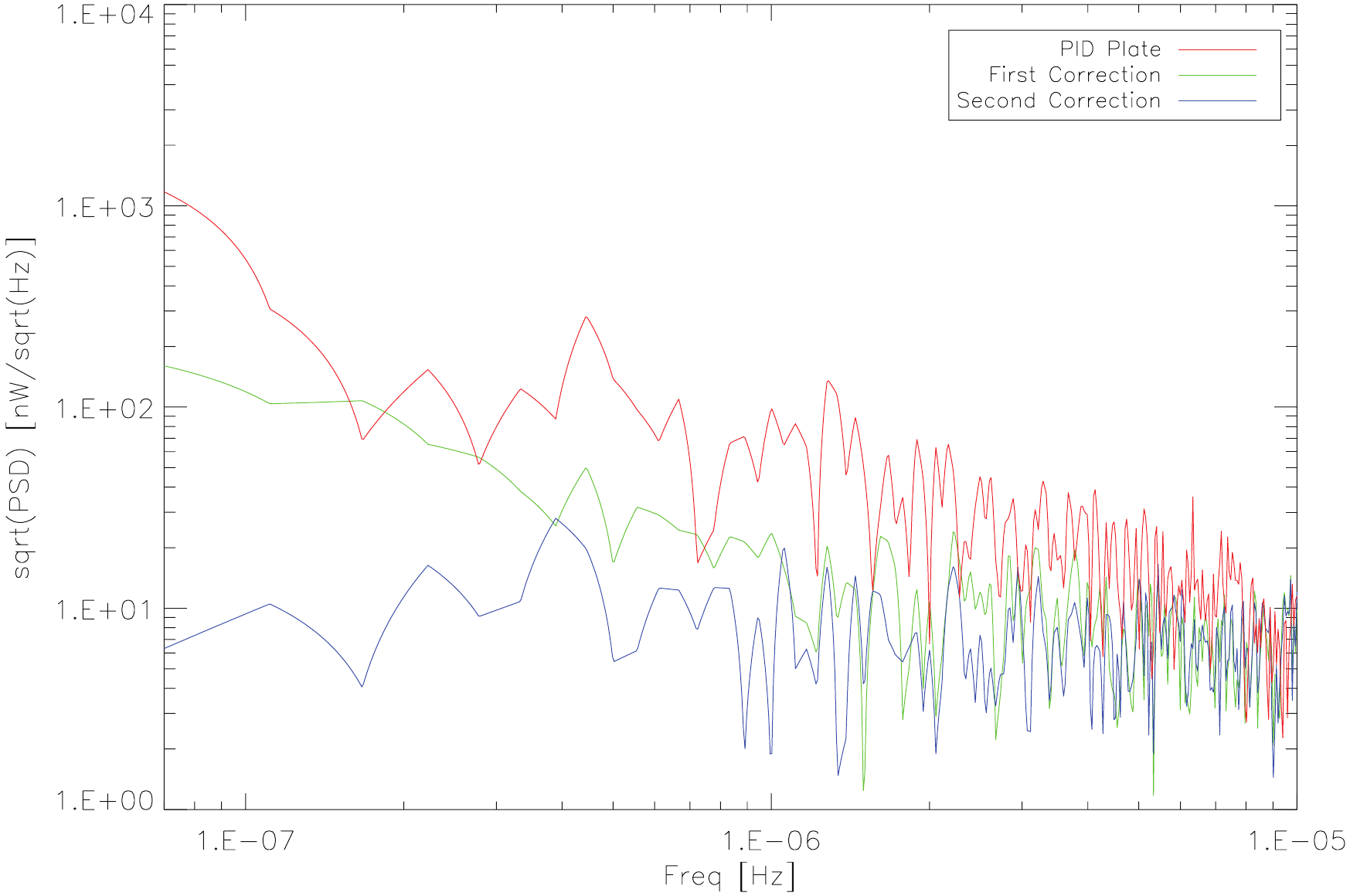}
}
\caption{\textit{Left --- }Variation with time of the power dissipated by the PID regulating the temperature of the bolometer plate (red curve; same as the green curve in Fig.~\ref{figure:PIDvsSREM}).  Subtracting the part that is correlated with the cosmic ray heat input as measured by the SREM gives the green curve.  Subtracting the part that is correlated with the PID on the dilution stage gives the blue curve.  The constancy of the blue curve shows that cosmic rays and variations in the dilution stage account for all of the very low frequency variation in heat input to the bolometer plate.  \textit{Right --- }Frequency domain versions of the same curves.}
\label{figure:correction}
\end{center}
\end{figure*}

The particle flux on the satellite is monitored by the \hbox{SREM} as discussed in Sect.~4.4.  Although mainly aimed at monitoring solar particles during flares, the SREM also detects Galactic cosmic rays, which dominate in periods of low solar activity.  Protons (electrons are only 1\% of the protons of a given energy and can be ignored) are detected in the range 20--600\,MeV, the range in which the energy deposited in matter is dominated by ionization. Figure~\ref{figure:PIDvsSREM} shows the excellent correlation between the active regulation of the temperature of the bolometer plate and the SREM at frequencies between $10^{-7}$ and $10^{-5}$\,Hz.  The correlation between the active regulation of the temperature of the dilution plate, which is also responding to slow changes in the flows of $^3$He and $^4$He, is less strong but still present.  If we subtract the two intrinsic sources of drift, we find a flat noise spectrum as expected from ground calibration (Fig.~\ref{figure:correction}). This illustrates that all dominant sources of fluctuations of the 100\,mK stage below a few  $\times 10^{-5}$\,Hz have been identified.

The noise spectra of the bolometer plate thermometer from ground calibration and flight at frequencies below $10^{-3}$\,Hz (Fig.~\ref{figure:thermo}) are consistent, as expected.  This part of the spectrum probably has a large contribution from the intrinsic fluctuations of the dilution stage, which are not expected to change between ground and flight calibrations.  This completes our understanding of the bolometer plate fluctuations down to this frequency. The difference between the curves in the frequency range 1--10\,mHz (Fig.~\ref{figure:correction} left) is at least partly due to glitches from particles hitting the thermometers.  After the contribution of the glitches is removed, the level of fluctuations is closer to the level observed in ground calibration, but still not identical. 

In flight, the 100\,mK temperature fluctuations induced by the modulation of Galactic cosmic rays dominate, but they do not affect the signal. In the frequency range between 16\,mHz and 300\,mHz, excess noise with respect to ground measurements is seen on the bolometers and thermometers; however, any common thermal mode affecting all thermometers and bolometers that is not fully corrected by the bolometer plate PID is easily removed using the dark bolometer signals (see \citealt{planck2011-1.7}). 

Showers of particles affecting more than 20~bolometers within a few milliseconds are also seen. (For fewer than 20 bolometers, the probability of physically unrelated but temporally correlated fluctuations becomes non-negligible). Their arrival is followed by an increase in the temperature of the bolometer plate by up to 10\muK.  The strongest showers occur at a rate of about one per day. Smaller temperature increases of order $0.01$--$0.1$\muK\ occur at a rate of one per hour. These showers are likely to be induced by high energy particles interacting with parts of the payload, depositing energy in the bolometers and also in the 100\,mK stage. The detailed physics of these events is not yet understood, but their phenomenological behaviour is well characterised. The temperature increases of the bolometer plate last up to one hour, in agreement with the time constant of the YHo link to the dilution plate. The bolometer plate itself, coated with copper and gold, becomes isothermal in milliseconds. Fluctuations induced by these particles on the bolometer plate are therefore removed along with the other common modes. However, the bolometer housings have time constants relative to the bolometer plate of 1--2\,s.  The bolometers and housings directly touched by the shower particles heat up suddenly, then cool down, while the bolometers and housings untouched by the shower heat up under the influence of the bolometer plate. This induces for a short time interval of a few seconds a non-uniform thermal behaviour of the bolometer housings. This systematic effect is not treated in the pipeline used for the early papers, as it is not a dominant effect; however, it is still one of the possible contributors to the remaining low frequency noise seen in flight, which was not present in ground tests (see \citealt{planck2011-1.5}).

%5.6
\subsection{Transients and seasonal effects}

\begin{figure*}
\begin{center}
\includegraphics[width=16.6cm]{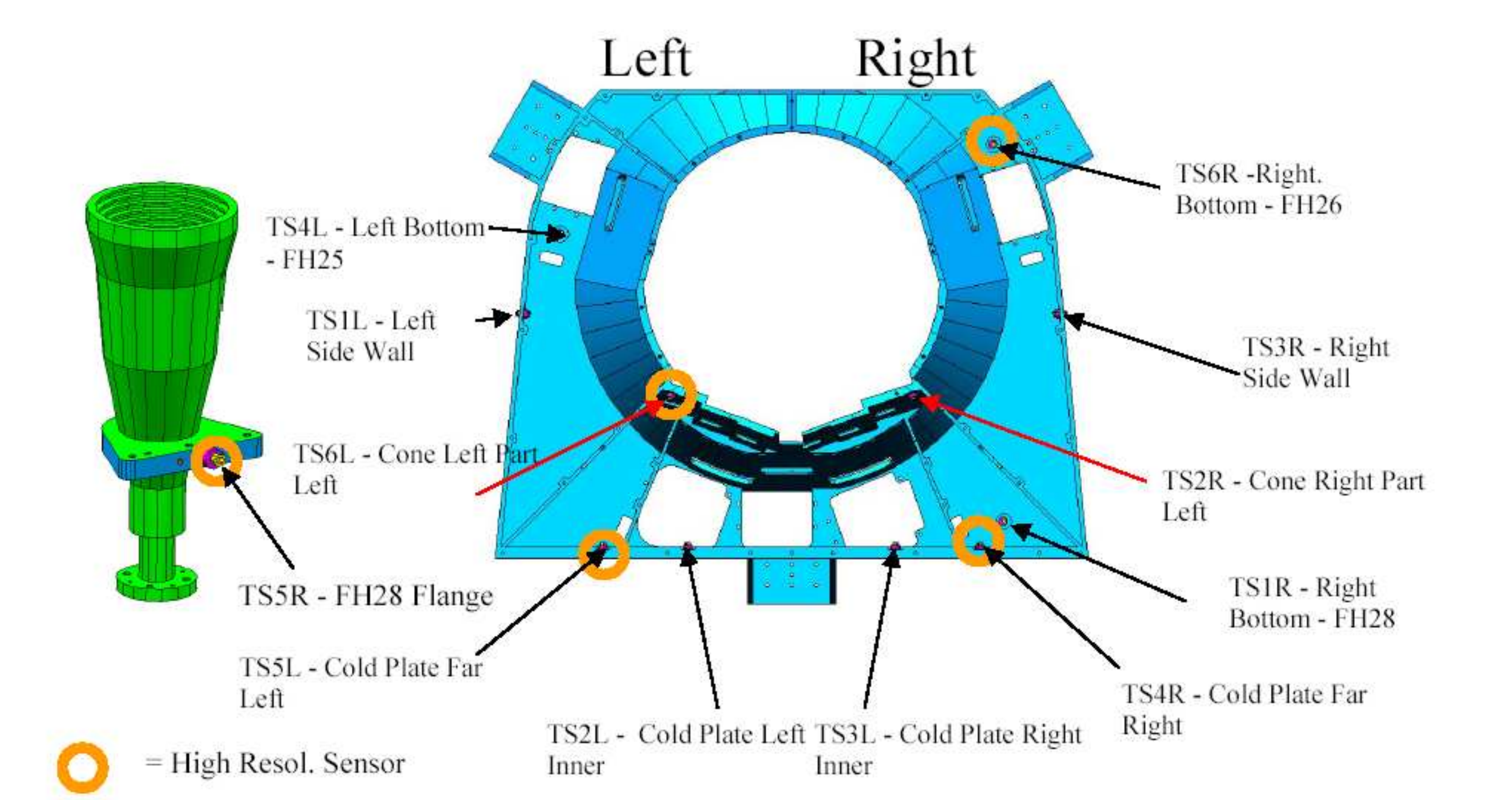}
\end{center}
\caption{Locations and names of temperature sensors in the LFI focal plane. }
\label{fig:LFI_sens}
\end{figure*}

\begin{figure*}
\begin{center}
\includegraphics[width=17cm]{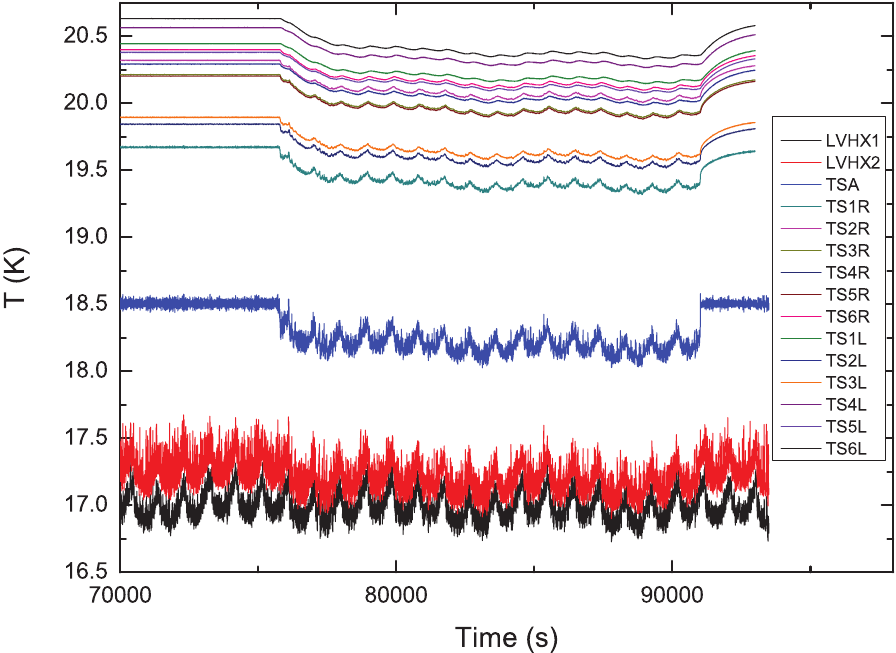}
\end{center}
\caption{Dynamic behaviour of temperature sensors when TSA control is switched off at roughly 76,000\,s.  Fluctuations at 1\,mHz are reduced more than 40\%, revealing longer-term fluctuations caused by the cycling of the sorption cooler.}
\label{fig:LFI_TSA_off}
\end{figure*}

At the end of the CPV phase a set of tests was performed to characterize the timescales and propagation of thermal fluctuations, as well as the response of the radiometers to changes in temperature.  The dynamic response of the LFI focal plane to turning off the TSA was measured, along with the transfer functions between the various thermal sensors in the LFI focal plane.  Figure~\ref{fig:LFI_sens} shows the locations of the sensors.  Figure~\ref{fig:LFI_TSA_off} shows the response of the system when the TSA is turned off.

\begin{figure*}
\centering
\includegraphics[width=9cm]{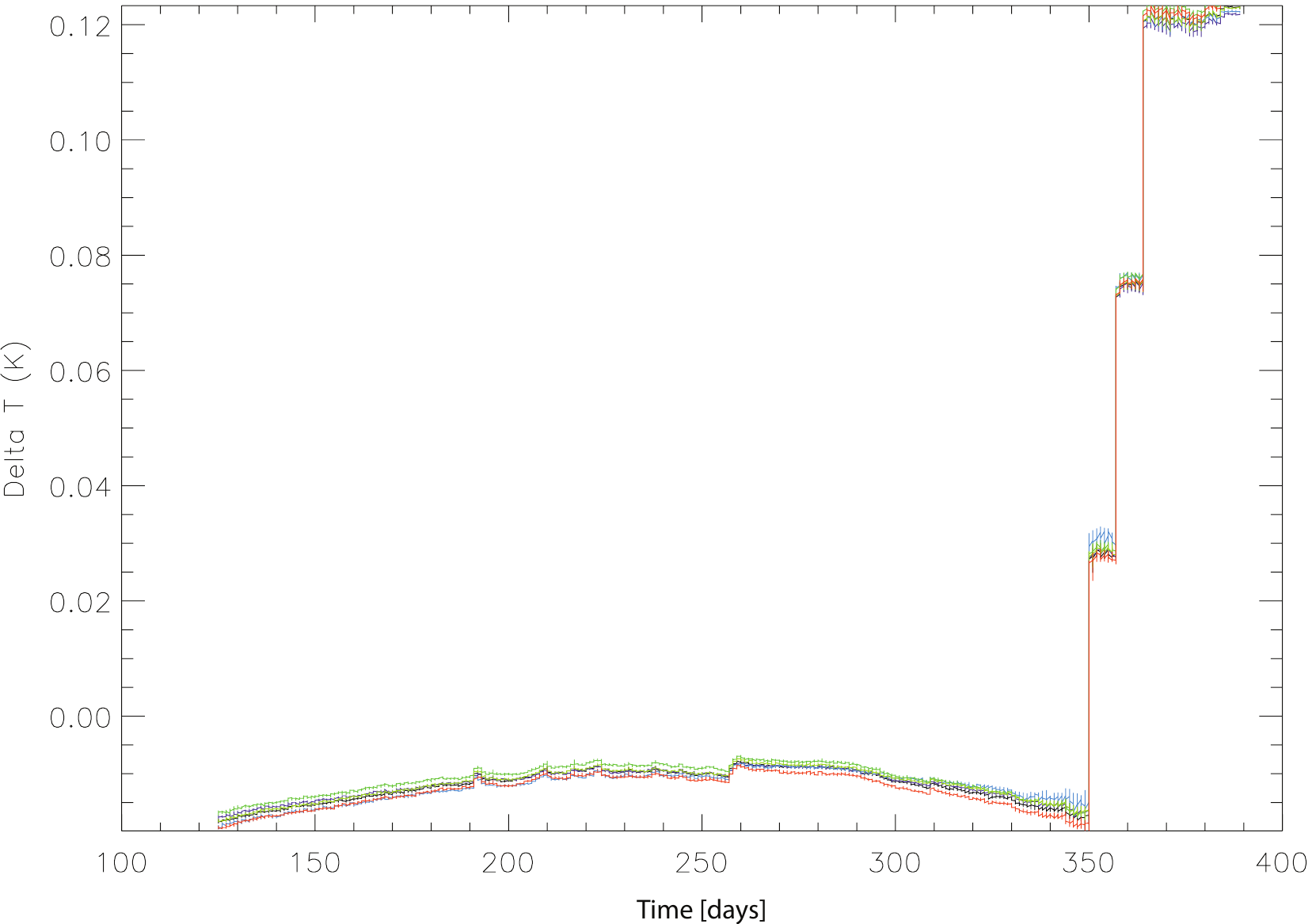}
\includegraphics[width=9.2cm]{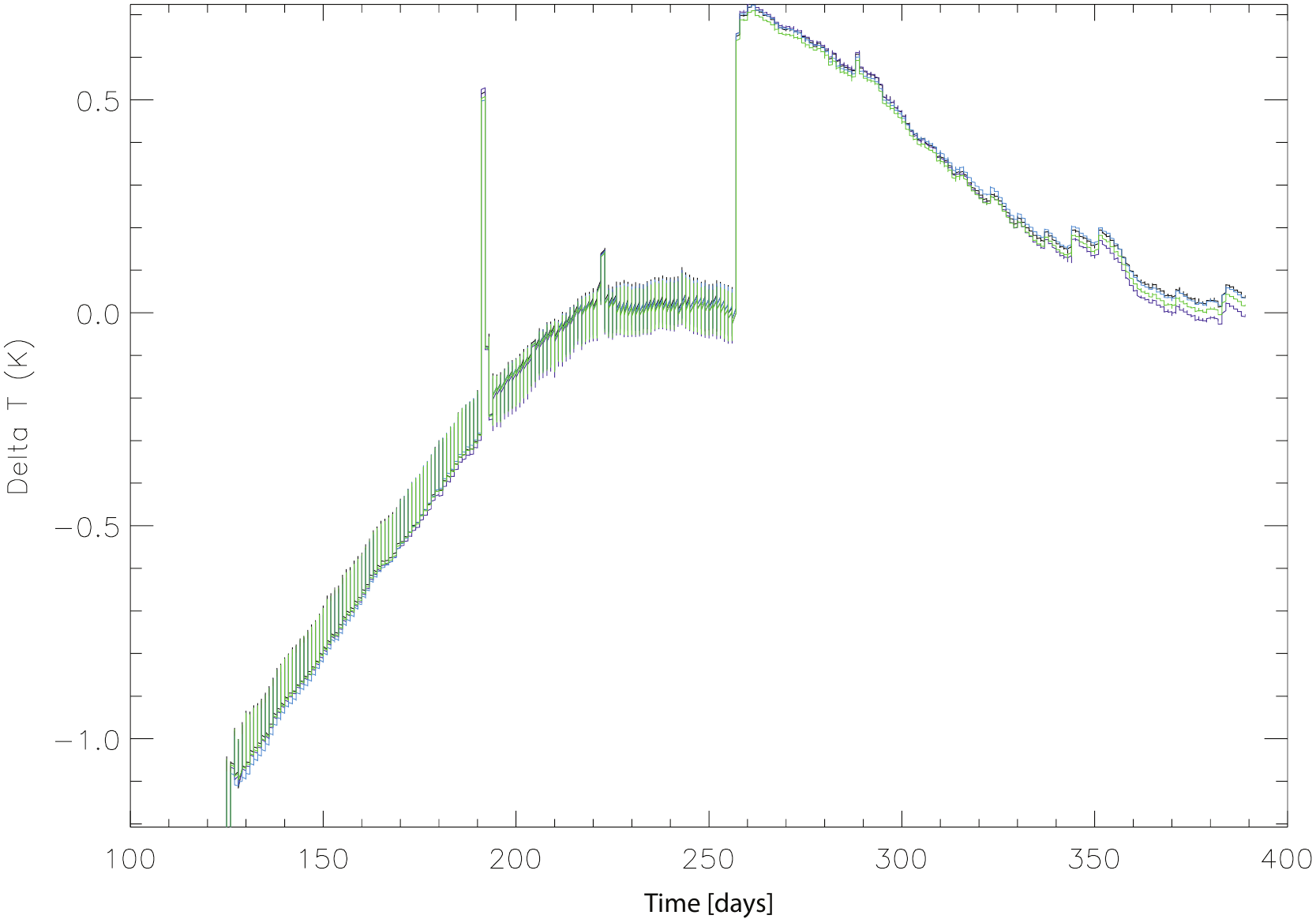}
\caption{\textit{Left --- }Temperature of LFI focal plane sensors TSL1 to TSL6 (see Fig.~\ref{fig:LFI_sens}) in the period from 125 to 389 days after launch.  Until intentional changes in the TSA setpoint started around day 350, the temperature shows only the slow drift correlated with distance from the Sun.  See also Fig.~\ref{fig:seasonal_variations}.  \textit{Right --- } Temperature of the LFI backend sensors (left backend block sensors) in the period from 125 to 389 days after launch.  Until day 259, the transponder was on only during downlinks.  This approximately diurnal variation drove a large periodic variation in temperature.  After day 259, the transponder was left on continuously, and the daily temperature variations became much smaller.}
\label{fig:t_sum}
\end{figure*}

\begin{figure}
\begin{center}
\includegraphics[width=9cm]{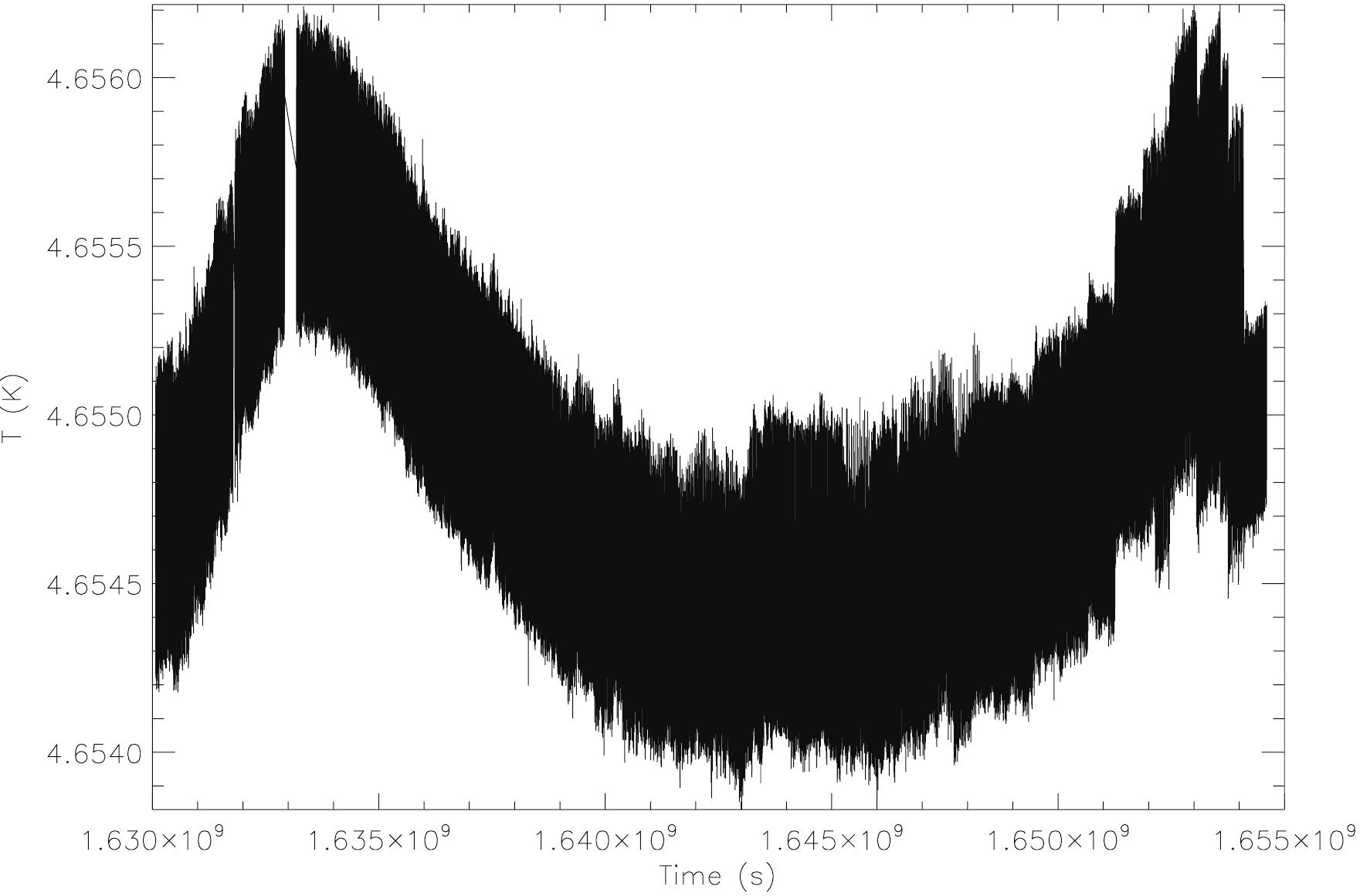}
\end{center}
\caption{Temperature of the lower part of the HFI outer shield, closest to the 30 and 44\,GHz reference loads, from the start of the first sky survey on 13 August 2009 to day 389.}
\label{fig:tl1}
\end{figure}

\begin{figure*}
\begin{center}
\includegraphics[width=0.76\textwidth]{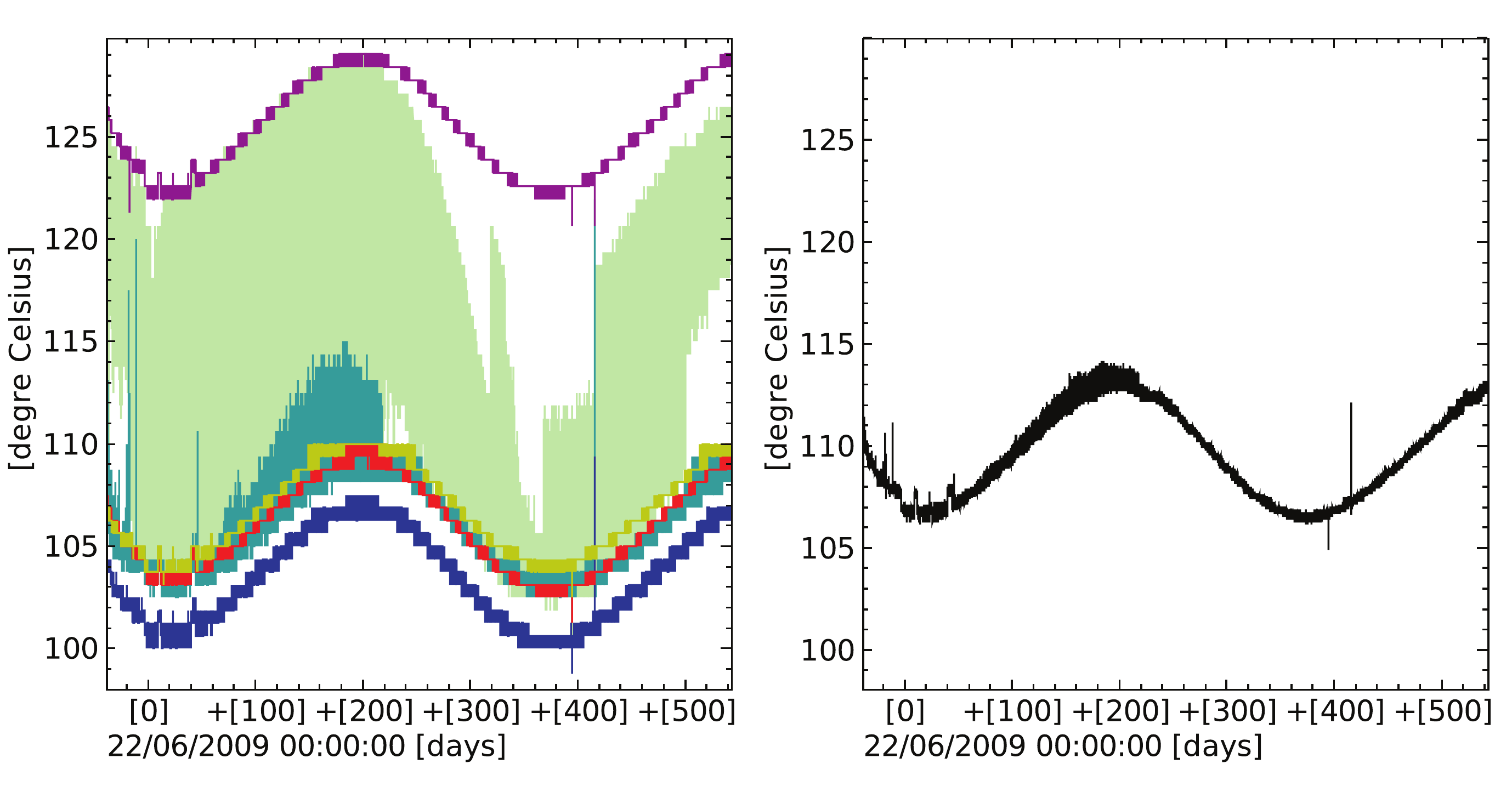}
\includegraphics[width=0.76\textwidth]{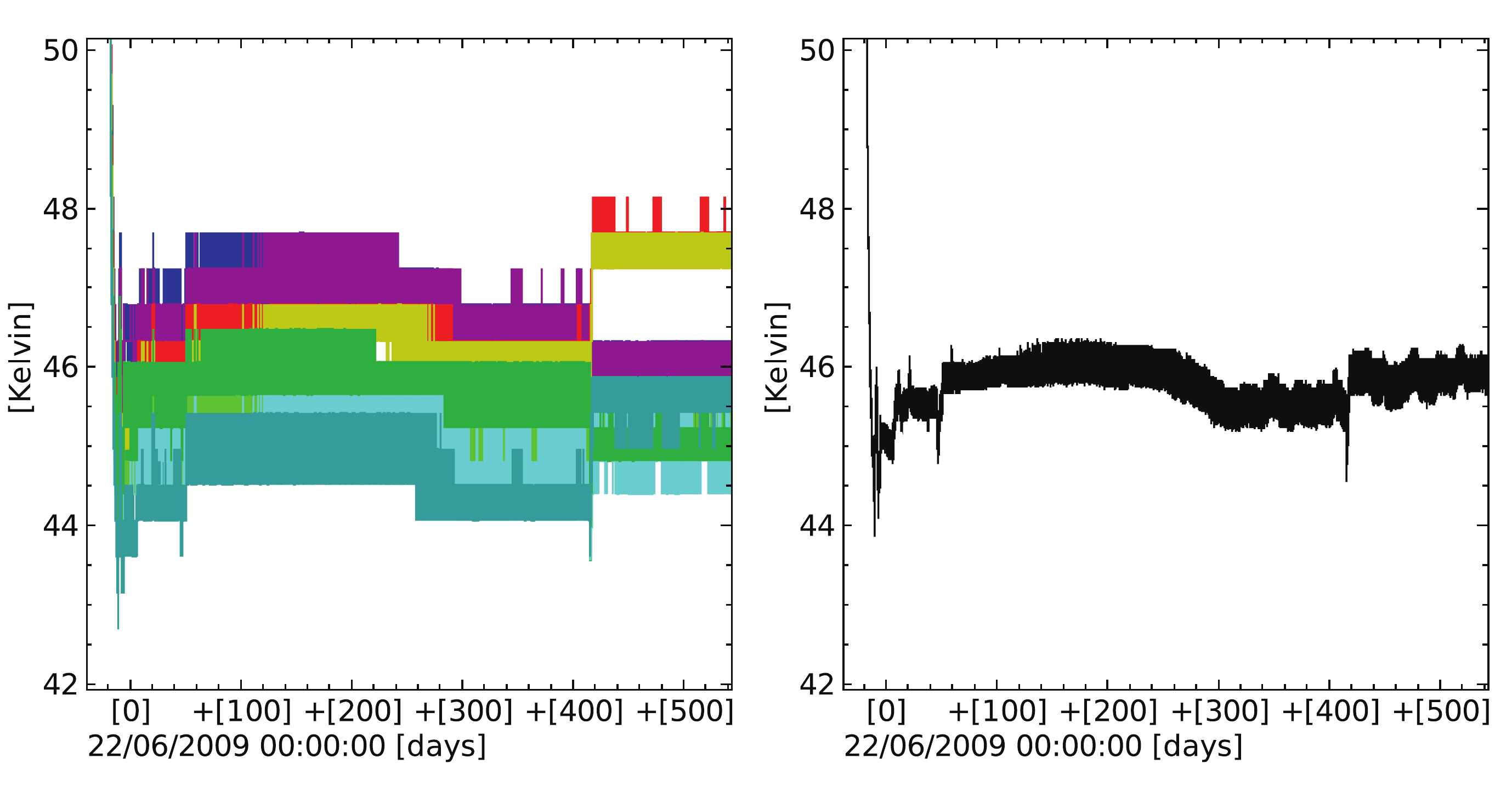}
\includegraphics[width=0.76\textwidth]{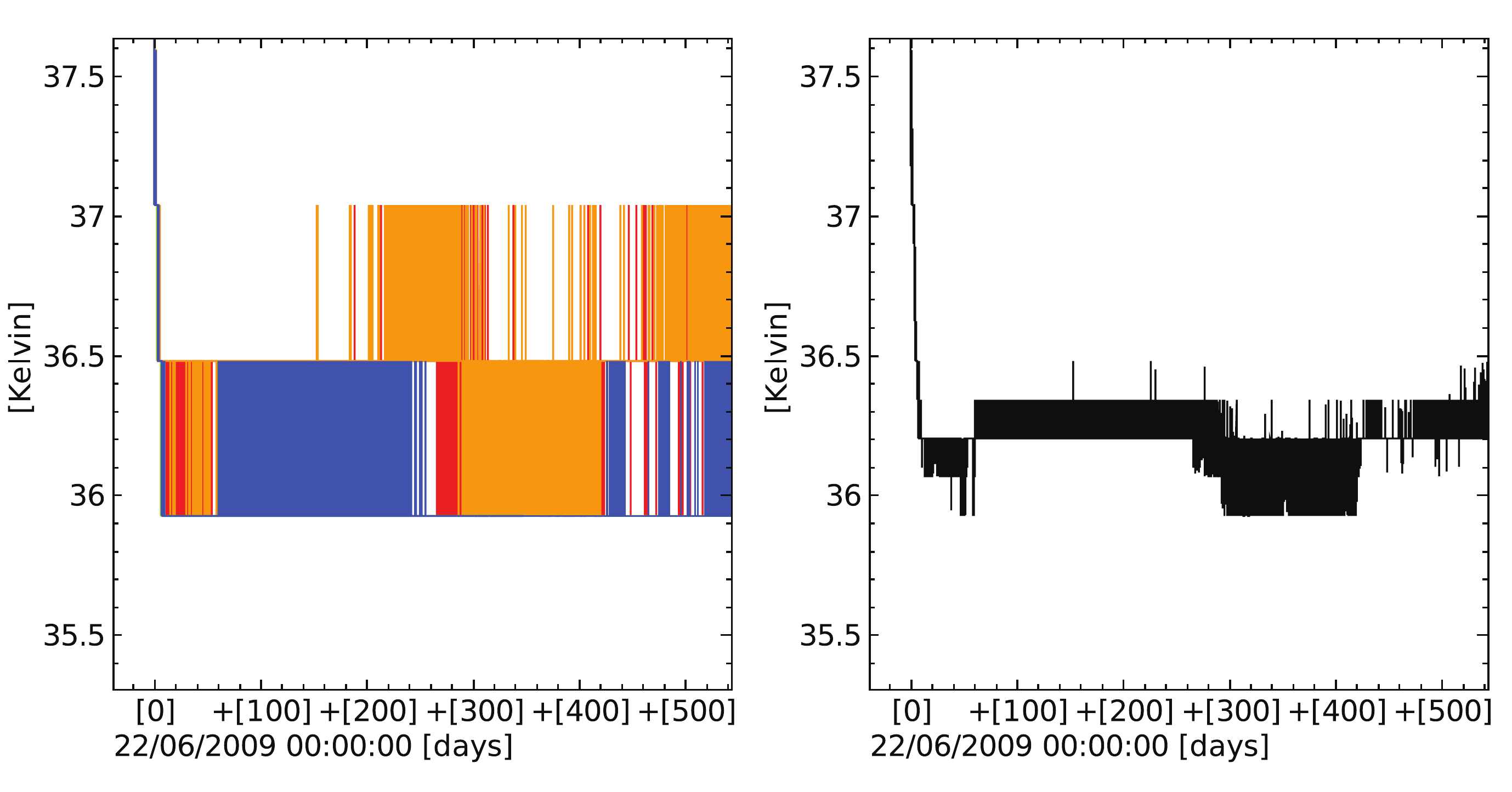}
\caption{Seasonal variation in the temperature of the solar panel ({\it top\/}), V-groove 3 ({\it middle\/}), and the primary mirror ({\it bottom\/}).  Individual readings are shown for multiple temperature sensors on each part of the structure ({\it left, all three panels\/}).  Quantization effects are clearly visible for V-groove 3 and the primary mirror.  These are reduced but not eliminated by averaging the relevant sensors ({\it black line in all three panels\/}).}
\label{fig:seasonal_variations}
\end{center}
\end{figure*}

The temperature of the LFI focal plane is quite stable over time (Fig.~\ref{fig:t_sum} left), showing a slow drift caused by the orbital variation in the Earth's distance from the Sun.  Starting at about day 350, changes in the TSA control set-point were needed in order to keep the stability within acceptable limits.
This orbit-induced seasonal variation appears in the temperatures of many components of \Planck, as can be seen in Figs.~\ref{fig:t_sum} left, \ref{fig:t_sum} right, \ref{fig:tl1}, and \ref{fig:seasonal_variations}.

The temperature of the LFI BEU, mounted in the SVM, was much less stable initially, showing a strong daily variation driven by the transponder.  As described in Sect.~5.1, this daily variation was effectively eliminated starting with day 259 by keeping the transponder on continuously.  After that, the temperature showed only the common SVM seasonal drift, plus some discontinuities related to major sorption cooler operations (Fig. \ref{fig:t_sum} right).

The LFI reference loads \citep{valenziano2009} are thermally connected to the HFI outer shield.  Temperature fluctuations in the lower part of the shield are transmitted to the 30 and 44\,GHz reference targets.  Temperature fluctuations in the upper part of the shield are almost entirely damped by the PID controls (Sect.~2.4.5).  The temperature shows a 1\,mK peak-to-peak fluctuation amplitude at the typical time scales of the sorption cooler, plus the overall seasonal drift (Fig. \ref{fig:tl1}).

%SECTION 5.7
\subsection{Interactions between systems and instruments }
\label{sec:interactions}

%5.7.1
\subsubsection{Behavior of liquid hydrogen at 20\,K}

\begin{figure*}
\begin{center}
\includegraphics[width=13cm]{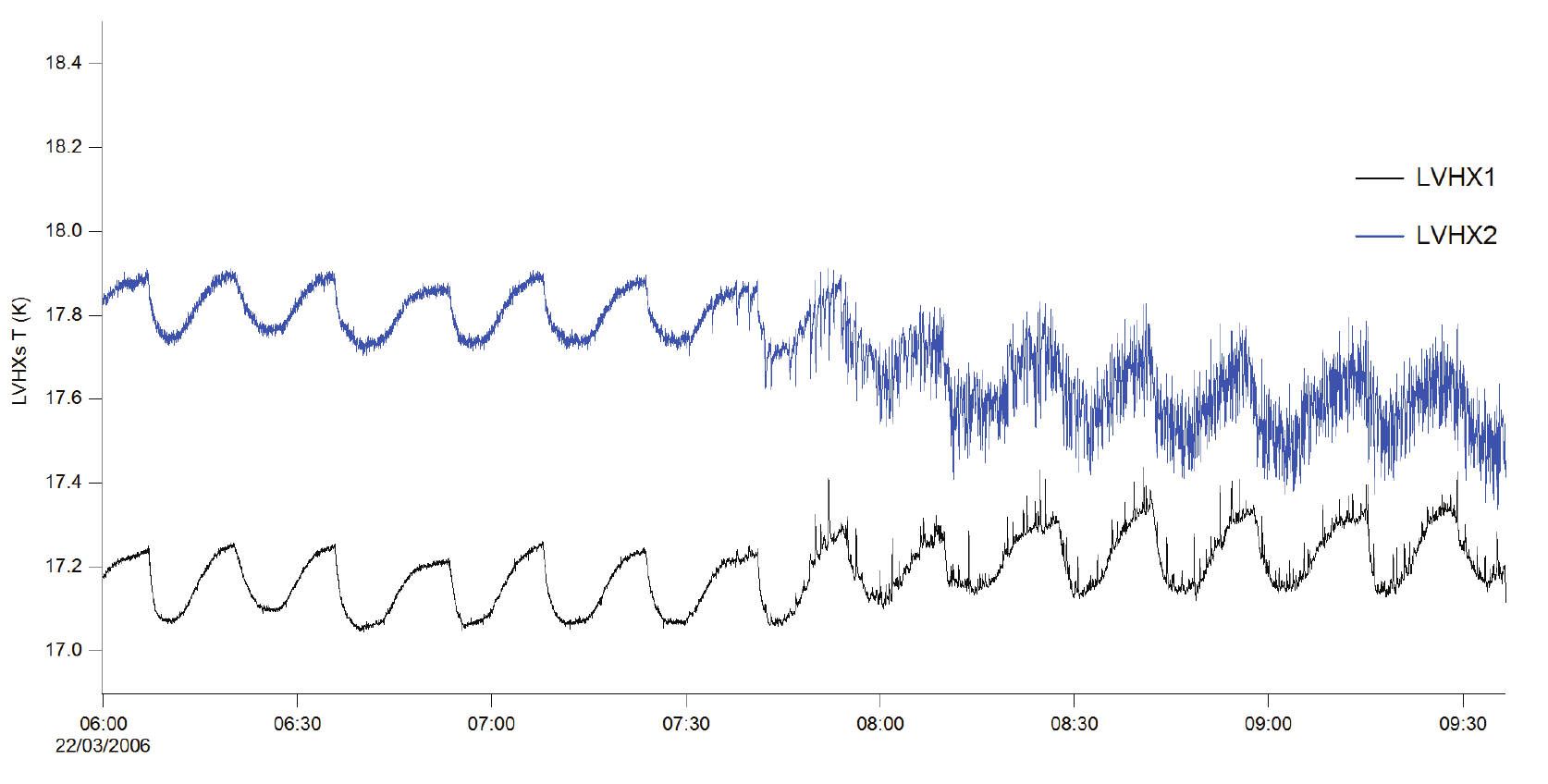}
\caption{Temperature fluctuations for two different operating regimes of the sorption cooler cold end.  \textit{Left --- }The cold end is in a balanced state and the liquid interface is inside the LVHX2 body.  \textit{Right --- }The cold end is in an unbalanced state and the liquid interface moves into the counterflow heat exchanger beyond the LVHX2 body.  The flow becomes erratic because of plug-flow events. }
\label{fig:SCStemperature_fluctuations}
\end{center}
\end{figure*}

Temperature and temperature fluctuations are shown for two different regimes in the cold end.  On the left side of Fig.~\ref{fig:SCStemperature_fluctuations}, the fluctuations are much cleaner, i.e., with lower fluctuation levels, especially at high frequencies, than on the right side.  This behavior was observed during ground testing, and is well understood. The cold-end, in the left regime, is in a balanced state (heat lift = power dissipation) where the liquid interface is drawn into the LVHX2 body.  In contrast, the cold end in the right regime is in an unbalanced state (heat lift $>$ power dissipation) in which excess liquid forms.  The increased fluctuations are due to plug-flow events as the liquid interface moves into the counter-flow heat exchanger just past the LVHX2 body.

%5.7.2
\subsubsection{Behavior of liquid helium at 4\,K and 1.4\,K}

Liquid $^4$He is generated at the 4\,K precool and at the 1.4\,K \hbox{JT}. 
Instabilities developed during the CQM CSL test when the 1.4\,K stage was too cold (Sect.~3.3), and in flight  when the 4\,K stage was too cold (Sect.~4.3).  As discussed earlier, liquid helium was overproduced and filled the pipes between the 1.4\,K stage and the 4\,K stage, and unstable evaporation then generated large temperature fluctuations, including of the 100\,mK stage.  It is clear that because of instabilities associated with excess liquid helium, adjustment of the coolers for maximum cooling performance and margins in operation of the cooling chain  does not necessarily lead to the best overall configuration.  It was essential for Planck that the PIDs have enough power to warm both the 1.4 and 4\,K stages to the optimal operating point far away from the unstable evaporation region, even with the lowest isotope flow for the dilution cooler.

%SECTION 6
\section{Cryo operations}

Adjustments to the sorption cooler are necessary throughout the mission due to the gradual loss of hydrogen-absorbing capacity of the compressor hydride material \citep{borders2004} with cycling and time at high temperature.   As the hydride degrades, the input power is increased while the cycle-time is decreased.  In terms of degradation, it is ultimately the input power that limits the sorption cooler lifetime, as both the desorption and heat-up channels are limited to 250\,W each.  Adjustments are made as needed on a weekly schedule, based on analysis of the pressures and temperatures of the compressor elements and the various heat exchangers.

The HFI cryochain parameters are set once for the entire mission.  HFI operations essentially consist in monitoring the housekeeping parameters during the daily telecommunication periods. The only moving part of the satellite payload is the \HeJT\ cooler compressor; its stroke amplitude, head temperature, currents, transducers, high and low gas pressures, and flow rates are monitored carefully.  Dilution cooler helium pressures and temperatures are also monitored.  The helium remaining in the tanks, which determines the HFI lifetime, is calculated from measurements of temperature, pressure, and flow rate along the dilution cooler pipes. The power put on the various PIDs (at the 4\,K box, 1.4\,K filters, dilution plate, and bolometer plate stages) is monitored as well.  As a side effect, this allows detection of solar flares through their heating effect on the bolometer plate.  Finally all temperature sensors at the different subsystems are closely monitored.

%SECTION 7
\section{Lessons learned for future missions, and conclusion}
\label{sec:lessons}

The \Planck\ thermal system combines passive radiative cooling and thermal isolation with three mechanical coolers --- two representing new technologies used for the first time to cool scientific instruments in space --- in an elegant overall thermal design.  The system works as designed, and it works very well, delivering the thermal environment required by \Planck's state-of-the-art instruments.  

We conclude here with some ``lessons learned,'' many incorporated in \Planck, which we hope will be useful for future missions.  Some may seem obvious in retrospect; however, trade-offs must be made during construction without full knowledge of the consequences, and prior experience is an invaluable guide.
The main points are:

\begin{itemize}

\item In cryogenic missions with stringent temperature stability requirements,  temperature sensors on the spacecraft (SVM) and telescope should have an accuracy significantly better than on \Planck\ (typically 0.1\,K resolution). 

\item The design and lifetime of the sorption cooler gas-gap can affect the power requirements and
lifetime of the sorption cooler. Additional margin in this area would have been helpful at little
cost to the mass of the system.

\item Temperature stability requirements should be specified for all spacecraft components and panels not according to their own operating requirements, but rather for their effect on the thermal stability of the cryogenic system.

\item The initial definition of margins in a long cryogenic chain, such as \Planck's, can lead to overspecification if many new systems are considered for which  performance is not yet well established.  When each stage works nominally, the system ends up with a large margin.  It would be better to stretch the range over which the coolers work, if possible, rather than to add an additional stage.

\item Excess cooling capacity provides margin in the operation of the cryo chain, but can lead to decreased stability by unstable evaporation of excess cryogenic fluids (helium or hydrogen).  Enough adjusment of cooling and PID power is needed to cope with the configuration where all coolers are at their best performance.

\item Tests of cryogenic space systems are long and expensive. Precooling loops like the one installed in the dilution cooler to shorten the cooldown time in tests can have serious drawbacks if they become failure hazards (it is difficult to remove the helium in the loop and test it).  Heat switches are another mechanism to optimize cool down time, but they require heat input at the lower temperature end, decreasing the heat lift of the corresponding cooler. This is a drawback which has to be taken into account (see next point).

\item The system definition should carefully consider not only the margins for the planned operating point, but also margins along the entire cooldown path.  Coolers have little power when they are far from their optimal operating temperature, and the system could be stuck at a temperature well above the one at which it can operate.  The \Planck\ \HeJT\ cooler, for example, had its minimum margin when starting to cool down from 20\,K, when the heat switch to lower stages was turned on.

\item Sub-kelvin stages have been found on \Planck\ to be, as expected, very sensitive to heat input by microvibration from mechanical compressors (there is no other source of microvibrations on \Planck). It has not been possible to determine precisely the frequency range responsible for the heating. 

\item An end-to-end thermal model is necessary, but that does not imply making a single unified model containing the full complexity of all stages.  Such a model would be very complicated.  Detailed models of subsystems with empirical interface models to be used in a global model was the philosophy used in \Planck, and it worked well.

\end{itemize}

The \Planck\ thermal architecture is the first implementation in space of a combination of active and passive cooling for a CMB mission.  The architecture allows the simultaneous cooling of the LFI radiometers to 20\,K and the HFI bolometers to 0.1\,K, thereby enabling unprecedented instrumental sensitivity.  The thermal system has operated successfully to date and is expected to continue to perform beyond the nominal mission duration. Thermal fluctuations are present in the various temperature stages, particularly near 20\,K, but these have not compromised the systematic error budget of the instruments. The wide range of scientific discovery enabled by \Planck\ with its unique thermal architecture is evidenced by the accompanying papers describing early scientific results.

\begin{acknowledgements}
    \Planck\ is too large a project to allow full acknowledgement of all contributions by individuals, institutions, industries, and funding agencies. The main entities involved in the mission operations are as follows. The European Space Agency operates the satellite via its Mission Operations Centre located at ESOC (Darmstadt, Germany) and coordinates scientific operations via the \Planck\ Science Office located at ESAC (Madrid, Spain). Two Consortia, comprising around 50 scientific institutes within Europe, the USA, and Canada, and funded by agencies from the participating countries, developed the scientific instruments LFI and HFI, and continue to operate them via Instrument Operations Teams located in Trieste (Italy) and Orsay (France). The Consortia are also responsible for scientific processing of the acquired data. The Consortia are led by the Principal Investigators: J.-L. Puget in France for HFI (funded principally by CNES and CNRS/INSU-IN2P3) and N. Mandolesi in Italy for LFI (funded principally via ASI). NASA's US \Planck\ Project, based at JPL and involving scientists at many US institutions, contributes significantly to the efforts of these two Consortia. A description of the Planck Collaboration and a list of its members, indicating which technical or scientific activities they have been involved in, can be found at 
(\url{http://www.rssd.esa.int/index.php?project=PLANCK&page=Planck_Collaboration}). The Planck Collaboration acknowledges the support of: ESA; CNES and CNRS/INSU-IN2P3-INP (France); ASI, CNR, and INAF (Italy); NASA and DoE (USA); STFC and UKSA (UK); CSIC, MICINN and JA (Spain); Tekes, AoF and CSC (Finland); DLR and MPG (Germany); CSA (Canada); DTU Space (Denmark); SER/SSO (Switzerland); RCN (Norway); SFI (Ireland); FCT/MCTES (Portugal); and DEISA (EU).  
    
We acknowledge the use of thermal models from Thales  for the payload, IAS for the HFI, JPL for the sorption cooler, and Laben for the \hbox{LFI}.  Some of the results in this paper have been derived using the HEALPix package \citep{gorski2005}.  The HFI team wishes to thank warmly the Herschel-Planck project team under the leadership of Thomas Passvogel for their time, effort, and competence in solving the crises following failures of several parts of the cyrochain during Planck system tests.  We acknowledge very useful discussions on the thermal behaviour of Planck during the system tests from the CSL team, who went far beyond their formal responsibilities.

\end{acknowledgements}

\bibliographystyle{aa}

\bibliography{Planck_bib,scs_bib.bib,various_ref.bib,supplement_bib.bib,main_bib.bib,ref_MP.bib,WHcooler_bib.bib}

\raggedright
\end{document}